\documentclass[twocolumn,superscriptaddress,floatfix,letterpaper,nofootinbib]{revtex4-1}

\pdfoutput=1
\allowdisplaybreaks

\usepackage[all]{xy}

\usepackage{euscript}
\newcommand\Rep{\EuScript{R}\mathrm{ep}}
\newcommand\sRep{\mathrm{s}\EuScript{R}\mathrm{ep}}

\newcommand{\w}{{\rm w}}
\newcommand{\Tor}{\text{Tor}}
\newcommand{\Ext}{\text{Ext}}
\newcommand{\Hom}{\text{Hom}}
\newcommand{\Kan}{\text{Kan}}

\newcommand{\gSq}{\mathbb{Sq}}
\newcommand{\Sq}{\text{Sq}}
\newcommand{\Bs}{\beta}
\newcommand{\RZ}{{\mathbb{R}/\mathbb{Z}}}
\newcommand\se[1]{\overset{\scriptscriptstyle #1}{=}}
\newcommand\ase[1]{\overset{\scriptscriptstyle #1}{\approx}}
\newcommand\hcup[1]{\underset{{\scriptscriptstyle #1}}{\smile}}

\newcommand\toZ[1]{\lfloor #1 \rfloor}

\newcommand{\fSPT}{\text{fSPT}}

\begin{document}

\begin{titlepage}

\title{Fermion decoration construction of symmetry protected trivial orders\\
 for fermion systems with any symmetries $G_f$ and in any dimensions}

\author{Tian Lan} 
\affiliation{Institute for Quantum Computing,
  University of Waterloo, Waterloo, Ontario N2L 3G1, Canada}

\author{Chenchang Zhu} 
\affiliation{Mathematics Institute, Georg-August-University,
G\"ottingen, G\"ottingen 37073, Germany}

\author{Xiao-Gang Wen}
\affiliation{Department of Physics, Massachusetts Institute of
Technology, Cambridge, Massachusetts 02139, USA}

\begin{abstract} 
We use higher dimensional bosonization and fermion decoration to construct
exactly soluble interacting fermion models to realize fermionic symmetry
protected trivial (SPT) orders (which are also known as symmetry protected
topological orders) in any dimensions and for generic fermion symmetries $G_f$,
which can be a non-trivial $Z_2^f$ extension (where $Z_2^f$ is the
fermion-number-parity symmetry).  This generalizes the previous results from
group superconhomology of Gu and Wen (arXiv:1201.2648), where $G_f$ is assumed
to be a trivial $Z_2^f$ extension.  We find that the SPT phases from fermion
decoration construction can be described in a compact way using higher groups.

\end{abstract}


\maketitle

\end{titlepage}

{\small \setcounter{tocdepth}{1} \tableofcontents }

\section{Introduction} 

We used to think that different phases of matter all come from spontaneous
symmetry breaking \cite{L3726,L3745}. In last 30 years, we started to realize
that even without symmetry and without symmetry breaking, we can still have
different phases of matter, due to a new type of order -- topological order
\cite{Wrig,WNtop} (\ie patterns of long range entanglement
\cite{KP0604,LW0605,CGW1038}).  

If there is no symmetry breaking nor topological order, it appears that systems
must be in the same trivial phase.  So it was a surprise to find that even
without symmetry breaking and without topological order, systems can still have
distinct phases, which are called Symmetry-Protected Trivial (SPT), or
synonymously, Symmetry-Protected Topological (SPT) phases
\cite{GW0931,CLW1141}.  The realization of the existence of SPT orders and the
fact that there is no topological order in 1+1D \cite{FNW9243,VCL0501} lead to
a classification of all 1+1D gapped phases of bosonic and fermionic systems
with any symmetries \cite{CGW1107,SPC1139,FK1103,CGW1128}, in terms of
projective representations \cite{PBT1039}.  It is the first time, after Landau
symmetry breaking, that a large class of interacting phases are completely
classified.

\begin{table*}[tb]
 \centering
 \begin{tabular}{ |c|l|l|l|p{3.0in}| }
\hline
 $G_f$ & $1+1$D & $2+1$D & $3+1$D & ~~~~~~~~~~~~~~~ Realization  \\
\hline
$Z_2\times Z_2^f$ & $\Z_2$ \blue{[$\Z_2$] ($\Z_2$)} & $\Z_4$ \blue{[$\Z_8$] ($\Z$)} & ~1~ \blue{[~1~] (~1~)} & Double-layer superconductors with layer symmetry \\ 
\hline
$Z_4^f$ & ~1~ \blue{[~?~] (1)} & ~1~ \blue{[~?~] (1)} & ~1~ \blue{[~?~] (~1~)}  &  Charge-$2e$ superconductors
with a $180^\circ$ spin rotation symmetry or charge-$4e$ superconductors \\
\hline
{$Z_2^T\times Z_2^f$} & $\Z_4$ \blue{[$\Z_4$] ($\Z$)} & ~1~ \blue{[~1~] (~1~)} & ~1~  \blue{[~1~] (~1~)} & Charge-$2e$ superconductors with coplanar spin order \\
\hline
{$Z_4^{T,f}$} & $\Z_2$ \blue{[$\Z_2$] ($\Z_2$)} & ~1~ \blue{[$\Z_2$] ($\Z_2$)} & $\Z_4$  \blue{[$\Z_{16}$] ($\Z$)} &  Charge-$2e$ superconductors with spin-orbital coupling  \\
\hline
$(U_1^f \rtimes_\phi Z_4^{T,f})/Z_2$ & ~1~ \blue{[~1~] (1)} & ~1~ \blue{[$\Z_2$] ($\Z_2$)} & $\Z_2^3$  \blue{[$\Z_2^3$] ($\Z_2$)} & Insulator with spin-orbital coupling \\
\hline
$SU_2^f$ & ~1~ \blue{[~1~] (1)} & ~$\Z$~ \blue{[~$\Z$~] ($\Z$)} & ~1~  \blue{[~1~] (~1~)} & Charge-$2e$ spin-singlet superconductor \\
\hline
$Z_2\times Z_4\times Z_2^f$ & $\Z_2^3$ \blue{[?] ($\Z_2^7$)} & {\footnotesize $8\cdot 16$ \blue{[?] ($\Z^{7}$)}} & {\footnotesize $2\cdot 4$  \blue{[$\Z_2\times \Z_4$] (1)}} & \\
\hline
 \end{tabular}
 \caption{
A table for $G_f$-symmetric fermionic SPT orders obtained via the fermion
decoration.\cite{GW1441}  We either list the group that describes the SPT
phases or the number of SPT phases (both including the trivial one).
Those fermionic SPT phases have no topological order, \ie they become trivial
if we break the symmetry down to $Z_2^f$.  The numbers in [~] are results from
spin-cobordism approach \cite{KTT1429,FH160406527,WY180105416}.  The numbers in
(~) are for non-interacting fermionic SPT phases,
\cite{K0986,SRF0825,W11116341}.  Note that the numbers given in
\Ref{K0986,SRF0825,W11116341,KTT1429} are for SIT orders which include both
fermionic SPT orders and invertible fermionic topological orders, while the
above numbers only include fermionic SPT orders.  The number, for example,
$8\cdot 16$ means that the $256$ SPT phases can be divided into 8 classes with
16 SPT phases in each class.  The SPT phases in the same class only differ by
stacking bosonic SPT phases from fermion pairs.  The last column indicates how
to realize those fermionic SPT phases by electronic systems.
}
\label{tbSPT}
\end{table*}

In higher dimensions, the SPT orders, or more generally symmetric invertible
topological (SIT) orders,\footnote{As a state with no topological order, an SPT
order must become trivial if we igore the symmetry.  An SIT order may becomes a
non-trivial invertible topological order if we ignore the symmetry.  An
invertible topological order is a topological order with no non-trivial bulk
topological excitations, but only non-trivial boundary states
\cite{KW1458,F1478,K1467,FH160406527}.} in bosonic systems can be
systematically described by group cohomology theory
\cite{CGL1314,VS1306,W1477}, cobordism theory \cite{K1459,FH160406527}, or
generalized cohomology theory \cite{FH160406527,GJ171207950}.  The SPT and SIT
orders in fermionic systems can be systematically described by group
super-cohomology theory \cite{GW1441,CG150101313,GK150505856,KT170108264,WG170310937}, or
spin cobordism theory \cite{KTT1429,FH160406527,WY180105416}.  In 2+1D, the  SPT orders in
bosonic or fermionic systems can also be systematically classified by the
modular extensions of $\Rep(G_b)$ or $\sRep(G_f)$ \cite{LW160205946}.  Here
$\Rep(G_b)$ is the symmetric fusion category formed by representations of the
boson symmetry $G_b$ where all representations are bosonic, and $\sRep(G_f)$ is
the symmetric fusion category formed by representations of the fermion symmetry
$G_f=Z_2^f\gext G_b$ where the representations with non-zero $Z_2^f$ charge are
fermionic. ($Z_2^f \gext G_b$ denote an extension of $G_b$ by the
fermion-number-parity symmetry $Z_2^f$.)

For SPT orders in fermionic systems, the  modular extension approach in 2+1D
can handle generic fermion symmetry $G_f=Z_2^f \gext G_b$.  However, in higher
dimensions, the group super-cohomology theory can only handle a special form of
fermion symmetry $G_f=Z_2^f\times G_b$.
In this paper, we will develop a more general group super-cohomology theory for
SPT orders of fermion systems based on the decoration
construction\cite{CLV1407} by fermions,\cite{GW1441} which covers generic
fermion symmetry $G_f$ beyond $Z_2^f\times G_b$.  The symmetry group $G_f$ can
also include time reversal symmetry, and in this case, the fermions can be
time-reversal singlet or Kramers doublet.  Our approach works in any
dimensions.  But our theory does not covers the fermionic SPT orders obtained
by decorating  symmetry line-defects\cite{CG150101313,KT170108264,WG170310937}
with Majorana chains (\ie the $p$-wave topological superconducting chains
\cite{K0131}). 

Our theory is constructive in nature. We have constructed exactly soluble local
fermionic path integrals (in the bosonized form) to realize the fermionic SPT
orders systematically.  The simple physical results of this paper is summarized
in Table \ref{tbSPT}.  A mathematical summary of the results is represented
Section \ref{sumB} (and in Section \ref{sum} where more details are given).
However, one needs to use mathematical language of cohomology or higher group
to state the results precisely.

We note that there are seven non-trivial fermionic $(U_1^f \rtimes_\phi
Z_4^{T,f})/Z_2$-SPT phases in 3+1D, while non-interacting fermions only realize
one of them. Other SPT phases are obtained by stacking the bosonic $(U_1
\rtimes_\phi Z_2^{T})$-SPT phases formed by electron-hole pairs.  
%We also find
%a fermionic $Z_2^3\times Z_2^{f})/Z_2$-SPT phase in 3+1D.  Such a fermionic SPT
%phase cannot be realized by non-interacting fermions nor by bosons formed by
%fermion pairs.

\section{Notations and conventions}

Let us first explain some notations used in this paper.  We will use
extensively the mathematical formalism of cochains, coboundaries, and cocycles,
as well as their higher cup product $\hcup{k}$, Steenrod square $\Sq^k$, and
the Bockstrin homomorphism $\Bs_n$. A brief introduction can be found in
Appendix \ref{cochain}.  We will abbreviate the cup product $a\smile b$ as $ab$
by dropping $\smile$.  We will use a symbol with bar, such as $\bar a$ to
denote a cochain on the classifying space $\cB$ of a group or higher group.  We
will use $a$ to denote the corresponding pullback cochain on space-time
$\cM^{d+1}$: $a = \phi^* \bar a$, where $\phi$ is a homomorphism of complexes
$\phi: \cM^{d+1} \to \cB$.  In this paper, when we refer $\RZ$-valued cocycle
or coboundary we really mean $\RZ$-valued almost-cocycle and almost-coboundary
(see Appendix \ref{almco}.

We will use $\se{n}$ to mean equal up to a multiple of $n$, and use $\se{\dd }$
to mean equal up to $\dd f$ (\ie up to a coboundary).  We will use 
$\toZ{x}$ to denote the largest integer smaller than or equal to $x$, and
$\<l,m\>$ to denote the greatest common divisor of $l$ and $m$ ($\<0,m\>\equiv
m$).  

Also, we will use $Z_n=\{1,\ee^{\ii \frac{2\pi}{n}},\ee^{\ii 2
\frac{2\pi}{n}},\cdots,\ee^{\ii (n-1) \frac{2\pi}{n}} \}$ to denote an Abelian
group, where the group multiplication is ``$*$''.  We use
$\Z_n=\{\toZ{-\frac{n}2+1},\toZ{-\frac{n}2+1}+1,\cdots,\toZ{\frac n 2}\}$ to denote an
integer lifting of $Z_n$, where ``+'' is done without mod-$n$.  In this sense,
$\Z_n$ is not a group under ``+''.  But under a modified equality $\se{n}$,
$\Z_n$ is the $Z_n$ group under ``+''.  Similarly, we will use
$\RZ=(-\frac12,\frac12]$ to denote an $\R$-lifting of $U_1$ group.  Under a
modified equality $\se{1}$, $\RZ$ is the $U_1$ group under ``+''.  In this
paper, there are many expressions containing the addition ``+'' of
$\Z_n$-valued or $\RZ$-valued, such as $a^{\Z_n}_1+a^{\Z_n}_2$ where
$a^{\Z_n}_1$ and $a^{\Z_n}_2$ are $\Z_n$-valued.  Those  additions ``+'' are
done without mod $n$ or mod 1.  In this paper, we also have expressions like
$\frac1n a^{\Z_n}_1$.  Such an expression convert a $\Z_n$-valued $a^{\Z_n}_1$
to a $\RZ$-valued $\frac1n a^{\Z_n}_1$, by viewing the $\Z_n$-value as a
$\Z$-value. (In fact, $\Z_n$ is a $\Z$ lifting of $Z_n$.)

We introduced a symbol $\gext$ to construct fiber bundle $X$ from the fiber $F$
and the base space $B$:
\begin{align}
pt\to  F \to X=F\gext B \to B\to pt .
\end{align}
We will also use $\gext$ to construct group extension of $H$ by $N$
\cite{Mor97}:
\begin{align}
1 \to  N \to N\gext_{e_2,\al} H \to H\to 1 .
\end{align}
Here $e_2 \in H^2[H;Z(N)]$ and $Z(N)$ is the center of $N$.  Also $H$ may have
a non-trivial action on $Z(N)$ via $\al: H \to \text{Aut}(N)$.
$e_2$ and $\al$ characterize different group extensions.

We will also use the notion of higher group in some part of the paper. Here we
will treat a $d$-group $\cB(\Pi_1,1;\Pi_2,2;\cdots)$ as a special one-vertex
triangulation of a manifold $K$ that satisfy $\pi_{n}(K)=\Pi_n$ (with unlisted
$\Pi_n$ treated as 0, see \Ref{ZLW} and Appendix \ref{hgroup}).  We see that
$\Pi_1$ is a group and $\Pi_n$, $n\geq 2$, are Abelian groups.  The 1-group
$\cB(G,1)$ is nothing but an one-vertex triangulation of the classifying space
of $G$.  We will abbreviate $\cB(G,1)$ as $\cB G$.  (More precisely, the so
called one-vertex triangulation is actually a simplicial set.)

\section{A brief mathematical summary} \label{sumB}

In this paper, we use a higher dimensional
bosonization\cite{W161201418,KT170108264} to describe local fermion systems in
$d+1$-dimensional space-time via a path integral on a random space-time lattice
(which is called a space-time complex that triangulate the space-time
manifold).  This allows us to construct exactly soluble path integrals on
space-time complexes based on fermion decoration construction\cite{GW1441} to
systematically realize a large class of fermionic SPT orders with a generic
fermion symmetry $G_f = Z_2^f \leftthreetimes G_b$.  This generalizes the
previous result of \Ref{GW1441,CG150101313} that only deal with fermion symmetry of form
$G_f = Z_2^f \times G_b$.  The constructed models are exactly soluble since the
partition functions are invariant under any re-triangulation of the space-time.

The constructed exactly soluble path integrals and the corresponding fermionic
SPT phases are labeled by some data.  Those data can be described in a compact
form using terminology of higher group ${\cal B}(\Pi_1,1;\Pi_2,2;\cdots)$  (see
Appendix \ref{hgroup} for details).  We note that, for a $d$-group
$\cB(G,1;Z_2,d)$ (\ie a complex with only one vertex), its links are labeled by
group elements $g \in G$.  This gives rise to the so called canonical $G$-valued
1-cochain $\bar a$ on the complex $\cB(G,1;Z_2,d)$. On each $d$-simplex in
$\cB(G,1;Z_2,d)$ we also have a $Z_2$ label. This gives us the  canonical
$\Z_2$-valued $d$-cochain $\bar f_d$ on the complex $\cB(G,1;Z_2,d)$.
Now, we can are ready to state our results:
\begin{enumerate}
\item \textbf{Characterization data without time reversal}:
For unitary symmetry $G_f$, the fermionic SPT phases obtained via fermion
decoration are described by a pair $(\vphi,\nu_{d+1})$, where $\vphi:
{\cal B}(G_f\leftthreetimes SO_\infty,1) \to {\cal B}_f(SO_\infty,1;Z_2,d)$ is
a homomorphism between two higher groups
 and $\nu_{d+1}$ is a
$\R/\Z$-valued $d+1$-cochain on ${\cal B}(G_f\leftthreetimes SO_\infty,1)$ that
trivializes the pullback of a $\R/\Z$-valued $d+2$-cocycle $\bar
\omega_{d+2}=\frac12 \Sq^2 \bar f_d + \frac12 \bar f_d \bar {\text{w}}_2 $ on
${\cal B}_f(SO_\infty,1;Z_2,d)$, {\it i.e.} $-\text{d} \nu_{d+1} = \text{d}
\vphi^* \bar \omega_{d+2}$.\\ 
Here $\bar f_d$ is the canonical $d$-cochain on
${\cal B}_f(SO_\infty,1;Z_2,d)$, and $\bar \w_n$ is the $n^\text{th}$
Stiefel-Whitney class constructed from canonical $d$-cochain $\bar a$ on ${\cal
B}_f(SO_\infty,1;Z_2,d)$.  Note that $\bar a$ can be viewed as the $SO_\infty$
connection of a $SO_\infty$ bundle over ${\cal B}_f(SO_\infty,1;Z_2,d)$.
\item \textbf{Characterization data with time reversal}:
In the presence of time reversal symmetry $G_f=G_f^0\leftthreetimes Z_2^T$, we
find that the fermionic SPT phases obtained via fermion decoration are
described by a pair $(\vphi,\nu_{d+1})$, where $\vphi: {\cal
B}(G_f^0\leftthreetimes O_\infty,1) \to {\cal B}_f(O_\infty,1;Z_2,d)$ is a
homomorphism between two higher groups  and $\nu_{d+1}$ is a
$\R/\Z$-valued $d+1$-cochain on ${\cal B}(G_f^0\leftthreetimes O_\infty ,1)$
that trivializes the pullback of a $\R/\Z$-valued $d+2$-cocycle $\bar
\omega_{d+2}=\frac12 \Sq^2 \bar f_d + \frac12 \bar f_d (\bar {\text{w}}_2
+\bar {\text{w}}_1^2) $ on ${\cal B}_f(O_\infty,1;Z_2,d)$, {\it i.e.}
$-\text{d} \nu_{d+1} = \text{d} \vphi^* \bar \omega_{d+2}$.  
\item \textbf{Model construction and SPT invariant}:
Using the data $(\vphi,\nu_{d+1})$, we can write down the explicit
re-triangulation invariant path integral that describes a local fermion model
(in bosonized form) that realized the corresponding SPT phases (see
\eqn{ZfSPTSOhg}, \eqn{ZfSPTSO}, and \eqn{ZfSPTO}).  We can also  write down the
SPT invariant\cite{W1447,HW1339,K1459,W1477,KTT1429}  that characterize the
resulting fermionic SPT phase (see \eqn{SPTinvSO1}, \eqn{SPTinvSO}, and
\eqn{SPTinvO}). Those bosonized fermion path integrals, \eqn{ZfSPTSOhg},
\eqn{ZfSPTSO}, and \eqn{ZfSPTO},  and the corresponding SPT invariants,
\eqn{SPTinvSO1}, \eqn{SPTinvSO}, and \eqn{SPTinvO}, are the main results of
this paper.  
\item \textbf{Equivalence relation}:
Only the pairs $(\vphi,\nu_{d+1})$ that give rise to
distinct SPT invariants correspond to distinct SPT phases.
The pairs $(\vphi,\nu_{d+1})$ that give rise to
the same SPT invariant are regarded as equivalent.
In particular, two homotopically connected $\vphi$'s are equivalent and
two $\nu_{d+1}$'s differ by a coboundary are equivalent.
\end{enumerate}

The data $(\vphi,\nu_{d+1})$ cover all the fermion SPT states obtained
via fermion decoration.  But they do not include the fermion SPT states
obtained via decoration of chains of 1+1D topological $p$-wave superconducting
states, but may include some fermion SPT states obtained via decoration of
sheets of 2+1D topological $p$-wave superconducting state.

\section{Exactly soluble models for bosonic SPT phases} 
\label{bSPT}

Let us first review the construction of exactly soluble models for bosonic SPT
orders with on-site symmetry group $G_b$ \cite{CGL1314}.  The same line of
thinking will also be used in our discussions of fermionic SPT phases.

\begin{figure}[t]
\begin{center}
\includegraphics[scale=0.7]{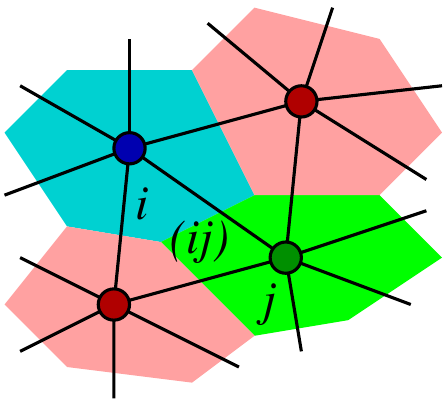} \end{center}
%1
\caption{ (Color online) 
The vertices $i$ and $j$ mark the regions with order parameter
$g_i$ and $g_j$. The link labeled by $(ij)$ connects the  vertices $i$ and $j$.
}
\label{domain}
\end{figure}

\subsection{Constructing path integral}

We start with a phase that breaks the $G_b$ symmetry completely.  Then we
consider the quantum fluctuations of the $G_b$-symmetry-breaking domains which
restore the symmetry.  We mark each domain with a vertex (see Fig.
\ref{domain}), which form a space-time complex $\cM^{d+1}$  (see Appendix
\ref{cochain} for details).  So the quantum fluctuations of the domains are
described by the degrees of freedom given by $g_i \in G_b$ on each vertex $i$.
In other words, the degrees of freedom is a $G_b$-valued 0-cochain field $g \in
C^0(\cM^{d+1};G_b)$.  

In order to probe the SPT orders in a universal way, we can add the symmetry
twist to the system (\ie gauging the symmetry)
\cite{LG1209,HW1339,W1447,K1459}. This is done by adding a fixed $G_b$-valued
1-cochain field $A \in C^1(\cM^{d+1};G_b)$ with $A^{G_b}_{ij}\in G_b$ on each link
$(ij)$ that connect two vertices $i$ and $j$.  The cochain field $A$ satisfy
the flat condition
\begin{align}
\label{flatA}
(\del A^{G_b})_{ijk}\equiv A^{G_b}_{ij}A^{G_b}_{jk}A^{G_b}_{ki}=1, 
\end{align}
where we have assumed $A^{G_b}_{ij}=(A^{G_b}_{ji})^{-1}$.  Those flat 1-cochain
field will be called 1-cocycle field.  The collection of  those $G_b$-valued
1-cocycle fields will be denoted by $ Z^1(\cM^{d+1},G_b)$.  We stress that the
1-cocycle field $A$ is a fixed background field that do not fluctuate.

From the dynamical $g$ and non-dynamical background field $A^{G_b}$, we can construct
an effective dynamic $G_b$-valued 1-cocycle field $a^{G_b}$ whose values on
links are given  by 
\begin{align}
a^{G_b}_{ij}=g_i A^{G_b}_{ij}g_j^{-1}.  
\end{align}
Using such an effective dynamic
field, we can construct our model as
\begin{align}
&Z(\cM^{d+1}, A^{G_b}) 
= 
\sum_{ g \in C^0(\cM^{d+1};{G_b}) } 
\hskip -1em
\ee^{\ii 2\pi \int_{\cM^{d+1}} \om_{d+1}( a^{G_b} )}
,
\end{align}
where $\om_{d+1}( a^{G_b} )$ is $\R/\Z$-valued $d$-cochain: $\om_{d+1}( a^{G_b} ) \in
C^{d+1}(\cM^{d+1};\R/\Z)$, whose value
on a $d+2$-simplex $(i_0i_1\cdots i_{d+1})$ is
a function of
$a^{G_b}_{i_0i_1},a^{G_b}_{i_0i_2},a^{G_b}_{i_1i_2},\cdots$:
$(\om_{d+1})_{i_0i_1\cdots i_{d+1}}= \om_{d+1}(a^{G_b}_{i_0i_1},a^{G_b}_{i_0i_2},a^{G_b}_{i_1i_2},\cdots)$. 
Note that $a^{G_b}$ also satisfies the flat condition
\begin{align}
\label{flat}
(\del a^{G_b})_{ijk}\equiv a^{G_b}_{ij}a^{G_b}_{jk}a^{G_b}_{ki}=1. 
\end{align}
So we say $a^{G_b} \in Z^1(\cM^{d+1};{G_b})$.  In this case the function
$\om_{d+1}(a^{G_b}_{i_0i_1},a^{G_b}_{i_0i_2},a^{G_b}_{i_1i_2},\cdots)$ only
depends on $a^{G_b}_{i_0i_1},a^{G_b}_{i_1i_2},a^{G_b}_{i_2i_3}$, \etc, since
other variables are determined from those variables:
\begin{align}
\label{omdomd}
(\om_{d+1})_{i_0i_1\cdots i_{d+1}} 
&=\om_{d+1}(a^{G_b}_{i_0i_1},a^{G_b}_{i_1i_2},a^{G_b}_{i_2i_3},\cdots).
\end{align}  

We note that the assignment of $a^{G_b}_{ij} \in {G_b}$ on each link $(ij)$ can
be viewed as a map $\phi$ (a homomorphism of complexes) from space-time complex
$\cM^{d+1}$ to $\cB G_b$ which is a simplicial complex that model the
classifying space $B{G_b}$ of the group ${G_b}$.  The $d+1$-cochain $\om_{d+1}(
a^{G_b} )$ can be viewed as a pull back of a cochain $\bar\om_{d+1}( \bar
a^{G_b} )$ in the classifying space $\cB{G_b}$: $\bar\om_{d+1}(\bar  a^{G_b} )
\in C^{d+1}(\cB{G_b};\R/\Z)$.    Here $\bar\om_{d+1}(\bar  a^{G_b} )$ is a
function of the canonical 1-cochain $\bar  a^{G_b}$.  Note that links in
$\cB{G_b}$ are labeled by elements of $G_b$, which give rise to the canonical
1-cochain $\bar  a^{G_b}$ on $\cB{G_b}$ (see \Ref{ZLW} and Appendix \ref{hgroup}).  The above
can be written as 
\begin{align}
 a^{G_b} &= \phi^* \bar  a^{G_b},\ \ \ \
\nonumber\\
 \om_{d+1}( a^{G_b} ) &= \phi^* \bar\om_{d+1}(\bar  a^{G_b} ).
\end{align}

\subsection{Making path integral exactly soluble}
\label{extpath}

To make the model exactly soluble, we require $\om_{d+1}$ to be a pullback of a
cocycle $\bar\om_{d+1}$ in the classifying space $B{G_b}$:
\begin{align}
 \dd \bar\om_{d+1}(\bar a^{G_b} ) \se{1}0 \ \ \text{ or } \ \ 
\bar\om_{d+1}(\bar a^{G_b} ) \in Z^{d+1}(\cB{G_b};\R/\Z).
\end{align}
Why the above condition make our model exactly soluble?  Let us compare two
action amplitudes $\ee^{\ii 2\pi \int_{\cM^{d+1}} \om_{d+1}( a^{G_b} )}$ and
$\ee^{\ii 2\pi \int_{\cM^{d+1}} \om_{d+1}( a^{G_b\prime} )}$ for two different
field values $a^{G_b}$ and $a^{G_b\prime}$.  We like to show that if $a^{G_b}$
and $a^{G_b\prime}$ can homotopically deform into each other, then $\ee^{\ii
2\pi \int_{\cM^{d+1}} \om_{d+1}( a^{G_b} )} =\ee^{\ii 2\pi \int_{\cM^{d+1}}
\om_{d+1}( a^{G_b\prime} )}$.  But $a^{G_b}$ and $a^{G_b\prime}$ are discrete
fields on discrete lattice.  It seems that they can never homotopically deform into
each other, in the usual sense.

To define the homotopical deformation for  discrete fields on discrete lattice,
we try to find a flat connection $\t a^{G_b}$ on a complex $\cM^{d+1}\times I$ in
one higher dimension, such that $\t a^{G_b}=a^{G_b}$ on one boundary of
$\cM^{d+1}\times I$, and $\t a^{G_b}=a^{G_b\prime}$ on the other boundary of
$\cM^{d+1}\times I$.  If such a field $\t a^{G_b}$ exists, then we say $a^{G_b}$
and $a^{G_b\prime}$ can homotopically deform into each other.  In this case, we
find that
\begin{align}
 \frac{\ee^{\ii 2\pi \int_{\cM^{d+1}} \om_{d+1}( a^{G_b\prime} )}}{\ee^{\ii 2\pi \int_{\cM^{d+1}} \om_{d+1}( a^{G_b} )}}
&= \ee^{\ii 2\pi \int_{\cM^{d+1}\times I} \dd \om_{d+1}( \t a^{G_b} )}
\nonumber\\
&= \ee^{\ii 2\pi \int_{\phi(\cM^{d+1}\times I)} \dd \bar\om_{d+1}( \bar a^{G_b} )},
\end{align}
where $\phi$ is a homomorphism from $\cM^{d+1}\times I$ to $\cB G_b$.
Therefore $\ee^{\ii 2\pi \int_{\cM^{d+1}} \om_{d+1}( a^{G_b} )} = \ee^{\ii 2\pi
\int_{\cM^{d+1}} \om_{d+1}( a^{G_b\prime} )}$ if $\dd \bar\om_{d+1}(\bar
a^{G_b}) \se{1}0$ and if $a^{G_b}$ can homotopically deform into $a^{G_b\prime}$
without breaking the flat condition \eqn{flat}.  

Let us view  $a^{G_b}$ and $a^{G_b\prime}$ to be equivalent, if they can
homotopically deform into each other.  This define the ``gauge equivalent''
configurations.
It turns out that two configurations $a^{G_b}$ and $a^{G_b\prime}$ are   equivalent
if and only if they are connected by
a gauge transformation
\begin{align}
a^{G_b\prime}_{ij}
= h_i a^{G_b}_{ij} h_j^{-1}, \ \ \ h_i \in G_b.
\end{align}
Thus the action amplitude $\ee^{\ii 2\pi
\int_{\cM^{d+1}} \om_{d+1}( a^{G_b} )}$ only depends on the equivalent classes
of the field configurations $a^{G_b}$.  Since the homotopical deformations
represent the most but a finite kinds of the fluctuations of $a^{G_b}$, there
is only a finite number of  equivalent classes.  Thus the action amplitude
$\ee^{\ii 2\pi \int_{\cM^{d+1}} \om_{d+1}( a^{G_b} )}$ only dependents on a
finite number of equivalent classes of $a^{G_b}$, and the partition function
$Z(\cM^{d+1})$ can be calculated exactly.  In this case, we say the model is
exactly soluble.  We see that different $\om_{d+1}( a^{G_b} ) \in
\cZ^{d+1}(G_b;\R/\Z)$ characterize different exactly soluble models.  

The space $\cZ^{d+1}(G_b;\R/\Z)$ is not connected.  Each connected piece
describes models in the same phase.  Thus the different phases of the models
are given by $\pi_0[\cZ^{d+1}(G_b;\R/\Z)] = H^{d+1}(\cB G_b,\R/\Z)$.  Those
models do not spontaneously break the ${G_b}$-symmetry.  Without symmetry twist
$A^{G_b}_{ij}=1$, the volume-independent partition function $Z^\text{top}(\cM^{d+1})$
\cite{KW1458,WW180109938} of the above models is $Z^\text{top}(\cM^{d+1})=1$
for any closed oriented manifolds.  So the models have no topological orders,
and realize only SPT orders.  We see that the SPT orders described by those
models are labeled by $H^{d+1}(\cB G_b,\R/\Z)$.

\subsection{Including time reversal symmetry $Z_2^T$} \label{Z2TbSPT}

In the above, we did not include time reversal symmetry described by group
$Z_2^T$.  In the presence of time reversal symmetry the full symmetry group
$G_b$ is an extension of $Z_2^T$: $G_b = G_b^0 \gext  Z_2^T$, where $G_b^0$ is
the unitary on-site symmetry.  In this case, the dynamical variable is still
$g_i \in G_b$, the symmetry twist is still describe by $A^{G_b}_{ij} \in G_b$.  But
the symmetry twist satisfy a constraint: The natural projection $G_b \to Z_2^T$
reduce the $G_b$ connection $A^{G_b}_{ij}$ to a $Z_2^T$ connection $A^T_{ij} \in
Z_2^T$.  $A^T_{ij}$ describes a $Z_2$ bundle over the space-time $\cM^{d+1}$.
The tangent bundle of space-time $\cM^{d+1}$ give rise to a $O_{d+1}$ bundle
over $\cM^{d+1}$.  From the $O_{d+1}$ bundle we can get its determinant
bundle, which is also a $Z_2$ bundle over $\cM^{d+1}$.  Such a $Z_2$ bundle
must be the homotopic equivalent to the $Z_2$ bundle described by the $Z_2^T$
connection $A^T_{ij}$.  In other words, 
\begin{align}
A^T \se{2,\dd} \w_1, 
\end{align}
where $\w_n$ is the $n^\text{th}$ Stiefel-Whitney class of $\cM^{d+1}$.  In
addition to the above constraint, we also require the Lagrangian $ 2\pi
\om_{d+1}(a^{G_b})$ to have a time reversal symmetry. As pointed out in
\Ref{CGL1314}, this can be achieved by requiring $\om_{d+1}(a^{G_b})$ to
be a $\RZ$-valued cocycle where $Z_2^T$ has a non-trivial action on the
value $\RZ \to -\RZ$.

We see that, to include time reversal symmetry, we need to extend a space-time
symmetry, space-time refection $Z_2^T$, by the internal symmetry $G_b^0$ to
obtain the full symmetry group $G_b$.  As shown in \Ref{CGL1314,K1459}, in this
case, the SPT states are labeled by the elements in
$H^{d+1}(\cB G_b,(\R/\Z)_T)$ where time reversal in $G_b$ has a non-trivial
action on the value $T: \R/\Z \to -\R/\Z$.  Also, the differential
operator $\dd$ should be understood as the one with this non-trivial action.

\subsection{Classification of bosonic SPT phases}
\label{CbSPT}

However, $H^{d+1}(\cB G_b,(\R/\Z)_T)$ fail to cover all bosonic SPT orders
\cite{VS1306}.  It misses the SPT orders obtained by decorating \cite{CLV1407}
symmetry defects with the invertible bosonic topological orders
\cite{KW1458,K1459,FH160406527}.  This problem can be fixed if we replace
${G_b}$  by $G_{bO}=G_b^0 \gext  O_\infty$, where $O_n$ is the $n$-dimensional
orthogonal group.  This generalizes the extension $ G_b=G_b^0 \gext  Z_2^T $
discussed in the last section to include time reversal symmetry.  The time
reversal symmetry is included in $ O_\infty$ as the disconnected component,
which is denoted by $Z_2^T$.

After replacing ${G_b}$  by ${G_{bO}}$, we can obtain local bosonic models that
realize more general SPT states, as well as the bosonic SIT orders
\cite{W1477}:
\begin{align}
&Z(\cM^{d+1}) 
= 
\sum_{ g \in C^0(\cM^{d+1};G_{bO}) } 
\hskip -1em
\ee^{\ii 2\pi \int_{\cM^{d+1}} \om_{d+1}( a^{G_{bO}} )}
,
\end{align}
where $(a^{G_{bO}})_{ij}=g_i A^{bO}_{ij}g_j^{-1}$ and $A^{bO}_{ij} \in G_{bO}$.
Also $\om_{d+1}( a^{G_{bO}} )$ is the pullback of $\bar \om_{d+1}( \bar
a^{G_{bO}} ) \in \cZ^{d+1}[\cB G_{bO};(\R/\Z)_T]$.  

We like to stress that the $A^{bO}_{ij}$ is not the most general connection of
a $G_{bO}$ bundle on $\cM^{d+1}$.  We may project $A^{bO}_{ij} \in G_{bO}$ to
$A^{O}_{ij} \in O_\infty$ via the natural projection $ G_{bO} \to O_\infty$.
The resulting $A^{O}_{ij}$ describes a $O_\infty$ bundle on $\cM^{d+1}$, and
such a $O_\infty$ bundle must be the tangent bundle of $\cM^{d+1}$ (extended by
a trivial bundle).

What is the physical meaning of the above statement ``the  resulting
$A^{O}_{ij}$ describes the tangent bundle of $\cM^{d+1}$''?  We can cover the
space-time manifold $\cM^{d+1}$ with meany simple patches.  On each patch,
$(a^{G_{bO}})_{ij}=g_i A^{bO}_{ij} g_j^{-1}$.  On the overlaping region of two
patches, $g_i,A^{bO}_{ij}$ and $g_i',A^{bO\prime}_{ij}$ on the two patches are
related
\begin{align}
 g_i' = g_i h_i, \ 
A^{bO\prime}_{ij} =
h_i^{-1}A^{bO}_{ij}h_j, \  h_i \in G_{bO}=G_b^0 \gext  O_\infty .
\end{align}
We require that the projection $h_i \to h^{O}_i \in O_\infty$ corresponds to
the $O_{d+1}$ space-time rotation.  Thus the field $g_i$ is not a scalar field,
since it transforms non-trivially under the space time rotation $g_i \to g_i
h_i$, $h_i \in G_{bO}=G_b^0 \gext  O_\infty$.  Therefore, requiring that ``the
resulting $A^{O}_{ij}$ describes the tangent bundle of $\cM^{d+1}$'' means that
\emph{The field $g_i$ in the path integral transforms non-trivially under the
space time rotation, as well as the internal symmetry transformation: $g_i\to
g_i h_i$, $h_i \in G_{bO}=G_b^0 \gext  O_\infty$ }.

The different exactly soluble models are labeled by the elements in
$H^{d+1}[\cB G_{bO} ;(\R/\Z)_T]$, where $G_{bO}$ has a non-trivial action on
the value $\RZ$, to ensure the time-reversal symmetry of the Lagrangian
\cite{CGL1314}.  However, since the $O_\infty$ part of $A^{bO}_{ij}$ is only
the connection of tangent bundle of $\cM^{d+1}$, different cohomology classes
may give rise to the same volume-independent partition function for those
limited choices of $A^{bO}_{ij}$. As a result, different cohomology classes in
$H^{d+1}[\cB G_{bO} ;\R/\Z]$ may give rise to the same ${G_b}$-SPT order.
Thus, $H^{d+1}[\cB G_{bO} ;\R/\Z]$ provides a many-to-one label of bosonic SPT
orders in any dimensions and for any on-site symmetries \cite{W1477}.

As pointed out in \Ref{W1477}, adding $O_\infty$ has the same effect as
decorating symmetry membrane-defects with $E_8^3$ quantum Hall states
\cite{VS1306,CLV1407}.  In other words, our models not only contain the
fluctuating field $g_i$, they also contain the fluctuations of membrane object
formed by $E_8^3$ quantum Hall states.  The $E_8$ quantum Hall state is
described by the following 8-layer $K$-matrix wave function \cite{Wtoprev} in
2D space (with coordinate $z=x+\ii y$)
\begin{align}
\Psi(\{z_i^I\})  = 
\prod_{i<j,I} (z_i^I-z_j^I)^{K_{II}}
\prod_{i,j,I<J} (z_i^I-z_j^J)^{K_{IJ}} \ee^{- \sum\frac{ |z_i^I|^2}{4}}
\end{align}
\begin{align}
\label{E8}
K={\footnotesize \begin{pmatrix}
\mathbf{2}&\mathbf{1}&0&0&0&0&0&0\\ 
\mathbf{1}&\mathbf{2}&\mathbf{1}&0&0&0&0&0\\ 
0&\mathbf{1}&\mathbf{2}&\mathbf{1}&0&0&0&0\\ 
0&0&\mathbf{1}&\mathbf{2}&\mathbf{1}&0&0&0\\
0&0&0&\mathbf{1}&\mathbf{2}&\mathbf{1}&0&\mathbf{1}\\ 
0&0&0&0&\mathbf{1}&\mathbf{2}&\mathbf{1}&0\\ 
0&0&0&0&0&\mathbf{1}&\mathbf{2}&0\\ 
0&0&0&0&\mathbf{1}&0&0&\mathbf{2}\\
\end{pmatrix} } \, , 
\end{align}
It has gapless chiral edge excitation with chiral central charge $c=8$.  The
$E_8^3$ membrane is formed by stacking three $E_8$ quantum Hall state together.

Another important question to ask is that whether all the ${G_b}$-SPT states
can be obtained this way.  It is known that $G_{bO}$ cocycle models (or
$H^{d+1}(\cB G_{bO};\R/\Z)$) do not produce all the H-type\cite{KW1458} bosonic
invertible topological orders in 2+1D. But $H^{d+1}(\cB G_{bO} ;\R/\Z)$ may
classify all the L-type\cite{KW1458} SPT phases with ${G_b}$ symmetry in a
many-to-one way.
 
Similarly, if we do not have time reversal symmetry, more general bosonic SPT
states can be constructed by choosing the dynamical variables on each vertex to
be $g_i \in G_{bSO}=G_b\gext SO_\infty$.
In this case, the effective variable in the link
will be $a^{G_{bSO}} \in G_{bSO}$. The corresponding local bosonic models
that can produce more general $G_b$-SPT phases for bosons are given by
\begin{align}
&Z(\cM^{d+1}) 
= 
\sum_{ g \in C^0(\cM^{d+1};G_{bSO}) } 
\hskip -1em
\ee^{\ii 2\pi \int_{\cM^{d+1}} \om_{d+1}( a^{G_{bSO}} )}
,
\end{align}
where $\om_{d+1}( a^{G_{bSO}} )$ is the pullback of $\bar \om_{d+1}(
\bar a^{G_{bSO}} ) \in H^{d+1} (\cB G_{bSO};\R/\Z)$.  The cohomology classes in
$H^{d+1} (\cB G_{bSO};\R/\Z)$ give rise to a many-to-one classification of
bosonic SPT orders which contain decoration of bosonic invertible topological
orders, such as the $E_8^3$ states.

\section{ 
Bosonizantion of fermions in any dimensions with any symmetry
$G_f=Z_2^f \gext G_b$
}
\label{bosonization}

In this section, we will construct exactly soluble models to realize fermionic
SPT phases.  We will use high dimensional bosonization
\cite{W161201418,KT170108264} and use the approach in the last section to
construct exactly soluble path integrals.  Our discussion in this section is
similar to that in \Ref{GK150505856}, but with a generalization at one point,
so that the formalism can be applied to study fermionic SPT phases with generic
fermionic symmetry $G_f=Z_2^f \gext G_b$ beyond $G_f=Z_2^f \times G_b$.

\subsection{3+1D cochain models for fermion}

Let us first construct a cochain model that describes fermion system in 3+1D.
A world line of the fermions is dual to a $\Z_2$-valued 3-cocycle $f_3$.  So we
will use $f_3$ as the field to describe the dynamics of the fermions.  To make
$f_3$ describe fermionic particles, we do a 3+1D statistics
transmutation\cite{W161201418,KT170108264}, \ie by adding a term $\ee^{\ii \pi
\int_{\cM^4} b^2 + b \hcup1 \dd b},\ \dd b=f_3$ in the action amplitude. So our
model has the form
\begin{align}
\label{fermZ}
Z(\cM^4) 
&= 
\hskip -1em
\sum_{ f_3 \in B^3(\cM^4;\Z_2) } 
\hskip -1em
\ee^{\ii 2\pi \int_{\cM^4} \cL( f_3 )+  \frac12 [b^2 + b \hcup1 \dd b]}
\nonumber \\
&= 
\hskip -1em
\sum_{ f_3 \in B^3(\cM^4;\Z_2) } 
\hskip -1em
\ee^{\ii 2\pi \int_{\cM^4} \cL( f_3 )+  \frac12 \gSq^2b}
,
\end{align}
where the path integral is a summation of the coboundaries $f_3 \in
B^3(\cM^4;\Z_2)$.  Here $\cL( f_3 )$ is a $\R/\Z$-valued 4-cochain that depends
on the field $f_3$, which can be viewed as the Lagrangian density of our model.
Different choices of $\cL( f_3 )$ will give rise to different models.  In the
above, $\gSq^2$ defined in \eqn{Sqdef} is a square operation acting on cochains
.  It coincides with the Steenrod square when acting on cohomology classes
\begin{align}
 \gSq^k x = \Sq^k x , \ \ \ \ \ \ \text{ if }  \dd x =0.
\end{align}

The term $\frac12 \gSq^2 b$ is included to make $f_3$ to
describe fermions.  Here $b$ is a 2-cochain that is a function of $f_3$ as
determined by $ \dd b = f_3$.  However, there are many different $b$'s that
satisfy $ \dd b = f_3$. We hope those different $b$ all give rise to the same
action amplitude \cite{GK150505856}.  To see  this, let us change $b$ by a
2-cocycle $b_0\in Z^2(\cM^4;\Z_2)$. We find that (using \eqn{Sqplus})
\begin{align}
\gSq^2 (b+b_0) -\gSq^2 b 
&\se{2,\dd} 
\Sq^2 b_0 \se{2,\dd} (\w_2+\w_1^2) b_0,
\end{align}
where we have used $\Sq^2 b_0 \se{2,\dd} (\w_2+\w_1^2) b_0$ on $\cM^4$ (see
Appendix \ref{Rswc}).  We see that different solutions of $b$ will all give
rise to the same action amplitude, provided that $\cM^4$ is a Pin$^-$ manifold
satisfying $\w_2+\w_1^2\se{2,\dd}0$ (see Appendix~\ref{spinstructure}).

There is another more general way to see why many different $b$'s that satisfy
$ \dd b \se{2} f_3$ give rise to the same action amplitude \cite{GK150505856}.
We note that (see \eqn{Sqd}) 
\begin{align}
 \dd \gSq^2 b\se{2}  \gSq^2 \dd b
\se{2} \Sq^2  f_3 .
\end{align}
So \eqn{fermZ} can be rewritten as
\begin{align}
\label{fermZext}
&Z(\cM^4) 
= 
\hskip -1em
\sum_{ f_3 \in Z^3(\cM^4;\Z_2) } 
\hskip -1em
\ee^{\ii 2\pi \int_{\cM^4} \cL( f_3 )+\ii \pi \int_{\cN^5} \Sq^2  f_3}
,
\end{align}
where $\cN^5$ is an extension of $\cM^4$: $\cM^4=\prt \cN^5$ and we have assumed that
$f_3$ can be extended to $\cN^5$.  The above expression directly depends on
$f_3$. We do not need to solve $b$ to define the action amplitude.  This is the
better way to write 3+1D statistics transmutation.

However, in order for \eqn{fermZext} to define a path integral in 3+1D, the
action amplitude $\ee^{\ii 2\pi \int_{\cM^4} \nu_4( f_3 )+\ii \pi \int_{\cN^5}
\Sq^2  f_3}$ must not depend\cite{GK150505856} on how we extend $f_3$ from
$\cM^4$ to $\cN^5$.  This requires that
\begin{align}
\label{N5cond}
 \int_{\cN^5} \Sq^2f_3 \se{2} 0
\end{align}
for any $f_3 \in Z^3(\cN^5;\Z_2)$ and for any closed $\cN^5$.  We note that on
$\cN^5$, $\Sq^2f_3=(\w_1^2+\w_2) f_3$ (see Appendix \ref{Rswc}).  So the
condition \eqn{N5cond} cannot be satisfied.  To fix this problem, we restrict
$\cN^5$ and $\cM^4$ to be Pin$^-$ manifold where $\w_1^2+\w_2 \se{2,\dd} 0$ (see
Appendix \ref{spinstructure}).  So the fermionic path integral can be defined
on Pin$^-$ manifold \cite{GK150505856} $\cM^4$.  

When  $\cL(f_3)$ respect the time reversal symmetry, we may ask if the fermion
is a time-reversal singlet or a Kramers doublet?  The fact that the path
integral can be defined on Pin$^-$ manifold implies that the fermions are
time-reversal singlet \cite{KTT1429,W161201418}. 

We also like to remark that the two path integrals \eqn{fermZ} and
\eqn{fermZext} are not exactly the same.  In \eqn{fermZ} the summation is
over the coboundaries $f_3 \in B^3(\cM^4;\Z_2)$, while in \eqn{fermZext} the
summation is over the cocycles $f_3 \in Z^3(\cM^4;\Z_2)$, which is what we really
want for a fermion path integral.

\subsection{Bosonization of fermion models in any dimensions}
\label{bfanyD}

The above bosonization of fermion models also works in other dimensions.  In
$d+1$D space-time, the fermion world line is described by $d$-cocycles $f_d \in
Z^d(\cM^{d+1},\Z_2)$, after the Poincar\'e duality.  The bosonized fermion model
is given by
\begin{align}
\label{fermZextD}
&
Z(\cM^{d+1}) 
= 
\hskip -2.5em
\sum_{ f_d \in Z^d(\cM^{d+1};\Z_2) } 
\hskip -2.5em
\ee^{\ii 2\pi \int_{\cM^{d+1}} \cL( f_d,w_2 )+\ii \pi \int_{\cN^{d+2}} 
\Sq^2 f_d + f_d w_2}
.
\end{align}
Here  we have generalized the discussion in \Ref{GK150505856} by including an
extra term $f_d w_2$ where $w_2$ is a fixed $\Z_2$-valued $2$-cocycle
background field  on $\cM^{d+1}$. On $\cN^{d+2}$, $w_2$ is also a $\Z_2$-valued
$2$-cocycle which is an extension of $w_2$ on $\cM^{d+1}$: $w_2\in
Z^2(\cN^{d+2};\Z_2)$.  As we will see later that such a generalization allow us
to study fermionic SPT phases with generic fermionic symmetry $G_f=Z_2^f \gext
G_b$.  

At the moment, we only mention that, in the presence of time reversal
symmetry,  when $w_2=0$, the model is well defined on space-time with a Pin$^-$
structure, \ie $\w_1^2+\w_2 \se{2,\dd} 0$ (see Appendix \ref{spinstructure}),
which means that the fermions are time-reversal singlet \cite{W161201418}.
When  $w_2=\w_1^2$, the model is well defined on space-time with a Pin$^+$
structure, \ie $\w_2 \se{2,\dd} 0$, which means that the fermions are Kramers
doublet \cite{W161201418}.

Let us discuss in detail how the path integral \eqn{fermZextD} is calculated.
First the path integrals is a summation over all fields $f_d$ on a space-time
manifold $\cM^{d+1}$ that satisfy $ \dd f_d \se{2} 0$.  $w_2$ is treated as a
non-dynamical background field on $\cM^{d+1}$ satisfying $\dd w_2 \se{2}0$.  To
evaluate the action amplitude, we choose a $\cN^{d+2}$ such that $\prt \cN^{d+2} =
\cM^{d+1}$. 
Also the field $f_d,\ w_2$ can be extend to $\cN^{d+2}$ so that they still
satisfy $\dd f_d \se{2} 0$ and $\dd w_2 \se{2} 0$ on $\cN^{d+2}$. Therefore,
$\cN^{d+2}$ depends on $f_d, w_2$ on $\cM^{d+1}$. 

However, even after we find the extension $\cN^{d+2}$ that satisfy the above
conditions, still $\int_{\cN^{d+2}} \Sq^2 f_d+f_d w_2$ may not be well defined,
since it may depend on how we extend $f_d, w_2$ into $\cN^{d+2}$.  To see this,
we shrink $\cM^{d+1}$ to a point and consider a closed $\cN^{d+2}$.  In this case,
$f_d,\ w_2$ can be any cocycles on $\cN^{d+2}$, and the integration
$\int_{\cN^{d+2}} \Sq^2 f_d+f_d w_2$ may depend on the choices of  cocycles
$f_d,\ w_2$ that we choose for the extension.

To fix this
problem, we replace $\int_{\cN^{d+2}} \Sq^2 f_d +f_dw_2$ by
\begin{align}
\label{MNw2}
\int_{\cM^{d+1}} f_da^{\Z_2^f}  +
 \int_{\cN^{d+2}} \Sq^2 f_d +f_d(\w_2^N+(\w_1^N)^2)  ,
\end{align}
where $a^{\Z_2^f}$ is the (twisted) spin structure on $\cM^{d+1}$, which is a
$\Z_2$-valued 1-cochain that satisfies
\begin{align}
\dd a^{\Z_2^f} \se{2} \w_2^M+(\w_1^M)^2 +w_2.
\end{align}
Here $\w_i^M$ is the Stiefel-Whitney classes for the tangent bundle of
$\cM^{d+1}$ and $\w_i^N$ is the Stiefel-Whitney classes for the tangent bundle of
$\cN^{d+2}$.  Also, we have assumed that $\cM^{d+1}$ satisfies $\w_2^M+(\w_1^M)^2
+w_2 \se{2,\dd}0$.  If $\cM^{d+1}$ has an extension $\cN^{d+2}$ such that the
twisted  spin structure $a^{\Z_2^f}$ on $\cM^{d+1}$ can be extended to $\cN^{d+2}$, so
that on $\cN^{d+2}$
\begin{align}
\dd a^{\Z_2^f} \se{2} \w_2^N+(\w_1^N)^2 +w_2,
\end{align}
then
\begin{align}
&\ \ \ \ \int_{\cM^{d+1}}f_d a^{\Z_2^f}  +
 \int_{\cN^{d+2}} \Sq^2 f_d +f_d(\w_2^N+(\w_1^N)^2)  
\nonumber\\
&=\int_{\cN^{d+2}} \Sq^2 f_d +  f_dw_2
\end{align}
But even when $\cM^{d+1}$ has an extension $\cN^{d+2}$ that does not have a twisted
spin structure, the expression \eqn{MNw2} is still well defined in the sense
that it does not depend on how we extend $f_d$ into $\cN^{d+2}$. This is
because $\Sq^2 f_d +f_d(\w_2^N+(\w_1^N)^2) $ is always a coboundary on
$\cN^{d+2}$ (see Appendix \ref{Rswc}).

To summarize, the above discussions motivate us to guess that the bosonized
fermion model is more generally given by
\begin{align}
\label{Zf}
Z(\cM^{d+1},a^{\Z_2^f}) 
&= 
\hskip -1.5em
\sum_{ f_d \in Z^d(\cM^{d+1};\Z_2) } 
\hskip -1.5em
\ee^{\ii 2\pi \int_{\cM^{d+1}} \cL( f_d ,w_2)+\frac12 f_d a^{\Z_2^f}}
\nonumber\\
&\ \ \ \  \ \ \ \ \ \ \ \ \
 \ee^{\ii \pi \int_{\cN^{d+2}} \Sq^2 f_d +f_d[\w_2^N+(\w_1^N)^2]  },
\nonumber\\
\dd a^{\Z_2^f} &\se{2} \w_2^M+(\w_1^M)^2 +w_2
.
\end{align}
We see that to make the fermion model well defined, the path integral will
depend on the twisted spin structure of the space-time $\cM^{d+1}$.  The above
expression generalizes the one in \Ref{GK150505856} by including an extra term
$w_2$ in $\dd a^{\Z_2^f}$ which defines a twisted spin structure.  As we will see
later that such a generalization allows us to study fermionic SPT phases with
generic fermionic symmetry $G_f=Z_2^f \gext G_b$.
 
Let us compare two closely related expressions \eqn{fermZextD} and \eqn{Zf}.
In order to make the path integral \eqn{fermZextD} well defined, we need to
choose $\cN^{d+2}$ to satisfy $\w_2^N+(\w_1^N)^2 +w_2 \se{2} 0$.  In this case,
the integrant of $\int_{\cN^{d+2}}$, \ie the term $\Sq^2 f_d + f_dw_2$, is a
coboundary (see Appendix \ref{Rswc}).  In contrast, the path integral \eqn{Zf}
is well defined even when $\cN^{d+2}$ does not satisfy $\w_2^N+(\w_1^N)^2 +w_2
\se{2} 0$, as long as $\cM^{d+1} =\prt \cN^{d+2}$ and $f_d$ can be extended to
$\cN^{d+2}$. This is because the integrant of $\int_{\cN^{d+2}}$, the term $\gSq^2
f_d + f_d[\w_2^N+(\w_1^N)^2]$, is always a coboundary.

We like to remark that in this paper, we will only consider the so called
\emph{regular extension} of closed manifold $\cM^{d+1}$ to $\cN^{d+2}$.  To define
regular extension, we will first extend $\cM^{d+1}$ to $\cM^{d+1}\times I$ and then
extend one boundary of $\cM^{d+1}\times I$ to $\t \cN^{d+2}$.  In other words, we
will glue  one boundary of $\cM^{d+1}\times I$ to the boundary of $\t \cN^{d+2}$
which is $\cM^{d+1}= \prt \t \cN^{d+2}$.  The regular extension $\cN^{d+2}$ is the
union $\cN^{d+2} = \cM^{d+1}\times I \cup \t \cN^{d+2}$.  The regular extension has a
property that $\w_i^N = \w_i^M$ when restricted to the boundary.  So in the
rest of the paper, we will drop the superscripts in $\w_i^N,\ \w_i^M$.

\subsection{Bosonized fermion models with $G_f=Z_2^f \gext G_b$
symmetry in $(d+1)$-dimensions}
\label{decoferm}

In this section, we try to construct models that describe fermionic SPT orders
in d+1D with $G_f$ symmetry.  The fermion symmetry group $G_f=Z_2^f\gext G_b$
is a central extension of $G_b$ by fermion-number-parity symmetry $Z_2^f$.
Such a central extension is characterized by a 2-cocycle $e_2 \in
H^2(\cB G_b;\Z_2)$ (see Appendix \ref{cenext}). So we write, more precisely,
$G_f=Z_2^f\gext_{e_2} G_b$.

To construct fermionic models to realize $G_f$ symmetric SPT phases, we can
first break the boson symmetry $G_b$ completely.  We then consider the domain
fluctuations of the symmetry breaking state to restore the symmetry.  Just like
the bosonic model discussed in the Section \ref{bSPT}, such  domain
fluctuations are described by $g_i \in G_b$ on each vertex.  The fermion
world-lines are described by $d$-cocycle $f_d$ as in the last subsection.
After bosonization, such a fermion system is described by
\begin{align}
\label{fGext}
Z(\cM^{d+1}) 
= 
\hskip -5em
\sum_{ g \in C^0(\cM^{d+1};G_b); f_d \in Z^d(\cM^{d+1};\Z_2)} 
\hskip -5em
&\ee^{\ii 2\pi \int_{\cM^{d+1}} \cL( g, f_d) +\frac12 f_d a^{\Z_2^f}}
\nonumber\\
&\ee^{\ii \pi \int_{\cN^{d+2}} \Sq^2 f_d + f_d(\w_2+\w_1^2) }
,
\end{align}
where $\cN^{d+2}$ is a regular extension of $\cM^{d+1}$: $\cM^{d+1}=\prt \cN^{d+2}$,
and $a^{\Z_2^f}$ is a twisted spin structure
\begin{align}
\label{afw2s2}
 \dd a^{\Z_2^f} \se{2} \w_2 + \w_1^2 + w_2 .
\end{align}

Now let us try to couple the above model to the $G_b$-symmetry twist described
by 1-cocycle $A \in Z^1(\cM^4,G_b)$, which satisfies  
the flat condition
\begin{align}
\label{delA}
& (\del A^{G_b})_{ijk}\equiv A^{G_b}_{ij}A^{G_b}_{jk}A^{G_b}_{ki}=1
\end{align}
In the presence of the background $G_b$-connection, the fermion model becomes
\begin{align}
\label{fGextA}
Z(\cM^{d+1},A^{G_b}) 
= 
\hskip -5em &
\sum_{ g \in C^0(\cM^{d+1};G_b); f_d \in Z^d(\cM^{d+1};\Z_2)}  
\hskip -5em
\ee^{\ii 2\pi \int_{\cM^{d+1}} \cL( g, f_d, A^{G_b}) +\frac12 f_d a^{\Z_2^f}}
\nonumber\\
&\ \ \ \ \ \ \ \ \ \ \ \ \ \ \ \ \ \ \ \ \ \ \ \
\ee^{\ii \pi \int_{\cN^{d+2}} \Sq^2 f_d +f_d(\w_2+\w_1^2)}
,
\end{align}
The Lagrangian $\cL$ is a $\R/\Z$-valued $(d+1)$-cochain in
$C^{d+1}(\cM^{d+1},\R/\Z)$. 

Eqn. \eq{fGextA} is one of the main results of this paper.  It is a
bosonization of fermions with an arbitrary finite symmetry $G_f=Z_2^f \gext
G_b$ in any dimensions.

\section{Exactly soluble models for fermionic SPT phases: 
fermion decoration}
\label{fSPTferm}

Now, we are going to choose the Lagrangian $\cL$ such that the fermion model is
exactly soluble.  We first wrote
\begin{align}
  \int_{\cM^{d+1}} \cL( g, f_d, A^{G_b})=
  \int_{\cN^{d+2}} \dd \cL( g, f_d, A^{G_b}).
\end{align}
The fermion model is exactly soluble
when
\begin{align}
- \dd \cL( g, f_d, A^{G_b}) \se{1} \frac12 ( \Sq^2f_d + f_d w_2)  .
\end{align}
In this case, the action amplitude is always $1$.  But such an equation has no
solutions if we view $g$, $f_d$, and $w_2$ as independent cochains, since in
general $ \frac12 ( \Sq^2f_d + f_d w_2)$ is not a coboundary.

So to obtain an exactly soluble model, we further assume $f_d,\ w_2$ to be
functions of $g,A$:
\begin{align}
 f_d=n_d(g,A^{G_b}),\ \ \ \ \ \
 w_2=e_2(g,A^{G_b}).
\end{align}
This process is called decorating symmetry point-defects  (described by
$(g,A^{G_b})$) with fermion particles (described by $f_d$) \cite{GW1441}, or
simply fermion decoration.  This is also called trivializing the cocycle
$\Sq^2f_d + f_d w_2$ (see Section \ref{sum}).

We require
$n_d(g,A^{G_b})$ and $e_2(g,A^{G_b})$ to be $G_b$-gauge invariant
\begin{align}
& n_d(g_i,A^{G_b}_{ij})  = n_d(g_ih_i, h_i^{-1}A^{G_b}_{ij}h_j).
\nonumber\\
& e_2(g_i,A^{G_b}_{ij})  = e_2(g_ih_i, h_i^{-1}A^{G_b}_{ij}h_j).
\end{align}
We can use such a gauge transformation to set $g_i=1$:
$n_d(g_i,A^{G_b}_{ij})=n_d(1,a^{G_b}_{ij})$ 
and $e_2(g_i,A^{G_b}_{ij})=e_2(1,a^{G_b}_{ij})$ 
where
\begin{align}
 a^{G_b}_{ij} = g_iA^{G_b}_{ij}g_j^{-1}.
\end{align}
Thus $n_d$ and $e_2$ are cocycles $n_d \in H^d(\cB G_b,\Z_2)$ and $e_2 \in
H^2(\cB G_b,\Z_2)$.  Similarly, we can use the gauge transformation to set
$g_i=1$ in $\cL( g,A^{G_b}) = \cL( 1,a^{G_b}) \equiv \nu_{d+1}(
a^{G_b})$.  The exact solubility condition becomes
\begin{align}
\label{domSqnd}
 -\dd \nu_{d+1}(a^{G_b}) 
\se{1} 
\frac12 \Big( \Sq^2[n_d(a^{G_b})] + n_d(a^{G_b})e_2(a^{G_b})  \Big)
.
\end{align}
where $\nu_{d+1} \in C^{d+1}(\cB G_b;\R/\Z)$ is a cochain on $\cB G_b$.  With
proper choices of $n_d(a^{G_b})$ and $e_2(a^{G_b})$, $\Sq^2[n_d(a^{G_b})] +
e_2(a^{G_b}) n_d(a^{G_b})$ can be a coboundary and the above equation has
solutions.  The above is nothing but the twisted cocycle condition for group
super-cohomology first derived by Gu and Wen in \Ref{GW1441} (for the case
$e_2=0$ and $d=2,3$).

This way we obtain an exactly soluble local fermionic model
\begin{align}
\label{ZfSPT}
Z(\cM^{d+1},A^{G_b},A^{\Z_2^f})
&= 
\hskip -8em
\sum_{ 
\ \ \ \ \ \ \ \ \ \  \ \ \ \ \ \ \  \ \ \ \ \
g \in C^0(\cM^{d+1};G_b); f_d =n_d(a^{G_b})} 
\hskip -8em
\ee^{\ii 2\pi \int_{\cM^{d+1}} \nu_{d+1}( a^{G_b} ) +\frac12  f_d A^{\Z_2^f}}
\nonumber\\
& \ \ \ \ \ \ \ \ \ \
\ee^{\ii \pi \int_{\cN^{d+2}} \Sq^2 f_d + f_d(\w_2+\w_1^2) } ,
\nonumber\\
\dd A^{\Z_2^f} &\se{2} \w_2+\w_1^2 +e_2(A^{G_b})
.
\end{align}
The path integral sums over the $G^b$-valued 0-cochains $g$ and $\Z_2$-valued
$d$-cocycles $f_d$ on $\cM^{d+1}$, that satisfy $f_d =n_d(a^{G_b})$.  On
$\cN^{d+2}$, $f_d$ is a $\Z_2$-valued $d$-cocycles that satisfy $f_d
=n_d(a^{G_b})$ on the boundary.  In other words, $f_d $ may not be equal to
$n_d(a^{G_b})$ on $\cN^{d+2}$, where $n_d(a^{G_b})$ is not even defined.  The
$f_d =n_d(a^{G_b})$ condition on the boundary can be imposed by an energy
penalty term in the Lagrangian.  Thus $g$ and $f_d$ on $\cM^{d+1}$ are the
dynamical fields of the above local fermionic lattice model (in the bosonized
form).

The above model is well defined only if $\cM^{d+1}$ and $A$ satisfy
\begin{align}
 \w_2+\w_1^2 + e_2( A^{G_b}) 
 \se{2,\dd} 0 
,
\end{align}
so that the twisted spin structure $a^{\Z_2^f}$ can be defined.
This implies that the fermion in our model 
is described by a representation of $G_f=Z_2 \gext_{e_2} G_b$, where $G_f$ is 
a $Z_2^f$ extension of $G_b$ as determined by the 2-cocycle $e_2 \in
H^2(\cB G_b;\Z_2)$ (see \Ref{ZLW}).  

Eqn. \eq{ZfSPT} is the first main result of this paper:  \frmbox{Eqn.~\eq{ZfSPT}
describes a local fermionic system (in a bosonized form) where the full fermion
symmetry is $G_f=Z_2^f \gext_{e_2} G_b$. Such a fermionic model realizes a
fermionic SPT state obtained by fermion decoration construction.} The above
generalizes the previous results of \Ref{GW1441,GK150505856} from $G_f=Z_2^f
\times G_b$ to $G_f=Z_2^f\gext_{e_2} G_b$.

The exactly soluble model \eq{ZfSPT} systematically realizes a large class of
fermionic SPT phases with any on-site symmetry $G_f$ and in any dimensions.
This class of fermionic SPT phases is described by the data
(now written more precisely in terms of cochains on $\cB G_b$)
\begin{align}
\label{fSPTcl}
\bar e_2(\bar a^{G_b}_{01},\bar a^{G_b}_{12}) & \in Z^2(\cB G_b;\Z_2) ,
\nonumber\\
 \bar n_d(\bar a^{G_b}_{01},\cdots,\bar a^{G_b}_{d-1,d}) & \in Z^d(\cB G_b;\Z_2) ,
\\
 \bar \nu_{d+1}(\bar a^{G_b}_{01},\cdots,\bar a^{G_b}_{d,d+1})  
& \in C^{d+1}(\cB G_b;\RZ) ,
\nonumber\\
 -\dd \bar \nu_{d+1}(\bar a^{G_b}) & \se{1} \frac12 [ \Sq^2 \bar n_d(\bar a^{G_b}) +\bar n_d(\bar a^{G_b}) \bar e_2(\bar a^{G_b}) ] .
\nonumber 
\end{align}
$e_2$, $n_d$ and $\nu_{d+1}$ in \eqn{ZfSPT} are pullbacks of $\bar e_2$, $\bar
n_d$ and $\bar \nu_{d+1}$ in \eqn{fSPTcl}
by the homomorphism 
$\phi: \cM^{d+1} \to \cB G_b$:
\begin{align}
 e_2 \se{2} \phi^* \bar e_2, \ \
 n_d \se{2} \phi^* \bar n_d, \ \
 \nu_{d+1} \se{2} \phi^* \bar \nu_{d+1} .
\end{align}

For a fixed fermion symmetry $G_f$, the constructed SPT states are labeled by
$[ \bar n_d(\bar a^{G_b}), \bar \nu_{d+1}(\bar a^{G_b})]$ (where $\bar e_2$ is
fixed).  However, different pairs $[ \bar n_d(\bar a^{G_b}), \bar
\nu_{d+1}(\bar a^{G_b})]$ can some times label the same fermionic SPT states.
Those pairs that label the same SPT state are called equivalent.  The
equivalence relations are partially generated by the following two kinds of
transformations:
\begin{enumerate}
\item
 a transformation generated by a $d$-cochain 
$\bar \eta_d\in C^{d}(\cB G_b;\RZ)$:
\begin{align}
\label{gaugetom1}
  \bar n_d(\bar a^{G_b}) &\to \bar n_d(\bar a^{G_b}) ,
\nonumber\\
  \bar \nu_{d+1}(\bar a^{G_b}) &\to \bar \nu_{d+1}(\bar a^{G_b}) +\dd \bar \eta_d(\bar a^{G_b})   .
\end{align}
\item
 a transformation generated by a $(d-1)$-cochain $ \bar u_{d-1} 
\in C^{d-1}(\cB G_b;\Z_2)$ 
\begin{align}
\label{gaugend1}
  \bar n_d & (\bar a^{G_b}) \to \bar n_d(\bar a^{G_b}) +\dd \bar u_{d-1}(\bar a^{G_b}) ,
\nonumber \\
  \bar \nu_{d+1} & (\bar a^{G_b}) \to \bar \nu_{d+1}(\bar a^{G_b}) 
+ \frac{1}{2} \gSq^2  \bar u_{d-1}(\bar a^{G_b})  
\\
& 
+ \frac{1}{2} \bar u_{d-1}(\bar a^{G_b}) \hcup{d-2}  \bar n_d(\bar a^{G_b}) 
+ \frac12 \bar u_{d-1}(\bar a^{G_b}) \bar e_2(\bar a^{G_b}) 
.
\nonumber 
\end{align}
\end{enumerate}
More detailed discussions are given in the Section \ref{sum}. 
%We note that the
%above equivalent relations do not contain the one that change $e_2$ by a
%coboundary. The reason is explained in Appendix \ref{e2}.

\section{A more general construction for fermionic SPT states}

\subsection{With time reversal symmetry}

The action amplitude 
\begin{align}
\ee^{\ii 2\pi \int_{\cM^{d+1}} \cL( g, f_d, A^{G_b})
+\frac12 f_d a^{\Z_2^f}} \ee^{\ii \pi \int_{\cN^{d+2}} \Sq^2 f_d
+f_d(\w_2+\w_1^2)} 
\end{align}
in \eqn{fGextA} contains bosonic field $g$ and fermionic field $f_d$.  The
bosonic field couples to a $G_b$-connection $A^{G_b}$ and the bosonic field
couples to a spin connection $a^{\Z_2^f}$ (for the twisted spin structure).
More generally, the action amplitude has the following gauge invariance
\begin{align}
 \w_2+\w_1^2 \to \w_2+\w_1^2 + \dd u_1,\ \ \ \
a^{\Z_2^f} \to a^{\Z_2^f} + u_1,
\end{align}
where $u_1$ is a $\Z_2$ valued 1-cochain.
Such a gauge invariance ensure the proper coupling between the fermion current
and the spin connection.

To construct more general fermionic SPT states and to include time reversal
symmetry, like what we did in Section \ref{CbSPT}, we generalize $g$ and $f_d$
and allow them to couple to $A^{G_b}$ and $a^{\Z_2^f}$ as well as the
space-time connection $A^O \in O_\infty$.  We can package the three connections
$A^{G_b}_{ij}$, $a^{\Z_2^f}_{ij}$, and $A^O_{ij}$ into a single connection
$A^{G_{fO}}_{ij} \in G_{fO}$ where
\begin{align}
G_{fO}=G_f^0 \gext O_\infty.  
\end{align}
Here full fermion symmetry group is given by 
\begin{align}
G_f=G_f^0\gext Z_2^T .  
\end{align}
\ie $G_f^0$ is the fermion symmetry with time reversal removed.

So a more general action amplitude can have the following form
\begin{align}
\ee^{\ii 2\pi \int_{\cM^{d+1}} \cL( g, f_d, A^{G_{fO}}) } 
\ee^{\ii \pi \int_{\cN^{d+2}} \Sq^2 f_d
+f_d(\w_2+\w_1^2)}, 
\end{align}
where we may choose the field $g$ to have its value in $G_{fO}$.  Now $g_i$
transforms non-trivially under space-time transformation.  More precisely,
under space time transformation $O_{d+1}$, $g_i$ transforms as
\begin{align}
 g_i \to g_i h,\ \ \ h\in G_{fO}.
\end{align}
where $h$ has a property that under the natural projection $
G_{fO}\xrightarrow{\pi} G_{fO}/G_f^0 =O_\infty$, $h$ becomes $h^O =\pi(h) \in
O_\infty$, such that $h^O$ is in the $O_{d+1}$ subgroup of $O_\infty$.  The
symmetry twist is now described by $A^{G_{fO}}_{ij}\in G_{fO}$ on each link
$(ij)$, such that, under the projection $\pi$, $A^{G_{fO}}_{ij}$ become
$A^O_{ij}=\pi(A^{G_{fO}}_{ij}) \in O_{d+1} \subset O_\infty$ and $A^O_{ij}$ is
the connection that describe the tangent bundle of the space-time $M^{d+1}$.

The above action amplitude should be invariant under the following gauge
transformation
\begin{align}
 \w_2+\w_1^2 \to \w_2+\w_1^2 + \dd u_1,\ \ \ \
A^{G_{fO}}_{ij} \to A^{G_{fO}}_{ij}(-)^{(u_1)_{ij}}
\end{align}
It also has the following gauge invariance
\begin{align}
 g_i \to g_i h_i,\ \ \ \
A^{G_{fO}}_{ij} \to h_i^{-1} A^{G_{fO}}_{ij} h_j,\ \ \
g_i,h_i \in G_{fO}.
\end{align}
This leads to a more general bosonized fermion theory
\begin{align}
&\ \ \ \
Z(\cM^{d+1},A^{G_{fO}}) 
\\
&= 
\hskip -11em 
\sum_{ 
\ \ \ \ \ \ \ \ \ \ \ \ \ \ \ \ \ \ \ \ \ \ \ \ \ \ \ \ \ \ \
g \in C^0(\cM^{d+1};G_{fO}); f_d \in Z^d(\cM^{d+1};\Z_2)}  
\hskip -11em
\ee^{\ii 2\pi \int_{\cM^{d+1}} \cL( g, f_d, A^{G_{fO}}) }
\ee^{\ii \pi \int_{\cN^{d+2}} \Sq^2 f_d +f_d(\w_2+\w_1^2)}
.
\nonumber 
\end{align}
This allows us to introduce effective dynamical variables on the links given by
\begin{align}
 a^{G_{fO}}_{ij}=g_i A^{G_{fO}}_{ij} g_j^{-1},\ \ \ \ g_i \in G_{fO}.
\end{align}

Using the effective dynamical variables we can construct a local fermionic
exactly soluble models (in the bosonized form)
that describes the fermion decoration
\begin{align}
\label{ZfSPTO}
&\ \ \ \
Z(\cM^{d+1},A^{G_{fO}})
\\
&= 
\hskip -9em
\sum_{ 
\ \ \ \ \ \ \ \ \ \  \ \ \ \ \ \ \  \ \ \ \ \
g \in C^0(\cM^{d+1};G_{fO}); f_d =n_d(a^{G_{fO}})} 
\hskip -8em
\ee^{\ii 2\pi \int_{\cM^{d+1}} \nu_{d+1}( a^{G_{fO}} ) }
\ee^{\ii \pi \int_{\cN^{d+2}} \Sq^2 f_d + f_d(\w_2+\w_1^2) } .
\nonumber 
\end{align}
where $\cM^{d+1}$ is the boundary of $\cN^{d+2}$, $f_d$ on $\cN^{d+2}$ satisfy
$\dd f_d \se{2} 0$, and $f_d$ on $\cN^{d+2}$ is an extension of $f_d$ on
$\cM^{d+1}$.  The above model can realize more general fermionic SPT states,
which are constructed using the following data:
\begin{align}
\label{fSPTclO}
\bar n_d(\bar a^{G_{fO}}) &\in Z^d(\cB G_{fO};\Z_2);
\nonumber\\
\bar \nu_{d+1}(\bar a^{G_{fO}}) &\in C^{d+1}(\cB G_{fO};\R/\Z);
\nonumber\\
-\dd \bar \nu_{d+1}( \bar a^{G_{fO}} ) & \se{1} \frac12 \Sq^2 \bar n_d( \bar a^{G_{fO}} ) 
\nonumber\\
& + \frac12 \bar n_d( \bar a^{G_{fO}} ) [\bar \w_2(\bar a^O)+\bar \w_1^2(\bar a^O)]
\end{align}
where $\bar a^O_{ij} =\pi(\bar a^{G_{fO}}_{ij}) \in O_\infty$, and
$n_d,\ \nu_{d+1}$ in \eqn{ZfSPTO}
are the pullbacks of $\bar n_d,\ \bar \nu_{d+1}$.
Here we like to stress
that the time reversal transformation has a non-trivial action on the value of
$\bar \nu_{d+1}$: $\bar \nu_{d+1} \stackrel{T}{\to} - \bar \nu_{d+1}$.  Thus $\dd$ is
defined with this action and should be more precisely written as $\dd_{\w_1}$
(see Appendix \ref{cochain}).

Eqn. \eq{ZfSPTO} is the second main result of this paper:  \frmbox{For fermion
systems with full fermion symmetry $G_f=Z_2^f \gext G_b$ where $G_b$ contains
time reversal symmetry, the fermions transform as representations of
$G_{fO_{d+1}}=G_f^0\gext O_{d+1}$ (for imaginary time), under combined
space-time rotation and internal symmetry transformation.  Eqn. \eq{ZfSPTO} is
an exactly soluble model for such a fermionic system (in a bosonized form).
Such a fermionic model realizes a fermionic SPT state, labeled by a pair $[\bar
n_d( \bar a^{G_{fO}} ), \bar \nu_{d+1}( \bar a^{G_{fO}} )]$ (see \eqn{fSPTclO}).} Here we would
like to stress that \frmbox{to describe the symmetry of a fermion system, it is
not only important to specify the full fermion symmetry group $G_f$, it is also
important to specify the group $G_{fO_{d+1}}=G_f\gext O_{d+1}$ for the combined
space-time rotation and internal symmetry transformation.} Our model for
fermionic SPT state uses the information on how fermions transform under the
combined space-time rotation and internal symmetry transformation.

\subsection{Without time reversal symmetry}

If there is no time reversal symmetry, we can choose the dynamical
variable on each vertex to be $g_i \in G_{fSO} \equiv G_f \gext
SO_\infty$.  
Under space time transformation $SO_{d+1}$,
$g_i$ transforms as
\begin{align}
 g_i \to g_i h,\ \ \ h\in G_{fSO}.
\end{align}
$h$ also satisfy that under the natural projection $\pi: G_{fSO}\to G_{fSO}/G_f
=SO_\infty$, $h$ become $h^{SO}=\pi(h) \in SO_{d+1} \subset SO_\infty$.

The symmetry twist now is described by $A^{G_{fSO}}_{ij}\in G_{fSO}$ on each
link $(ij)$, such that $A^{SO}_{ij}=\pi(A^{G_{fSO}}_{ij}) \in SO_{d+1} \subset
SO_\infty$ and $A^{SO}_{ij}$ is the connection that describe the tangent bundle
of the space-time $M^{d+1}$.  The effective dynamical variables on the links
are given by
\begin{align}
\label{gAgSO}
 a^{G_{fSO}}_{ij}=g_i A^{G_{fSO}}_{ij} g_j^{-1}.
\end{align}

Using the effective dynamical variables we can construct a local fermionic
exactly soluble models (in the bosonized form)
\begin{align}
\label{ZfSPTSO}
&\ \ \ \
Z(\cM^{d+1},A^{G_{fSO}})
\\
&= 
\hskip -9em
\sum_{ 
\ \ \ \ \ \ \ \ \ \  \ \ \ \ \ \ \  \ \ \ \ \
g \in C^0(\cM^{d+1};G_{fSO}); f_d =n_d(a^{G_{fSO}})} 
\hskip -8em
\ee^{\ii 2\pi \int_{\cM^{d+1}} \nu_{d+1}( a^{G_{fSO}} ) }
\ee^{\ii \pi \int_{\cN^{d+2}} \Sq^2 f_d + f_d \w_2} .
\nonumber 
\end{align}
where $\cM^{d+1}$ is orientable and is the regular boundary of  orientable
$\cN^{d+2}$ and $\w_2$ is the second Stiefel-Whitney class on $\cN^{d+2}$.  The
above models are constructed using the following data:
\begin{align}
\label{fSPTclSO}
\bar n_d(\bar a^{G_{fSO}}) &\in Z^d(\cB G_{fSO};\Z_2),
\nonumber\\
\bar \nu_{d+1}(\bar a^{G_{fO}}) &\in C^{d+1}(\cB G_{fSO};\R/\Z);
\nonumber\\
-\dd \bar \nu_{d+1}( \bar a^{G_{fSO}} ) & \se{1} \frac12 \Sq^2 \bar n_d( \bar a^{G_{fSO}} ) 
\nonumber\\
& + \frac12 \bar n_d( \bar a^{G_{fSO}} ) \bar \w_2(\bar a^{SO}) ,
\end{align}
where $\bar a^{SO}_{ij} \in SO_\infty$ is obtained from $\bar a^{G_{fSO}}_{ij}$
by the natural projection $G_{fSO}\to SO_\infty$.
Again $n_d$ and $\nu_{d+1}$ in \eqn{ZfSPTSO}
are the pullbacks of $\bar n_d$ and $\bar \nu_{d+1}$.

Eqn. \eq{ZfSPTSO} is the third main result of this paper: \frmbox{For fermion
systems with full fermion symmetry $G_f=Z_2^f \gext G_b$ where $G_b$ contains
no time reversal symmetry, the fermions transform as representations of
$G_{fSO_{d+1}}=G_f\gext SO_{d+1}$ (for imaginary time), under combined
space-time rotation and internal symmetry transformation.  Eqn. \eq{ZfSPTSO} is
an exactly soluble model for such a fermionic system (in a bosonized form).
Such a fermionic model realizes a fermionic SPT state, labeled by a pair $[\bar
n_d( \bar a^{G_{fSO}} ), \bar \nu_{d+1}( \bar a^{G_{fSO}} )]$ (see
\eqn{fSPTclSO}).}

\section{Summary in terms of higher groups}
\label{sum}

The data \eq{fSPTclSO} that characterizes the exactly soluble model \eq{ZfSPTSO}
has a higher group description.  The higher group description is more compact,
and allow us to see the equivalence relation between the data more clearly.
We will first state this higher group description, and explain it later.

\subsection{Without time reversal symmetry}

The exactly soluble model \eq{ZfSPTSO} and the related fermionic SPT state
is characterized by the following data:
\begin{enumerate}
\item
A particular higher group $\cB_f(SO_\infty,1;Z_2,d)$, determined by its
$\Z_2$-valued canonical cochain $\dd \bar f_d \se{2}0$ (see \Ref{ZLW} and
Appendix \ref{hgroup}).
\item
A particular $\RZ$-valued $(d+2)$-cocycle 
\begin{align}
\label{bomd2}
\bar \om_{d+2}\se{1} \frac12 \Sq^2 \bar f_d +
\frac12 \bar f_d \bar \w_2(\bar a^{SO})
\end{align}
on the higher group $\cB_f(SO_\infty,1;Z_2,d)$.
\item
Different trivialization homomorphisms $ \vphi: \cB G_{fSO} \to
\cB_f(SO_\infty,1;Z_2,d)$, where $G_{fSO}=G_f\gext SO_\infty$.
($\vphi$ corresponds to $n_d$ in \eqn{fSPTclSO}.)
\item
Different choices of the trivialization $\bar\nu_{d+1}(\bar a^{G_{fSO}})$ that
satisfy $-\dd \bar\nu_{d+1}(\bar a^{G_{fSO}}) \se{1} 
\vphi^* \bar \om_{d+2}(\bar f_d,\bar a^{SO}) $.  
\end{enumerate}

To understand the above result, we note that the first two pieces of data
determine the term on $\cN^{d+2}$: $\ee^{\ii \pi \int_{\cN^{d+2}} \Sq^2 f_d +
f_d \w_2}$, which is always fixed.  In fact, $\frac12 \Sq^2 f_d + \frac12 f_d
\w_2$ can be viewed as the pullback of $\bar\om_{d+2}$ by a simplicial 
homomorphism $\phi_N: \cN^{d+2} \to  \cB_f(SO_\infty,1;Z_2,d)$:
\begin{align}
 \frac12 \Sq^2 f_d + \frac12 f_d \w_2(a^{SO}) = 
\phi_N^* \bar\om_{d+2}(\bar f_d, \bar a^{SO}).
\end{align}
Here $a^{SO}$ is the pullback of $ \bar a^{SO}$: $a^{SO} = \phi_N^* \bar
a^{SO}$.  We require $\phi_N$ to be a  homomorphism such that $a^{SO} =
\phi_N^* \bar a^{SO}$ is the connection that describes the tangent bundle of
$\cN^{d+2}$.

The next two pieces of data determine the term on $\cM^{d+1}$: $\ee^{\ii 2\pi
\int_{\cM^{d+1}} \nu_{d+1}( a^{G_{fSO}} ) }$. First, $ n_d( 
a^{G_{fSO}} )$  and $\w_2(a^{SO})$ in \eqn{fSPTclSO} are the pullback
of $\bar f_d$ and $\bar\w_2(\bar a^{SO})$ on $\cB_f(SO_\infty,1;Z_2,d)$ by the
trivialization homomorphism $\vphi: \cB G_{fSO} \to \cB_f(SO_\infty,1;Z_2,d)$.
(To be more precise, $ n_d( a^{G_{fSO}} )$  and $ \w_2( a^{SO})$ in
\eqn{fSPTclSO} are pullbacks of $\bar n_d( \bar a^{G_{fSO}} )$  and $\bar
\w_2(\bar a^{SO})$ by $\phi_M: \cM^{d+1}\to \cB G_{fSO} $.) Thus
\begin{align}
&\ \ \
\frac12 \Sq^2 \bar n_d( \bar a^{G_{fSO}} ) + \frac12 \bar n_d( \bar a^{G_{fSO}} )\bar  \w_2(\bar a^{SO})
\nonumber\\
&=\vphi^*[\frac12\Sq^2\bar f_d+\frac12\bar f_d\bar\w_2(\bar a^{SO})]
\end{align}
We also require $\vphi^*[\frac12\Sq^2\bar f_d+\frac12\bar f_d\bar\w_2(\bar
a^{G_{SO}})]$ to be a coboundary on $\cB G_{fSO}$. \ie there is a cochain
$  \bar \nu_{d+1}( \bar a^{G_{fSO}} )$ on $\cB G_{fSO}$ such that
\begin{align}
-\dd \bar \nu_{d+1}( \bar a^{G_{fSO}} ) & \se{1} \frac12 \Sq^2 \bar n_d( \bar a^{G_{fSO}} ) 
\nonumber\\ & 
+ \frac12 \bar n_d( \bar a^{G_{fSO}} ) \bar\w_2(\bar a^{SO})
\end{align}

Therefore, the exactly soluble models \eq{ZfSPTSO} and the corresponding
fermionic SPT states are characterized by a pair $(\vphi, \bar\nu_{d+1})$, a
trivialization homomorphism and a trivialization.  The different trivialization
homomorphisms $\vphi$ correspond to different choices of $n_d(a^{G_{fSO}})$.
The different trivializations $\bar\nu_{d+1}$ differ by $d+1$-cocycles on $\cB
G_{fSO}$.

In fact, the exactly soluble models \eq{ZfSPTSO} can be written explicitly
using the higher group data:
\begin{align}
\label{ZfSPTSOhg}
Z(\cM^{d+1},A^{G_{fSO}})
&= 
\sum_{\phi_M} 
\ee^{\ii 2\pi \big(\int_{\cM^{d+1}} \phi_M^* \bar \nu_{d+1} 
+\int_{\cN^{d+2}} \phi_N^* \bar \om_{d+2} \big)} ,
\nonumber\\
& -\dd \bar \nu_{d+1} \se{1} \vphi^* \bar \om_{d+2},
\end{align}
Here $\phi_M$ is a simplicial homomorphism $\phi_M: \cM^{d+1} \to \cB G_{fSO}$,
such that $\phi_M^* \bar a^{G_{fSO}}$ is gauge equivalent to $A^{G_{fSO}}$:
\begin{align}
 \bar a^{G_{fSO}}_{ij} = g_i A^{G_{fSO}}_{ij} g_j^{-1}.
\end{align}
Also, $\phi_N$ is a simplicial homomorphism $\phi_N: \cN^{d+2} \to \cB
(G_{fSO},1;Z_2,d)$, such that,  when restrict to the boundary of $\cN^{d+2}$,
$\phi_N= \vphi \phi_M$.

\begin{figure}[t]
\begin{center}
\includegraphics[scale=0.5]{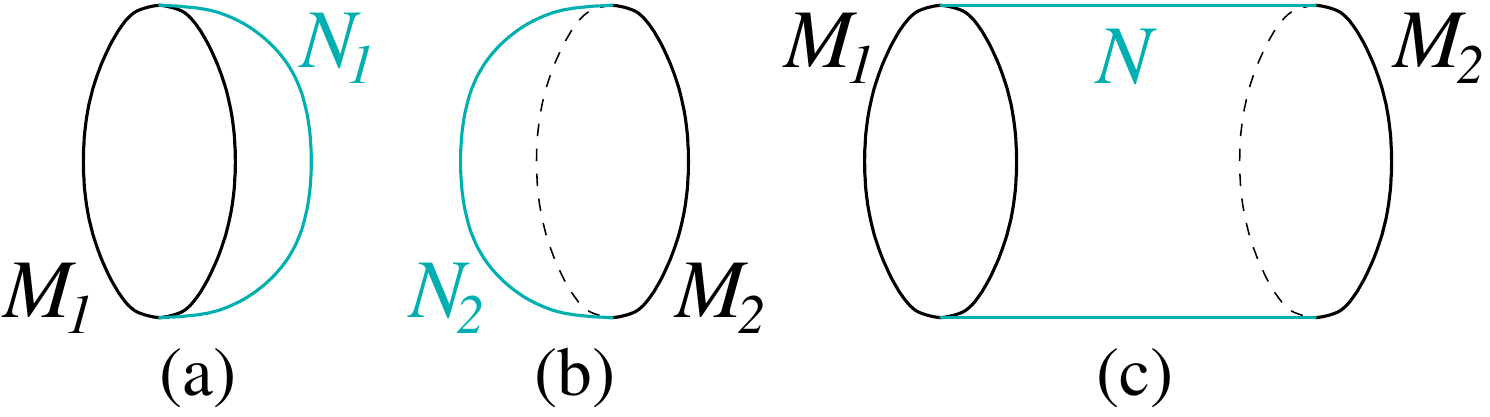} \end{center}
%2
\caption{ (Color online) Three space-time 
$\cM_1$, $\cM_2$, and $\cM_1\sqcup -\cM_2$, plus their extensions.
$\cN_1$, $\cN_2$, and $\cN$.  The ratio of the
action amplitudes (which are pure $U(1)$ phases) for space-time (a) and (b):
$Z(a)/Z(b)=Z(a)Z^*(b)$ is given by the action amplitude for space-time (c):
$Z(a)Z^*(b)=Z(c)$. 
} \label{M1M2} 
\end{figure}

\subsection{An exact evaluation of the partition function}

Let us examine the action amplitude $ \ee^{\ii 2\pi \int_{\cM^{d+1}} \phi_M^*
\bar\nu_{d+1} +\ii 2\pi \int_{\cN^{d+2}} \phi_N^* \bar \om_{d+2}}$ for different
$\phi_M$ but a fixed $\vphi$.  $\phi_N$ is chosen such that it is given by
$\vphi \phi_M$ at the boundary $\prt \cN^{d+2}$.  We know that $ \vphi^* \bar
\om_{d+2}$ is a coboundary on $\prt \cN^{d+2}$.  Thus the value of $ \ee^{\ii
\pi \int_{\cN^{d+2}} \phi_N^* \bar \om_{d+2}}$ only depends $\phi_N$ on the
boundary $\prt \cN^{d+2}$, \ie only depends on $\vphi \phi_M$.
Therefore, the action amplitude $ \ee^{\ii 2\pi \int_{\cM^{d+1}} \phi_M^*
\bar\nu_{d+1} +\ii 2\pi \int_{\cN^{d+2}} \phi_N^* \bar \om_{d+2}}$
is a function of $\phi_M$.

Let us call $\phi_{M_1}: \cM^{d+1}_1 \to \cB G_{fSO}$ and $\phi_{M_2}:
\cM^{d+1}_2 \to \cB G_{fSO}$ cobordant, if there exist a homomorphism $\vphi_N:
\cN^{d+2} \to \cB G_{fSO}$ such that $\prt \cN^{d+2} = \cM^{d+1}_1 \sqcup
\cM^{d+1}_2 $, $\vphi_N=\phi_{M_1}$ on the boundary $\cM^{d+1}_1$, and
$\vphi_N=\phi_{M_2}$ on the boundary $\cM^{d+1}_2$ (see Fig. \ref{M1M2}(c)).  We
can show that \emph{two cobordant $\phi_{M_1}$ and $ \phi_{M_2}$ have the same
action amplitude}.  This is because the difference (the ratio) of the action
amplitudes is given by (see Fig. \ref{M1M2})
\begin{align}
 \ee^{\ii 2\pi [
\int_{\cM^{d+1}_1} \phi_{M_1}^* \bar\nu_{d+1} 
-\int_{\cM^{d+1}_2} \phi_{M_2}^* \bar\nu_{d+1}] 
+\ii 2\pi \int_{\cN^{d+2}} \phi_N^* \bar \om_{d+2}}
\end{align}
where $\phi_N =\vphi \vphi_N$. Because $\vphi_N=\phi_{M_1}$ or
$\vphi_N=\phi_{M_2}$ on the two boundaries of $\cN^{d+2}$, we have
\begin{align}
& \ee^{\ii 2\pi [
\int_{\cM^{d+1}_1} \phi_{M_1}^* \bar\nu_{d+1} 
-\int_{\cM^{d+1}_2} \phi_{M_2}^* \bar\nu_{d+1}]}
\nonumber\\
=& 
 \ee^{\ii 2\pi [
\int_{\cM^{d+1}_1} \vphi_N^* \bar\nu_{d+1} 
-\int_{\cM^{d+1}_2} \vphi_N^* \bar\nu_{d+1}]}
\nonumber\\
=&
 \ee^{\ii 2\pi 
\int_{\cN^{d+2}} \dd \vphi_N^* \bar\nu_{d+1} }
\end{align}
Now the difference (the ratio) of the action amplitudes is given by
\begin{align}
& 
\ee^{\ii 2\pi \int_{\cN^{d+2}} \dd \vphi_N^* \bar\nu_{d+1} + \phi_N^* \bar \om_{d+2}}
\nonumber\\
=&
\ee^{\ii 2\pi \int_{\cN^{d+2}} \dd \vphi_N^* \bar\nu_{d+1} + \vphi_N^*\vphi^* \bar \om_{d+2}}
\nonumber\\
=&
\ee^{\ii 2\pi \int_{\cN^{d+2}} \vphi_N^* (\dd \bar\nu_{d+1} + \vphi^* \bar \om_{d+2})} =1 .
\end{align}

We like to remark that in the action amplitude $\ee^{\ii 2\pi \int_{\cM^{d+1}}
\phi_M^* \bar\nu_{d+1} + \ii 2\pi \int_{\cN^{d+2}} \phi_N^* \bar \om_{d+2}}$, the
homomorphism $\phi_M: \cM^{d+1} \to \cB G_{fSO}$ usually cannot be extended to
a homomorphism $\vphi_N: \cN^{d+2} \to \cB G_{fSO}$ (\ie $\phi_M$ is not
cobordant to a trivial homomorphism).  As a result the action amplitude is not
equal to 1. If the homomorphism $\phi_M: \cM^{d+1} \to \cB G_{fSO}$ can be
extend to $\cN^{d+2}$, the action amplitude will be equal to one.  If
$\phi_{M_1}$ and $\phi_{M_2}$ are cobordant, then $\phi_{M_1}$ and $\phi_{M_2}$
on $\cM_1$ and $\cM_2$ in  Fig. \ref{M1M2}(c) can be extended to $\cN$. In this
case, the action amplitude for space-time Fig. \ref{M1M2}(c) will be equal to
one.

In our exactly soluble model \eq{ZfSPTSOhg}, the homomorphism $\phi_M$ is
determined by $a^{G_{fSO}}$ which in turn is given by the dynamical fields
$g_i$ on vertices and the background field $A^{G_{fSO}}$ in the links (see
\eqn{gAgSO}).  For a fixed $A^{G_{fSO}}$, the different homomorphisms $\phi_M$
are all cobordant to each other, and the corresponding action amplitudes are
all equal to each other.  Therefore, the partition function for our model
\eq{ZfSPTSOhg} can be calculated exactly
\begin{align}
\label{ZfSPTSOZ}
&Z(\cM^{d+1},A^{G_{fSO}})
= 
\sum_{\phi_M } 
\ee^{\ii 2\pi \big( \int_{\cM^{d+1}} \phi_M^* \bar\nu_{d+1} 
+\int_{\cN^{d+2}} \phi_N^* \bar \om_{d+2} \big)} 
\nonumber\\
&= V^{N_v}
\ee^{\ii 2\pi \int_{\cM^{d+1}} \nu_{d+1}(A^{G_{fSO}}) + \ii \pi \int_{\cN^{d+2}} 
\Sq^2 f_d+f_d \w_2 }
\end{align}
where $V$ is the volume of
$G_{fSO}$ and $N_v$ the number of vertices in $\cM^{d+1}$.  We see that the
fermionic SPT state realized by \eq{ZfSPTSOhg} [which is labeled by a pair
$(\vphi, \bar\nu_{d+1})$] is characterized by the SPT invariant
\begin{align}
 \label{SPTinvSO}
& Z^\text{top}(\cM^{d+1},A^{G_{fSO}})
\nonumber\\
= &
\ee^{\ii 2\pi \int_{\cM^{d+1}} \nu_{d+1}(A^{G_{fSO}}) + \ii \pi \int_{\cN^{d+2}} 
\Sq^2 f_d+f_d \w_2 },
\end{align}
where $f_d \se{2} n_d(A^{G_{fSO}})$ on  $\prt \cN^{d+2}$ and 
\begin{align}
-\dd \nu_{d+1}( A^{G_{fSO}} ) & \se{1} \frac12 \Sq^2 n_d( A^{G_{fSO}} ) 
\nonumber\\ &
+ \frac12 n_d( A^{G_{fSO}} ) \w_2(A^{SO}) .  
\end{align}
Here $A^{SO}_{ij}=\pi(A^{G_{fSO}}_{ij}) \in SO_\infty$ and $A^{SO}$ must be the
connection for the tangent bundle on $\cM^{d+1}$.

Eqn. \eq{SPTinvSO} can also be rewritten as
\begin{align}
 \label{SPTinvSO1}
& Z^\text{top}(\cM^{d+1},A^{G_{fSO}})
\nonumber\\
= &
\ee^{\ii 2\pi \int_{\cM^{d+1}} \phi_M^* \bar\nu_{d+1} 
+ \ii 2\pi \int_{\cN^{d+2}} \phi_N^* \bar \om_{d+2}},
\nonumber\\
& \phi_N|_{\prt \cN^{d+2}} = \vphi \phi_M,
\end{align}
where the homomorphism $\phi_M: \cM^{d+1} \to \cB G_{fSO}$ is determined by the
background field $A^{G_{fSO}}$ via
\begin{align}
 A^{G_{fSO}} = \phi_M^* \bar a^{G_{fSO}}.
\end{align}
Also $\phi_N$ is a  homomorphism $\phi_N: \cN^{d+2} \to
\cB_f(SO_\infty,1;\Z_2,d)$, such that, on the boundary $\cM^{d+1}=\prt
\cN^{d+2}$, $\phi_N$ is given by $\vphi \phi_M$, where $\vphi$ is a
homomorphism $\vphi: \cB G_{fSO} \to \cB_f(SO_\infty,1;\Z_2,d)$.

\subsection{Equivalent relations between the labels  $(\vphi,
\bar\nu_{d+1})$}

It is possible that the SPT states labeled by different pairs $(\vphi,
\bar\nu_{d+1})$ and $(\vphi', \bar\nu_{d+1}')$, \ie by  different pairs
$[n_d(A^{G_{fSO}}), \nu_{d+1}(A^{G_{fSO}})]$ and $[n_d'(A^{G_{fSO}}),
\nu_{d+1}'(A^{G_{fSO}})]$, are the same SPT states since the two pair may
give rise to the same SPT invariant.  In this case, we say that the two pairs
are equivalent.  

What are the equivalent relations for the pairs $(\vphi, \bar\nu_{d+1})$ or
$[n_d(A^{G_{fSO}}), \nu_{d+1}(A^{G_{fSO}})]$?  Here is our proposal: $(\vphi,
\bar\nu_{d+1})$ and $(\vphi', \bar\nu_{d+1}')$ are equivalent if
\begin{enumerate}
\item 
There  exists a $\R/\Z$-valued $d+2$-cocycle $\bar \Om_{d+2}$ on  $I\times
\cB_f(SO_\infty,1;Z_2,d)$ such that when restricted on the two boundaries of
$I\times \cB_f(SO_\infty,1;Z_2,d)$, $\bar \Om_{d+2}$ becomes $\bar \om_{d+2}$ in
\eqn{bomd2}.
\item
There  exists a homomorphism $\Phi:
I\times \cB G_{fSO} \to I\times \cB_f(SO_\infty,1;Z_2,d)$
such that on the two boundaries, $\Phi$ reduces to $\vphi$ and $\vphi'$.
\item
There  exists a $\R/\Z$-valued $d+1$-cochain $\bar\mu_{d+1}$ on $I\times \cB
G_{fSO}$ such that $-\dd \bar\mu_{d+1} \se{1} \Phi^* \bar \Om_{d+2}$
and $\bar\mu_{d+1}$ reduces to $\bar\nu_{d+1}$ and  $\bar\nu_{d+1}'$ on the two boundaries.
\end{enumerate}

%\textbf{XG: Here is a stronger and more natural equivalence relation:
%\begin{enumerate}
%\item
%There exists a complex $\widehat \cC_f$ such that $\prt
%\widehat\cC_f =\cB_f(SO_\infty,1;Z_2,d)\sqcup
%-\cB_f(SO_\infty,1;Z_2,d)$.
%\item
%There exists a complex $\widehat \cC $ such that $\prt
%\widehat\cC  =\cB SO_\infty \sqcup
%-\cB SO_\infty$.
%\item 
%There  exists a $\R/\Z$-valued $d+2$-cocycle $\hat \om_{d+2}$ on $\widehat
%\cC_f$ such that when restricted on the two boundaries of
%$\widehat \cC_f$, $\hat \om_{d+2}$ becomes $\bar \om_{d+2}$
%in \eqn{bomd2}.
%\item
%There  exists a homomorphism $\hat \vphi:
%\widehat \cC  \to \widehat \cC_f$
%such that on the two boundaries, $\hat \vphi$ reduces to $\vphi$ and $\vphi'$.
%\item
%There  exists a $\R/\Z$-valued $d+1$-cochain $\hat \om_{d+1}$ on $\widehat \cC
%$ such that $-\dd \hat \om_{d+1} \se{1} \hat \vphi^* \hat \om_{d+2}$
%and $\hat \om_{d+1}$ reduces to $\bar\nu_{d+1}$ and  $\bar\nu_{d+1}'$ on the two boundaries.
%\end{enumerate}
%But the following proof does not go through. 
%}

\begin{figure}[t]
\begin{center}
\includegraphics[scale=0.5]{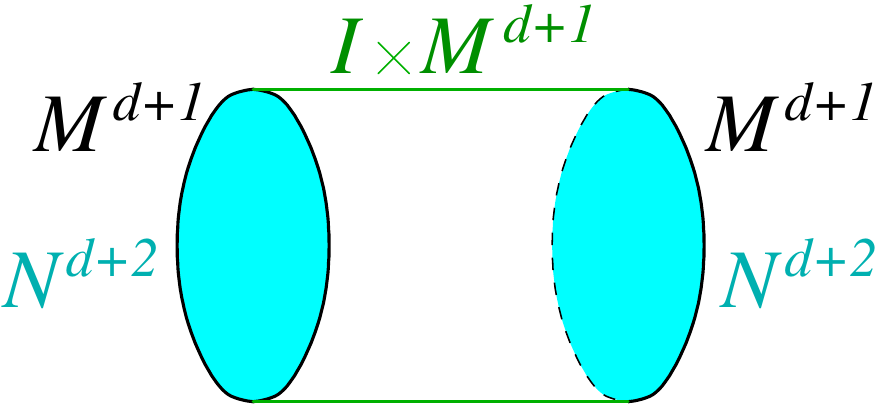} \end{center}
%3
\caption{ (Color online) The boundary of $I\times \cN^{d+2}$ is given by
$\cN^{d+2} \sqcup I\times \cM^{d+1} \sqcup \cN^{d+2}$.  The boundary of
$I\times \cM^{d+1}$ is given by $\cM^{d+1} \sqcup \cM^{d+1}$.
}
\label{MN}
\end{figure}

In the following, we like to show that equivalent $(\vphi, \bar\nu_{d+1})$ and
$(\vphi', \bar\nu_{d+1}')$ give rise to the same SPT invariant \eq{SPTinvSO}, and
hence the same SPT order.  Since $\dd \bar \Om_{d+2} \se{1}0$, we have (see
Fig. \ref{MN})
\begin{align}
\label{eee}
 1= & \ee^{\ii 2\pi \int_{I\times \cN^{d+2}} \t\phi_N^* \dd {\bar \Om}_{d+2} }
  =  
\ee^{\ii 2\pi \int_{I\times \cM^{d+2}} \t\phi_N^* \bar \Om_{d+2} } \times
\nonumber\\
&
\ee^{-\ii 2\pi \int_{\cN^{d+2}} \phi_N^* \bar \om_{d+2} }
\ee^{+\ii 2\pi \int_{\cN^{d+2}} (\phi_N')^* \bar \om_{d+2} },
\end{align}
where $\t\phi_N$ is a homomorphism $\t\phi_N: I\times \cN^{d+2} \to I\times
\cB_f(SO_\infty,1;Z_2,d)$ such that on the two boundaries $\t\phi_N$ is given
by $\phi_N: \cN^{d+2} \to  \cB_f(SO_\infty,1;Z_2,d)$ and $\phi_N': \cN^{d+2}
\to  \cB_f(SO_\infty,1;Z_2,d)$.  

We choose $\t\phi_N$ such that on the boundary of $\cN^{d+2}$, $\phi_N$ becomes
$\vphi \phi_M: \cM^{d+1} \to  \cB_f(SO_\infty,1;Z_2,d)$ and $\phi_N'$ becomes
$\vphi' \phi_M: \cM^{d+1} \to  \cB_f(SO_\infty,1;Z_2,d)$, where $\phi_M$ is a
homomorphisms $\phi_M: \cM^{d+1} \to  \cB G_{fSO}$, and $\vphi$ and $\vphi'$
are the homomorphisms $\vphi: \cB G_{fSO} \to \cB_f(SO_\infty,1;Z_2,d)$ and
$\vphi': \cB G_{fSO} \to \cB_f(SO_\infty,1;Z_2,d)$.  We note that $\phi_M$
determines the background field $A^{G_{fSO}}$.
Similarly, on the two boundaries of
$I\times \cM^{d+1}$, $\t\phi_N$ is also given by $\vphi\phi_M$ and
$\vphi'\phi_M$.

Since $\vphi$ and $\vphi'$ are given by $\Phi:  I\times \cB G_{fSO} \to I\times
\cB_f(SO_\infty,1;Z_2,d)$ when restricted on the two boundaries, therefore we
can choose $\t\phi_N$ such that on $I\times \cM^{d+1}$, $\t\phi_N$ is given by
$\Phi \t \phi_M$ where $ \t \phi_M$ is a homomorphism $ \t \phi_M: I\times
\cM^{d+1} \to I\times \cB G_{fSO}$.
Now, we find (see Fig. \ref{MN})
\begin{align}
 & \ee^{\ii 2\pi \int_{I\times \cM^{d+2}} \t\phi_N^* \bar \Om_{d+2} }
=\ee^{\ii 2\pi \int_{I\times \cM^{d+1}} \t \phi_M^*\Phi^* \bar \Om_{d+2} }
\nonumber\\
=& \ee^{-\ii 2\pi \int_{I\times \cM^{d+1}}  \t \phi_M^* \dd \bar\mu_{d+1} }
\nonumber\\
=& \ee^{-\ii 2\pi \int_{\cM^{d+1}}  \phi_M^*\bar\nu_{d+1} -\phi_M^*\bar\nu_{d+1}'}
\end{align}
The above reduces \eqn{eee} to
\begin{align}
 1= & 
\ee^{-\ii 2\pi \int_{\cM^{d+1}}   \phi_M^*\bar\nu_{d+1} -\phi_M^*\bar\nu_{d+1}' }
\times 
\nonumber\\
&
\ee^{-\ii 2\pi \int_{\cN^{d+2}} \phi_N^* \bar \om_{d+2} }
\ee^{+\ii 2\pi \int_{\cN^{d+2}} (\phi_N')^* \bar \om_{d+2} },
\end{align}
which complete our proof (see \eqn{SPTinvSO1}).

The fermionic SPT states can also be labeled by $[\bar n_d(\bar a^{G_{fSO}}),
\bar \nu_{d+1}(\bar a^{G_{fSO}})]$ where $\bar n_d(\bar a^{G_{fSO}})$ is a
$\Z_2$-valued $d$-cocycle and $\bar \nu_{d+1}(\bar a^{G_{fSO}})$ is a
$\R/\Z$-valued $d+1$-cochian on $\cB G_{fSO}$. They are functions of canonical
cocycle $\bar a^{G_{fSO}}$ on $\cB G_{fSO}$.  In terms of $[\bar n_d(\bar
a^{G_{fSO}}), \bar \nu_{d+1}(\bar a^{G_{fSO}})]$, the equivalence relations are
particially generated by the following two transformations (see \eqn{Sqgauge}):
\begin{enumerate}
\item
Transformation generated by $d$-cohain 
$\bar \eta_{d}(\bar a^{G_{fSO}}) \in C^{d-1}(\cB G_{fSO} ;\R/\Z)$
\begin{align}
\label{gaugetomSO}
\bar n_d(\bar a^{G_{fSO}}) \to &  \bar n_d(\bar a^{G_{fSO}})
\nonumber\\
\bar \nu_{d+1}(\bar a^{G_{fSO}}) \to & \bar \nu_{d+1}(\bar a^{G_{fSO}})
+\dd \bar \eta_d(\bar a^{G_{fSO}}) .
\end{align}
\item
Transformation
generated by $d-1$-cohain 
$\bar u_d \in C^{d-1}(\cB G_{fSO};\Z_2)$
\begin{align}
\label{gaugendSO}
& \bar n_d(\bar a^{G_{fSO}}) \to  \bar n_d(\bar a^{G_{fSO}}) + \dd \bar u_{d-1}(\bar a^{G_{fSO}})
\\
& \bar \nu_{d+1}(\bar a^{G_{fSO}}) \to  \bar \nu_{d+1}(\bar a^{G_{fSO}})
+\frac 12 \bar u_{d-1}(\bar a^{G_{fSO}}) \bar \w_2(\bar a^{SO})
\nonumber\\
& +\frac12 \gSq^2  \bar u_{d-1}(\bar a^{G_{fSO}})
+\frac 12 \dd \bar u_{d-1}(\bar a^{G_{fSO}}) \hcup{d-1} \bar n_d(\bar a^{G_{fSO}}) .
\nonumber 
\end{align}
\end{enumerate}
We can show the above two transformation generate equivalent relations since
they do not change the SPT invariant \eq{SPTinvSO}.

In fact, we have a more general equivalent relation in terms of
$[n_d(a^{G_{fSO}}), \nu_{d+1}(a^{G_{fSO}})]$ in \eqn{fSPTclSO}.  Here
$n_d(a^{G_{fSO}})=\phi_M^* \bar n_d(\bar a^{G_{fSO}})$ and
$\nu_{d+1}(a^{G_{fSO}})= \phi_M^* \bar \nu_{d+1}(\bar a^{G_{fSO}})$ are a
$d$-cocycle and a $d+1$-cochain on space-time $\cM^{d+1}$.  $[n_d(a^{G_{fSO}}),
\nu_{d+1}(a^{G_{fSO}})]$ and $[n_d'(a^{G_{fSO}}), \nu_{d+1}'(a^{G_{fSO}})]$
produce the same SPT invariant if they satisfy (see \eqn{Sqgauge}):
\begin{enumerate}
\item
Equivalence relation generated by $d$-cohain 
$\eta_{d} \in C^{d-1}(\cM^{d+1};\R/\Z)$
\begin{align}
\label{gaugetomSOM}
n_d'(a^{G_{fSO}}) \se{2}&  n_d(a^{G_{fSO}})
\nonumber\\
\nu_{d+1}'(a^{G_{fSO}}) \se{1} & \nu_{d+1}(a^{G_{fSO}})
+\dd \eta_d .
\end{align}
\item
Equivalence relation generated by $d-1$-cohain 
$u_d \in C^{d-1}(\cM^{d+1};\Z_2)$
\begin{align}
\label{gaugendSOM}
& n_d'(a^{G_{fSO}}) \se{2}  n_d(a^{G_{fSO}}) + \dd u_{d-1}
\\
& \nu_{d+1}'(a^{G_{fSO}}) \se{1}  \nu_{d+1}(a^{G_{fSO}})
+\frac 12 u_{d-1} \w_2(a^{SO})
\nonumber\\
& +\frac12 \gSq^2  u_{d-1}
+\frac 12 \dd u_{d-1} \hcup{d-1} n_d(a^{G_{fSO}}) .
\nonumber 
\end{align}
\end{enumerate}
Because $u_{d-1}$ and $\eta_d$ are cochains on $\cM^{d+1}$, which do not have
to be the pullbacks of  cochains $ \bar u_{d-1}(\bar a^{G_{fSO}})$ and $\bar
\eta_{d-1}(\bar a^{G_{fSO}})$ on $\cB G_{fSO}$.  So the above equivalent
relation is more general.

Using a similar method (see Fig. \ref{MN}), we can show that above
$[n_d(a^{G_{fSO}}), \nu_{d+1}(a^{G_{fSO}})]$ and $[n_d'(a^{G_{fSO}}),
\nu_{d+1}'(a^{G_{fSO}})]$ produce the same action amplitude (thus the same SPT
invariant).  So the above relations are indeed equivalent relations.

We first introduce a $\Z_2$-valued cochain $\Om_{d+2}$ on $I\times \cN^{d+2}$
\begin{align}
 \Om_{d+2} \se{1} \frac12 \gSq^2 F_d +  \frac12 F_d  \w_2
\end{align}
where $F_d$ is a  $\Z_2$-valued cocycle on $I\times \cN^{d+2}$
given by
\begin{align}
 F_d =f_d+\dd U_{d-1}
\end{align}
and $ U_{d-1}$ is a  $\Z_2$-valued cochain on $I\times \cN^{d+2}$.  We know
that the boundary of $I\times \cN^{d+2}$ has three pieces $\cN^{d+2}$, $I\times
\cM^{d+1}$ and $\cN^{d+2}$.  $ U_{d-1}$ is chosen such that it becomes $0$ on
one of the $\cN^{d+2}$, and becomes $u_{d-1}$ on the boundary of the second
$\cN^{d+2}$: $\cM^{d+1}=\prt \cN^{d+2}$.  Using \eqn{Sqd}, we see that
$\Om_{d+2}$ is actually a cocycle on $I\times \cN^{d+2}$. Therefore, we have
(see Fig. \ref{MN})
\begin{align}
\label{eee1}
 1= & \ee^{\ii 2\pi \int_{I\times \cN^{d+2}} \dd \Om_{d+2} }
  =  
\ee^{\ii \pi \int_{I\times \cM^{d+2}} \gSq^2 F_d +  F_d  \w_2 } \times
\nonumber\\
&
\ee^{-\ii \pi \int_{\cN^{d+2}}  \Sq^2 f_d +  f_d  \w_2 }
\ee^{+\ii \pi \int_{\cN^{d+2}}  \Sq^2 f_d' +  f_d'  \w_2 },
\end{align}
where $f_d'\se{2} f_d+\dd U_{d-1}$ on $\cN^{d+2}$.  On the boundary
$\cM^{d+1}=\prt \cN^{d+2}$, $f_d'\se{2} f_d+\dd U_{d-1}$ becomes $n_d'\se{2}
n_d+\dd u_{d-1}$ (see \eqn{gaugendSO}).

Next we calculate 
$\ee^{\ii \pi \int_{I\times \cM^{d+2}} \gSq^2 F_d +  F_d  \w_2 }$ 
(using \eqn{Sqgauge}):
\begin{align}
&\ \ \ \
 \ee^{\ii \pi \int_{I\times \cM^{d+2}} \gSq^2 (f_d+\dd U_{d-1}) 
+  (f_d+\dd U_{d-1}) \w_2 }
\nonumber\\ &
= \ee^{\ii \pi \int_{I\times \cM^{d+2}} \Sq^2 f_d 
+  f_d \w_2 
+ \dd \big( \gSq^2 U_{d-1} + \dd U_{d-1} \hcup{d-1} n_d +U_{d-1}\w_2 \big)
}
\nonumber\\ &
= \ee^{\ii 2\pi \int_{I\times \cM^{d+2}}  
 \dd \big( \nu_{d+1} + \frac12 \gSq^2 U_{d-1} + \frac12 \dd U_{d-1} \hcup{d-1} n_d +\frac12 U_{d-1}\w_2 \big)
}
\nonumber\\ &
= \ee^{\ii 2\pi \int_{\cM^{d+2}}  \nu_{d+1} }
 \ee^{-\ii 2\pi \int_{\cM^{d+2}}  \nu_{d+1}' }
\end{align}
Combined with \eqn{eee1}, we find
\begin{align}
&\ \ \ \
  \ee^{\ii 2\pi \int_{\cM^{d+2}}  \nu_{d+1} +\ii \pi \int_{\cN^{d+2}}  \Sq^2 f_d +  f_d  \w_2}
\nonumber\\ &
=
  \ee^{\ii 2\pi \int_{\cM^{d+2}}  \nu_{d+1}' +\ii \pi \int_{\cN^{d+2}}  \Sq^2 f_d' +  f_d'  \w_2}
 \end{align}
The action amplitudes are indeed the same.

We like to remark that $[n_d(a^{G_{fSO}}), \nu_{d+1}(a^{G_{fSO}})]$ and
$[n_d'(a^{G_{fSO}}), \nu_{d+1}'(a^{G_{fSO}})]$ not related by the above two
types of transformations may still describe the same SPT phase.
To really show $[n_d(a^{G_{fSO}}), \nu_{d+1}(a^{G_{fSO}})]$ and
$[n_d'(a^{G_{fSO}}), \nu_{d+1}'(a^{G_{fSO}})]$ describe different SPT phases,
we need to show they produce different SPT invariants.

\subsection{Stacking and Abelian group structure of fermionic SPT phases }
\label{stack}

We have seen that the higher dimensional bosonization is closely related to a
higher group $\cB_f(SO_\infty,1;Z_2,d)$, and a particular choice of a
$\RZ$-valued cocycle $\frac12 \Sq^2 \bar f_d + \frac12 \bar f_d \bar \w_2$ on
the higher group.  Such a choice of cocycle has an important additive property
\begin{align}
\label{addf}
&\ \ \ \
 \frac12 \Sq^2 (\bar f_d+\bar f_d') + \frac12 (\bar f_d+\bar f_d') \bar \w_2 
\nonumber\\
& \se{1,\dd}
   \Big(\frac12 \Sq^2 \bar f_d + \frac12 \bar f_d \bar \w_2\Big)
 + \Big(\frac12 \Sq^2 \bar f_d' + \frac12 \bar f_d' \bar \w_2\Big) .
\end{align}

This additive property insure that the fermionic SPT phases can also be added
so that the collection of fermionic SPT phases actually form an Abelian group.
The addition of two SPT states physically corresponds to stacking two SPT states
one on top another, which implies that the SPT phases should always have an Abelian group
structure.

For two SPT phases described by $(\bar n_d,\bar \nu_{d+1})$ and
$(\bar n_d',\bar \nu_{d+1}')$, they add following a twisted addition rule
\begin{align}
\label{addnnu}
  &(\bar n_d,\bar \nu_{d+1})+(\bar n_d',\bar \nu_{d+1}')\nonumber\\
  =&(\bar n_d+\bar n_d',\bar \nu_{d+1}+\bar \nu_{d+1}'+\frac12 \bar n_d\hcup{d-1}\bar n_d').
\end{align}
This allows us to extract the group $\fSPT_{d+1}(G_{fSO})$ given by the
stacking of fermionic SPTs with symmetry $G^f$. Clearly there is homomorphism
\begin{align*}
  \fSPT_{d+1}(G_{fSO})&\to H^d(\cB G_{fSO};\Z_2)\nonumber\\
  (\bar n_d,\nu_{d+1})&\mapsto \bar n_d.
\end{align*}

Not every $\bar n_d$ allows a solution of $\bar \nu_{d+1}$; the image of the
above homomorphism is a subgroup of $H^d(\cB G_{fSO};\Z_2)$, which will be
called the obstruction-free subgroup, denoted by $BH^d(\cB G_{fSO};\Z_2)$. The
kernel of the above homomorphism is a quotient group of $H^{d+1}(\cB
G_{fSO};\RZ)$.  This is because of the extra gauge transformation $\frac12
(\gSq^2\bar u_{d-1}+\bar u_{d-1}\w_2)$ when $\bar n_d=0$.  Let
$\Gamma\subset H^{d+1}(\cB G_{fSO};\RZ)$ be the subgroup generated by
$\frac12(\gSq^2\bar u_{d-1}+\bar u_{d-1} \bar \w_2),
\forall \bar u_{d-1}\in\cC^{d-1}(\cB G_{fSO};\Z_2)$, we have the following
exact sequence for group extension:
\begin{align}
  H^{d+1}(\cB G_{fSO};\RZ)/\Gamma
& \to\fSPT_{d+1}(G_{fSO})
\nonumber\\
& \to BH^{d+1}(\cB G_{fSO};\Z_2),
\end{align}
whose corresponding group 2-cocycle in $H^2[BH^d(\cB G_{fSO};\Z_2), H^{d+1}(\cB
G_{fSO};\RZ)/\Gamma]$ is given by $\frac12 \bar n_d\hcup{d-1}\bar n_d'$.

We know that bosonic topological orders form a commutative monoid under the
stacking operation $\boxtimes$ \cite{KW1458}.  However, For bosonic topological
order with emergent fermions, they cannot have inverse for the stacking
operation $\boxtimes$, and thus they are not invertible topological orders
\cite{KW1458,F1478,K1467,FH160406527}.  However,  we may modify the stacking
operation by allowing the pairs of fermions from the two stacked phases to condense (equivalently, identifying fermions from the two stacked
phases).  Such a modified stacking operation is discussed in detail
in \Ref{LW160205936,LW160205946}. We denote the modified stacking operation by
$\boxtimes_f$.  The fermionic SPT states form an Abelian group under the
modified stacking operation $\boxtimes_f$, as in \eqn{addnnu}.

%The bosonic topological order with emergent fermions also form a monoid under
%$\boxtimes_f$.  The \emph{simplest topological orders} mentioned in Section
%\ref{triv} are  bosonic topological orders with emergent fermions that have an
%inverse and correspond to the identity for the $\boxtimes_f$ operation.  This
%$\boxtimes_f$-identity topological order is the origin of the additive
%properties discussed in \eqn{addf}.  The additive property \eqn{addf} allows
%us to show that the bosonic topological orders with emergent fermions
%described by \eqn{ZbN} are $\boxtimes_f$-identity topological orders.

\subsection{With time reversal symmetry}

In the above, we discussed the situation without time-reversal symmetry. In the
presence of time reversal symmetry, we have the following result.  The exactly
soluble model \eq{ZfSPTO} and the related fermionic SPT state is characterized
by the following data:
\begin{enumerate}
\item
A particular higher group $\cB_f(O_\infty,1;Z_2,d)$, determined by its
$\Z_2$-valued canonical cochain $\dd \bar f_d \se{2}0$ (see \Ref{ZLW} and Appendix
\ref{hgroup}).
\item
A particular $\RZ$-valued $d+2$-cocycle on  
\begin{align}
\label{bomd2}
\bar \om_{d+2}\se{1} \frac12 \Sq^2 \bar f_d +
\frac12 \bar f_d [\bar \w_2(\bar a^{O})+\bar \w_1^2(\bar a^{O})]
\end{align}
on $\cB_f(O_\infty,1;Z_2,d)$.
\item
Different trivialization homomorphisms $ \vphi: \cB G_{fO} \to
\cB_f(O_\infty,1;Z_2,d)$, where $G_{fO}=G_f\gext O_\infty$.
\item
Different choices of the trivialization $\nu_{d+1}(\bar a^{G_{fO}})$ that
satisfy $-\dd \nu_{d+1}(\bar a^{G_{fO}}) \se{1} 
\vphi^* \bar \om_{d+2}(\bar f_d,\bar a^{O}) $.  
\end{enumerate}
We like remark that in the above, the time reversal symmetry
in $O_\infty$ acts non-trivially on the value of $\nu_{d+1}$
(see Appendix \ref{cochain}).

We can also use $[\bar n_d(\bar a^{G_{fO}}), \bar \nu_{d+1}(\bar a^{G_{fO}})]$ in \eqn{fSPTclO}
to label the fermion SPT states with time reversal symmetry.  The equivalence
relations are partially generated by the following two relations (see
\eqn{Sqgauge}):
\begin{enumerate}
\item
Equivalence relation generated by $d$-cohain $\bar \eta_{d} \in C^{d-1}(\cB
G_{fSO};\R/\Z)$
\begin{align}
\label{gaugetomO}
\bar n_d'(\bar a^{G_{fO}}) \se{2}&  \bar n_d(\bar a^{G_{fO}})
\nonumber\\
\bar \nu_{d+1}(\bar a^{G_{fO}}) \se{1} & \bar \nu_{d+1}'(\bar a^{G_{fO}})
+\dd \bar \eta_d(\bar a^{G_{fO}})
\end{align}
\item
Equivalence relation generated by $d-1$-cohain $\bar u_{d-1} \in C^{d-1}(\cB
G_{fSO};\Z_2)$
\begin{align}
\label{gaugendO}
& \bar n_d'(\bar a^{G_{fO}}) \se{2}  \bar n_d(\bar a^{G_{fO}}) + \dd \bar u_{d-1}(\bar a^{G_{fO}})
\\
& \bar \nu_{d+1}'(\bar a^{G_{fO}}) \se{1}  \bar \nu_{d+1}(\bar a^{G_{fO}})
+\frac 12 \dd \bar u_{d-1}(\bar a^{G_{fO}}) \hcup{d-1} \bar n_d(\bar a^{G_{fO}}) 
\nonumber\\
& +\frac12 \gSq^2  \bar u_{d-1}(\bar a^{G_{fO}})
+\frac 12 \bar u_{d-1}(\bar a^{G_{fO}}) [\bar \w_2(a^{O})+\bar \w_1^2(a^{O})]
\nonumber 
\end{align}
\end{enumerate}
The corresponding SPT invariant is given by
\begin{align}
 \label{SPTinvO}
& Z^\text{top}(\cM^{d+1},A^{G_{fO}})
\nonumber\\
= &
\ee^{\ii 2\pi \int_{\cM^{d+1}} \nu_{d+1}(A^{G_{fO}}) + \ii \pi \int_{\cN^{d+2}} 
\Sq^2 f_d+f_d (\w_2+\w_1^2) },
\nonumber\\
= &
\ee^{\ii 2\pi \int_{\cM^{d+1}} \phi_M^* \bar\nu_{d+1} 
+ \ii 2\pi \int_{\cN^{d+2}} \phi_N^* \bar \om_{d+2}},
\nonumber\\
& f_d|_{\prt \cN^{d+2}} = n_d,\ \ \ \phi_N|_{\prt \cN^{d+2}} = \vphi \phi_M
\end{align}

\section{1+1D fermionic SPT states}
\label{example1D}

In this and next a few sections, we are going to apply our theory to study some
simple fermionic SPT phases.  In 1+1D, the SPT
invariant dose not depend on $n_1$ (see \eqn{SPTinvSO}).  Thus the different
1+1D fermionic SPT states are labeled by cocycles $\nu_2$.  After quotient out
the equivalence relations, we find that 1+1D fermionic SPT states from fermion
decoration are classified by $H^2(\cB G_{fSO};\R/\Z)$ without time reversal
symmetry and by $H^2[\cB G_{fO};(\R/\Z)_T]$ with time reversal symmetry, where
$(\R/\Z)_T$ remind us that the time reversal symmetry in $G_{fO}$ has a
non-trivial action on $\R/\Z \stackrel{T}{\to} -\R/\Z$.

Since $G_{fSO} = G_f \gext SO_\infty$, $H^2[\cB G_{fO};(\R/\Z)_T]$ is given by
a quotient of a subset of
\begin{align}
& H^1[\cB  SO_\infty; H^1(\cB G_f;\R/\Z)]\oplus H^2(\cB SO_\infty;\R/\Z)
\oplus 
\nonumber\\ &
H^2(\cB G_f;\R/\Z)
\end{align}
(see Appendix \ref{LHS}).
Using the universal coefficient theorem \eqn{ucf}
and \eqn{HBSOZ}, we find that $H^1[\cB  SO_\infty; H^1(\cB G_f;\R/\Z)]=0$.
Also $H^2(\cB SO_\infty;\R/\Z)$ does not involve symmetry $G_f$ can only
correspond to fermionic invertible topological order.  In 1+1D, we believe that
fermion decoration construction produces all fermionic SPT states.  Thus 1+1D
fermionic SPT states are classified by a subset of $H^2(\cB G_f;\R/\Z)$ without
time reversal symmetry.  This is consistent with the result obtained by 1+1D
bosonization: \frmbox{1+1D fermionic SPT states with on-site symmetry $G_f$ are
classified by $H^2(\cB G_f;\R/\Z)$ without time reversal symmetry.}

With time reversal symmetry, $H^2[\cB G_{fO};(\R/\Z)_T]$ is
given by a quotient of a subset of
\begin{align}
& H^1[\cB  O_\infty; H^1(\cB G_f^0;\R/\Z)_T]\oplus H^2(\cB O_\infty;\R/\Z)
\oplus 
\nonumber\\ &
H^2(\cB G_f^0;\R/\Z)
\end{align}
where the time reversal symmetry in $O_\infty$ may have a non-trivial action on
$H^1(\cB G_f;\R/\Z)$ (see Appendix \ref{LHS}).  In the above, we have used the
fact that $G_{fO} = G_f^0\gext O_\infty$, where $G_f^0$ is the fermionic
symmetry group after removing the time reversal symmetry .  The above result is
consistent with the result from 1d bosonization: \frmbox{1+1D fermionic SPT
states with on-site symmetry $G_f$ are classified by $H^2[\cB G_{f};(\RZ)_T]$
with time reversal symmetry.}  We like to remark that the 1d topological
$p$-wave superconductor\cite{K0131} is a fermionic invertible topological
order. It is not a fermionic SPT state.

\section{Fermionic $Z_2\times Z_2^f$-SPT state}

In this section, we are going to study the simplest fermionic SPT phases, where
the fermion symmetry is given by $G_f=Z_2\times Z_2^f$ symmetry.  In this case,
$G_b=Z_2$ and $e_2=0$.  We will consider 2+1D and in 3+1D systems.  
A double-layer superconductor with layer exchange symmetry can realize
$Z_2\times Z_2^f$ symmetry.

Our calculation contains three steps: (1) we first calculate the $\Z_2$-valued
cocycle $\bar n_d$; (2) we then compute the $\R/\Z$-valued cochain $\bar
\nu_{d+1}$; (3) last we construct the corresponding SPT invariant trying to
identify distinct SPT phases labeled by $(\bar n_d,\bar \nu_{d+1})$.  Our
calculation also come with two flavors: (1) without extension of $SO_\infty$,
and (2) with extension of $SO_\infty$.  Since our approaches are constructive,
both of the above two flavors produce exactly soluble fermionic models that
realize various SPT states.  However, the approach with extension of
$SO_\infty$ is more complete, \ie it produces all the SPT phases produced by
the approach without extension of $SO_\infty$.

\subsection{2+1D}

\subsubsection{Without extension of $SO_\infty$}

\noindent\textbf{Calculate $\bar n_2$}:  
We note that cohomology ring $H^*(\cB Z_2;\Z_2)$ is generated by 1-cocycle
$\bar a^{\Z_2}$.  Thus $\bar n_2\in H^2(\cB Z_2;\Z_2) =\Z_2$ has two choices:
$\bar n_2 =\al_n (\bar a^{\Z_2})^2$, $\al_n=0,1$.

\noindent\textbf{Calculate $\bar \nu_3$}:  
Next, we consider $\dd \bar \nu_3$ in \eqn{fSPTcl}.  Since $\bar e_2=0$, only
the term $\frac12 \Sq^2 \bar n_2 = \frac12 \bar n_2^2$ is non-zero.  The term
$\frac12 \bar n_2^2=\frac{\al_n}2 (\bar a^{\Z_2})^4$ is a cocycle in $Z^4(\cB
Z_2;\RZ)$.  Since $H^4(\cB Z_2;\RZ)=0$, thus $\frac12 \bar n_2^2$ is always a
coboundary in $B^4(\cB Z_2;\RZ)$.  Therefore \eqn{fSPTcl}  has solution for all
choices of $\bar n_2$.  Noticing that (see \eqn{Sq1Bs2})
\begin{align}
&\ \ \ \
\frac12 \dd (\bar a^{\Z_2})^3
\nonumber\\
&=
\frac 12 (\dd \bar a^{\Z_2}) (\bar a^{\Z_2})^2
- \frac12 \bar a^{\Z_2} (\dd \bar a^{\Z_2}) \bar a^{\Z_2}
+ \frac12 (\bar a^{\Z_2})^2 \dd \bar a^{\Z_2}
\nonumber\\ &
=
  (\Bs_2 \bar a^{\Z_2}) (\bar a^{\Z_2})^2
- \bar a^{\Z_2} (\Bs_2 \bar a^{\Z_2}) \bar a^{\Z_2}
+ (\bar a^{\Z_2})^2 \Bs_2 \bar a^{\Z_2}
\nonumber\\ &
\se{2}
  (\Sq^1 \bar a^{\Z_2}) (\bar a^{\Z_2})^2
+ \bar a^{\Z_2} (\Sq^1 \bar a^{\Z_2}) \bar a^{\Z_2}
+ (\bar a^{\Z_2})^2 \Sq^1 \bar a^{\Z_2}
\nonumber\\ &
\se{2} (\bar a^{\Z_2})^4 ,
\end{align}
we find that the solution has a form
\begin{align}
\bar \nu_3 \se{1} \frac{\al_n}4 (\bar a^{\Z_2})^3 + \frac{\al_\nu}{2} (\bar a^{\Z_2})^3,
\ \ \ \al_\nu =0,1.
\end{align}
where $\frac{\al_\nu}{2} (\bar a^{\Z_2})^3 \in H^3(\cB Z_2;\RZ)$.

We like to remark that in order for $\frac14 (\bar a^{\Z_2})^3$ to be well
defined mod 1, we need to view the $\Z_2$-valued $\bar a^{\Z_2}$ as $\Z$-valued
with values 0 and 1.  Let us use ${\bar a^{\Z_2}}$ denote such a  map from
$\Z_2$-valued to $\Z$-valued.  Thus more precisely, we have
\begin{align}
\bar \nu_3 \se{1} \frac{\al_n}4 (\bar a^{\Z_2})^3 + \frac{\al_\nu}{2} (\bar a^{\Z_2})^3,
\ \ \ \al_\nu =0,1.
\end{align}
As a result, $Z_2\times Z_2^f$ SPT states are labeled by $(\al_n,\al_\nu)$.
Thus, there are 4 different $Z_2\times Z_2^f$ fermionic SPT states from fermion
decoration.  

\noindent\textbf{SPT invariant}:
The four obtained $Z_2\times Z_2^f$ fermionic SPT states labeled by
$\al_n,\al_\nu=0,1$ are realized by the following local fermionic model (in the
bosonized form as in \eqn{ZfSPT})
\begin{align}
\label{ZfSPTZ2Z2}
&\ \ \ \
Z(\cM^3,A^{G_b}) 
\\
&= 
\hskip -7em
\sum_{ 
\ \ \ \ \ \ \ \ \  \ \ \ \ \ \ \
g \in C^0(\cM^{4};\Z_2); 
f_2 \se{2} \al_n (a^{\Z_2})^2
} 
\hskip -7em
\ee^{\ii \pi \int_{\cM^3}  \frac{\al_n+2 \al_\nu}{2} (a^{\Z_2})^3 
+\al_n   (a^{\Z_2})^2 a^{\Z_2^f}
+\ii \pi \int_{\cN^{4}}  \Sq^2 f_2+f_2\w_2
} 
.
\nonumber 
\end{align}
Its SPT invariant is 
\begin{align}
\label{SPTinvZ2Z2}
&\ \ \ \
Z^\text{top}(\cM^3,A^{G_b}) 
\nonumber\\
= & 
\ee^{\ii 2\pi \int_{\cM^3}  \frac{\al_n+2 \al_\nu}{4} (A^{\Z_2})^3 
+\frac{\al_n}2   (A^{\Z_2})^2 A^{\Z_2^f}
} 
\ee^{
\ii \pi \int_{\cN^{4}}  \Sq^2 f_2+f_2\w_2   
} ,
\nonumber\\
&
f_2\big|_{\prt \cN^{4}}  \se{2} n_2, \ \ \ \ \dd  A^{\Z_2} \se{2} \w_2
.
\end{align}

\subsubsection{With extension of $SO_\infty$}

\noindent\textbf{Calculate $\bar n_2$}:  
First $G_{fSO}=Z_2\times Spin_\infty$, where $Z_2^f$ is contained in
$Spin_\infty$.  Since $\bar n_2 \in H^2[\cB (Z_2\times Spin_\infty);\Z_2] $,
$\bar n_2$ can be written as a combination of $\bar a^{\Z_2}$, $\bar\w_1$ and
$\bar\w_2$. However, for $Spin_\infty$ $\bar \w_1\se{2,\dd}\bar \w_2\se{2,\dd}0$.
Thus $\bar n_2$ is given by
\begin{align}
\bar n_2 \se{2} \al_n (\bar a^{\Z_2})^2 .
\end{align}
Thus $\bar n_2$ has two choices: $\al_n=0,1$.

\noindent\textbf{Calculate $\bar \nu_3$}:  
Next, we consider $\bar \nu_3(\bar a^{G_{fSO}})$ in \eqn{fSPTclSO} which becomes
\begin{align}
 -\dd \bar \nu_3 &\se{1} \frac12 \Sq^2 \bar n_2 +\frac12 \bar n_2 \dd \bar a^{\Z_2^f}
\nonumber\\ &
 \se{1} 
\frac{\al_n}2 (\bar a^{\Z_2})^4 
+\frac{\al_n}2 (\bar a^{\Z_2})^2 \dd \bar a^{\Z_2^f}
\end{align}
where we have labeled $\bar a^{G_{fSO}}_{ij} \in G_{fSO}= Z_2\times (Z_2^f\gext
SO_\infty)$ by a triple $(\bar a^{Z_2}_{ij}, \bar a^{Z_2^f}_{ij}, \bar a^{SO}_{ij})$.  We have
used the fact that $\bar \w_2(a^{SO})$ is a coboundary: $\bar \w_2\se{2} \dd
\bar a^{\Z_2^f}$.  The solution of the above equation has a form
\begin{align}
\bar  \nu_3(\bar a^{\Z_2},\bar a^{\Z_2^f}) 
&\se{1} 
\frac{\al_\nu}2 (\bar a^{\Z_2})^3 
+\frac{\al_n}4 (\bar a^{\Z_2})^3 
+\frac{\al_n}2 (\bar a^{\Z_2})^2  \bar a^{\Z_2^f}
\end{align}
where $\al_\nu =0,1$.

\noindent\textbf{SPT invariant}:
The SPT invariant is given by
\begin{align}
& \ \ \ \
Z^\text{top}(\cM^{3},A^{G_{fSO}})
\nonumber \\
& = 
\ee^{\ii 2\pi \int_{\cM^{3}} \frac{\al_\nu}2 (A^{\Z_2})^3 
+\frac{\al_n}4 {A^{\Z_2}}^3 
+\frac{\al_n}2 (A^{\Z_2})^2  A^{\Z_2^f}}
\nonumber\\
& \ \ \ \ \ee^{ \ii \pi \int_{\cN^{4}} \Sq^2 f_2 + f_2 \w_2},
\nonumber \\
& f_2\big|_{\prt \cN^4} \se{2} \al_n (A^{Z_2})^2 ,\ \ \ \ \
\dd A^{\Z_2} \se{2} \w_2.
\end{align}
where the background connection $A^{G_{fSO}}$ is labeled by a triple
$(A^{\Z_2},A^{\Z_2^f},A^{SO})$. As a result, $Z_2\times Z_2^f$ SPT states from
fermion decoration are labeled by $(\al_n,\al_\nu)$.  It turns out that all
those labels $(\al_n,\al_\nu)$ are inequivalent according to \eqn{gaugetomSO}
and \eqn{gaugendSO}.  Thus there are 4 different $Z_2\times Z_2^f$ fermionic
SPT states from fermion decoration.  

However, for non-interacting fermions, the 2+1D $Z_2\times Z_2^f$-SPT
phases are labeled by $\Z$.  After include interaction, $\Z$ reduces to $\Z_8$,
and there are 8 different fermionic $Z_2\times Z_2^f$-SPT
phases\cite{RZ1232,Q1302,YR1307,GL1369}.  The extra fermionic SPT phases must
come from the decoration of the topological $p$-wave superconducting
chains.\cite{KT170108264,WG170310937} In this paper, we only develop a generic
theory for fermion decoration, which misses some of the $Z_2\times Z_2^f$
fermionic SPT phases. We hope to develop a generic theory for the decoration of
the topological $p$-wave superconducting chains in future.

\subsection{3+1D}

\subsubsection{Without extension of $SO_\infty$}

\noindent\textbf{Calculate $\bar n_3$}:  
$\bar n_3\in H^3(\cB Z_2;\Z_2) =\Z_2$ has two choices: $\bar n_3
=\al_n (\bar a^{\Z_2})^3$, $\al_n=0,1$.  These two $\bar n_3$'s are not equivalent.

\noindent\textbf{Calculate $\bar \nu_4$}:  
$\bar \nu_4$ is obtained from \eqn{fSPTcl}, which has a form  
\begin{align}
 -\dd \bar \nu_4 \se{1} \frac{\al_n}2 \Sq^2 (\bar a^{\Z_2})^3.
\end{align}
It turns out that $  \frac12 \Sq^2 (\bar a^{\Z_2})^3  \se{1} \frac12 (\bar
a^{\Z_2})^5 $ (see \eqn{a3a5}),  which is a non-trivial element in $H^5(\cB
Z_2;\RZ)=\Z_2$.  Thus $\bar \nu_4$ has no solution when $\al_n=1$ and $\al_n$
must be zero.  When $\bar n_3=0$, $\bar \nu_4 $ has only one inequivalent
solution since $H^4(\cB Z_2;\RZ)=0$.

\subsubsection{With extension of $SO_\infty$}

\noindent\textbf{Calculate $\bar n_3$}:  From $G_{fSO}=Z_2\times Spin_\infty$,
we see that  $\bar n_3 \in H^3(G_{fSO};\Z_2)$ is generated by $\bar a^{\Z_2}$
and Stiefel-Whitney class $\bar \w_n$.  For $Spin_\infty$, $\bar \w_1=\bar
\w_2=0$. Also, $\Sq^1\bar \w_2 \se{2,\dd} \bar \w_1\bar \w_2 +\bar \w_3
\se{2,\dd} \bar \w_3$.  Since $\bar \w_2$ is a coboundary for $Spin_\infty$,
$\bar \w_3$ is also a coboundary.  Thus, $\bar n_3$ is given by
\begin{align}
 \bar n_3 \se{2} \al_n (\bar a^{\Z_2})^3 ,\ \ \ \ \ \al_n=0,1.
\end{align}

\noindent\textbf{Calculate $\bar \nu_4$}:  
$\bar \nu_4$ is obtained from 
\begin{align}
 -\dd \bar \nu_4 \se{1,\dd}
\frac{\al_n}2 \Sq^2 (\bar a^{\Z_2})^3.
\end{align}
Since $ \frac12 \Sq^2 (\bar a^{\Z_2})^3  \se{1} \frac12 (\bar a^{\Z_2})^5$ (see
\eqn{a3a5}) is the non-trivial element in $H^5(\cB
G_{fSO};\RZ)=\Z_2$.  Thus $\bar \nu_4$ has solution only when $\al_n=0$.

When $\bar n_3=0$, $\bar \nu_4 $ has a form
\begin{align}
 \bar \nu_4(\bar a^{G_{fSO}}) \se{1} 
 \frac{\al_{\nu,1}}2 (\bar a^{\Z_2})^4
+\frac{\al_{\nu,2}}2 \bar \w_4 +\al_\nu p_1,
\nonumber\\
\al_{\nu,1},\al_{\nu,2}=0,1 ,\ \ \ \
\al_\nu \in [0,1)
,
\end{align}
where $p_1$ is the first Pontryagin class.
However, $\frac12 (\bar a^{\Z_2})^4$ is a coboundary in  $B^4(\cB
G_{fSO};\RZ)$.  

Also, in 3+1D space-time $\cM^4$, $ \w_4 \se{2,\dd} 
\w_2^2+ \w_1^4$ (see Appendix \ref{Rswc4D}).  Since $\cM^4$ is orientable
spin manifold, $ \w_2^2\se{2,\dd} \w_1^4\se{2,\dd}0$, we also have
$ \w_4 \se{2,\dd}0$.  Last $\al_\nu$ is not quantized and different values
of $\al_\nu$'s are connected and belong to the same phase.  Thus the above
solutions are equivalent.  We find that there is only one trivial fermionic
$Z_2\times Z_2^f$ SPT phases in 3+1D from fermion decoration.  This agrees with
the result in \Ref{GW1441}.  

\Ref{KTT1429,FH160406527} showed that there is only one trivial fermionic $Z_2\times
Z_2^f$-SPT phases in 3+1D.  Our result is also consistent with that.

\section{Fermionic $Z_4^f$-SPT state}

In this section, we are going to study fermionic SPT phases with $G_f=Z_4^f$
symmetry in 2+1D and in 3+1D.  Such a symmetry can be realized by a
charge-$2e$ superconductor of electrons where the $Z_4^f$ symmetry is generated
by $180^\circ$ $S_z$-spin rotation.  Another way to realize the $Z_4^f$
symmetry is via charge-$4e$  superconductors of electrons.  This kind of
fermionic SPT states is beyond the approach in
\Ref{GW1441,KT170108264,WG170310937} which only deal with $G_f$ of the form
$G_f=\Z_2^f\times G_b$.

For fermion systems with bosonic symmetry $G_b=Z_2$, the full fermionic
symmetry $G_f$ is an extension of $G_b$ by $Z_2^f$.  The  bosonic symmetry
$G_b=Z_2$ has two extensions described by $\bar e_2\in H^2(\cB Z_2;\Z_2)=\Z_2$.
For $\bar e_2\se{2}0$, the extension is $G_f=Z_2^f \times Z_2$. The
corresponding fermion SPT phases are discussed in the last section.  For $\bar
e_2\se{2} (\bar a^{\Z_2})^2$, the extension is $G_f=Z_2^f \gext Z_2=Z_4^f$. We
will discuss the corresponding fermionic SPT phases in this section.

For $G_f=Z_4^f$, the group $G_{fSO}$ is an extension of $SO_\infty$ by $Z_4^f$:
\begin{align} G_{fSO}=Z_4^f\gext_{e_2} SO_\infty=(Z_2^f\gext Z_2)\gext
SO_\infty .  \end{align} The possible extensions of $SO_\infty$ by $Z_4$ are
labeled by $\bar e_2\in H^2(\cB SO_\infty; \Z_4)=\Z_2$ which is generated by
$\bar e_2\se{4} 2\bar \w_2(\bar a^{SO})$.

The links in the simplicial complex $\cB G_{fSO}$ are labeled by $\bar
a^{G_{fSO}}_{ij}\in G_{fSO}$ (see \Ref{ZLW} and Appendix \ref{hgroup}).  We may
label the elements  $\bar a^{G_{fSO}}_{ij} \in G_{fSO}$ by a pair $\bar
a^{SO}_{ij}\in SO_\infty$ and $\bar a^{Z_4^f}_{ij} \in \Z_4^f$:
\begin{align}
\label{fSOZ4}
 a^{G_{fSO}} = (a^{Z_4^f}, a^{SO}).
\end{align}
  This allows us
to introduce two projections $\pi(\bar a_{ij}^{G_{fSO}})=\bar a_{ij}^{SO}$ and
$\si(\bar a_{ij}^{G_{fSO}})=\bar a_{ij}^{Z_4^f}$ (see Appendix \ref{cenext}).
Thus we can also label the links using a pair $(\bar a_{ij}^{Z_4^f}, \bar
a_{ij}^{SO})$.  Although $\w_2(\bar a^{SO})$ is a cocycle in $C^2(\cB
SO_\infty;\Z_2)$, $2\w_2(\pi(\bar a^{G_{fSO}}))$, when viewed as a function of
$\bar a^{G_{fSO}}$, is a coboundary in $B^2(\cB G_{fSO};\Z_4)$.  In other
words, the two canonical 1-cochain on $\cB G_{fSO}$, $\bar a^{SO}_{ij}$ and
$\bar a^{Z_4^f}_{ij}$, are related by (see \eqn{e2sig})
\begin{align}
\dd \bar a^{Z_4^f} \se{4} 2  \bar \w_2(\bar a^{SO}) 
.
\end{align}
We can write $\bar a^{Z_4^f}$ as
\begin{align}
\label{Z4Z2Z2}
 \bar a^{Z_4^f} \se{4} \bar a^{Z_2} + 2\bar a^{\Z_2^f}
\end{align}
where $\bar a^{Z_2}$ and $\bar a^{\Z_2^f}$ are $\Z_2$-valued 1-cochains.  We see
\begin{align}
\label{aZ4fw2}
\bar a^{Z_4^f} \se{2} \bar a^{\Z_2},\ \ \ \
\dd \bar a^{\Z_2} \se{2} 0.  
\end{align}
Eqn. \eq{aZ4fw2} implies that
\begin{align}
\Bs_2  \bar a^{Z_2} \se{2} \bar \w_2(\bar a^{SO}) +\dd \bar a^{\Z_2^f}.
\end{align}
which can be rewritten as (see \eqn{Sq1Bs2})
\begin{align}
\label{a2w2}
\Sq^1  \bar a^{Z_2} \se{2} (\bar a^{Z_2})^2 \se{2} \bar \w_2(\bar a^{SO})
+\dd \bar a^{\Z_2^f}
 .
\end{align}

\subsection{2+1D}

\subsubsection{Without extension of $SO_\infty$}

\noindent\textbf{Calculate $\bar n_2$}:  
First, $\bar n_2\in H^2(\cB Z_2;\Z_2) =\Z_2$. It has two choices: $\bar n_2
=\al_n (\bar a^{\Z_2})^2$, $\al_n=0,1$.

\noindent\textbf{Calculate $\bar \nu_3$}:  
Similar to the last section, $\bar \nu_3$ satisfies
\begin{align}
-\dd \bar \nu_3 & \se{1}  
 \frac{\al_n^2}2 \Sq^2 (\bar a^{\Z_2})^2 
+ \frac{\al_n}2 (\bar a^{\Z_2})^2\bar e_2 
\se{1} 0 .
\end{align}
Thus, $\bar \nu_3$ has two solutions: 
\begin{align}
\bar  \nu_3 \se{1} \frac{\al_\nu}2 (\bar a^{\Z_2})^3,\ \ \ \al_\nu=0,1,
\end{align}
since $H^3(\cB Z_2;\RZ) =\Z_2$.

\noindent\textbf{SPT invariant}:
The four $Z_4^f$ fermionic SPT states labeled by $\al_n,\al_\nu=0,1$ 
have the following SPT invariant
\begin{align}
\label{SPTinvZ4f}
& \ \ \ \
Z^\text{top}(\cM^3,A^{Z_2},A^{\Z_2^f}) 
\nonumber\\
& = 
 \ee^{\ii 2\pi \int_{\cM^3}  \frac{\al_\nu}{2} (A^{\Z_2})^3 
+\frac{\al_n}2   (A^{\Z_2})^2 A^{\Z_2^f} } 
\ee^{ \ii \pi \int_{\cN^{4}}  \Sq^2 f_2+f_2\w_2   }, 
\nonumber\\
& \ \ \ \
\dd A^{\Z_2^f} \se{2} \w_2+(A^{\Z_2})^2
, \ \ \ \ f_2\big|_{\prt \cN^4} \se{2} \al_n (A^{\Z_2})^2
\end{align}
where the space-time $\cM^3$ is orientable and $\w_1\se{2}0$.  However, as we
will see below, the four $Z_4^f$ fermionic SPT states all belong to the same
phase.

On 2+1D space-time manifold, $\w_2+\w_1^2 \se{2,\dd}0$ (see Appendix
\ref{Rswc3D}).  The $Z_4^f$ fermionic symmetry requires the space-time $\cM^3$
to be a orientable manifold with $\w_2+(A^{\Z_2})^2 \se{2,\dd}0$ and $\w_1
\se{2,\dd}0$. Thus $(A^{\Z_2})^2$ is always a coboundary:
$(A^{\Z_2})^2 = \dd u_1$.  

Let us write $f_2=\t f_2+ \al_n \dd u_1$ where $\t f_2 \se{2}0$ on $\prt
\cN^4$, which implies $\ee^{ \ii \pi \int_{\cN^{4}}  \gSq^2 \t f_2+\t f_2\w_2} =
1$. The SPT invariant now becomes (see \eqn{Sqgauge})
\begin{align}
& \ \ \ \ 
Z^\text{top}(\cM^3,A^{Z_2},A^{\Z_2^f}) 
\nonumber\\
&
 = 
 \ee^{\ii 2\pi \int_{\cM^3}  \frac{\al_\nu}{2} A^{\Z_2}\dd u_1 +\frac{\al_n}2   \dd u_1 A^{\Z_2^f} } 
\ee^{ \ii \pi \int_{\cM^3} \al_n  (\gSq^2 u_1 + u_1\w_2) }
\nonumber\\
&=
\ee^{\ii \pi \int_{\cM^3}  \al_n  [ u_1 (\w_2+\dd u_1)  
+  (u_1 \dd u_1 + u_1\w_2)] }
 = 1.
\end{align}
We see that the SPT invariant is independent of $\al_n,\al_\nu$.  Thus at the
end, we get only one $Z_4^f$ fermionic SPT phase, which is the trivial SPT
phase.  

\subsubsection{With extension of $SO_\infty$}
\label{Z4f2d}

\noindent\textbf{Calculate $\bar n_2$}:  
With extension of $SO_\infty$,
in general, $\bar n_2(\bar a^{G_{fSO}}) \in H^2(\cB
G_{fSO};\Z_2)$ is given by [using the pair $(\bar a^{SO},\bar a^{Z_4^f})$ to
label $\bar a^{G_{fSO}}$]
\begin{align}
\bar n_2(\bar a^{G_{fSO}}) 
&\se{2} \al_{n,1}\bar \w_2(\bar a^{SO}) + \al_{n,2}(\bar a^{Z_2})^2,
\end{align}
$\al_{n,1},\al_{n,2}=0,1$.  
The above can be reduced to
\begin{align}
\label{n2aZ4}
\bar n_2(\bar a^{G_{fSO}}) \se{2} 
    \al_{n}(\bar a^{Z_2})^2
\end{align}
$\al_n=0,1$,  due to the relation \eq{a2w2}.  So $\bar n_2(\bar a^{G_{fSO}})$
has two choices, and $H^2(\cB G_{fSO};\Z_2)=\Z_2$.

\noindent\textbf{Calculate $\bar \nu_3$}:  
Next, we consider $\nu_3$ in \eqn{fSPTclSO} which becomes
\begin{align}
-\dd \bar \nu_3 &\se{1} \frac12 \Sq^2 \bar n_2 +\frac12 \bar n_2 \bar \w_2 \se{1}  
\frac {\al_n}2 (\bar a^{Z_2})^2 \dd \bar a^{\Z_2^f}
,
\end{align}
where we have used \eqn{a2w2} and \eqn{n2aZ4}.  We find that $\bar \nu_3$ is given
by
\begin{align}
\bar  \nu_3 
\se{1} \frac{\al_\nu}2 (\bar a^{\Z_2})^3 + \frac {\al_n}2 (\bar a^{Z_2})^2 \bar a^{\Z_2^f}, 
\ \ \ \ \ \al_\nu=0,1.
\end{align} 

\noindent\textbf{SPT invariant}:
This leads to the SPT invariant 
\begin{align}
&\ \ \ \
Z^\text{top}(\cM^3,A^{G_{fSO}}) 
\nonumber\\
& = \ee^{\ii 2\pi \int_{\cM^3}  \frac{\al_\nu}{2} (A^{\Z_2})^3 
+ \frac {\al_n}2 (A^{\Z_2})^2 A^{\Z_2^f}
} 
\ee^{
\ii \pi \int_{\cN^{4}}  \Sq^2 f_2+f_2\w_2   
} ,
\nonumber\\ & \ \ \ \
f_2 \big|_{\prt \cN^{4}} \se{2} \al_n (A^{\Z_2})^2,\ \ \
\dd A^{\Z_2^f} \se{2} \w_2+(A^{\Z_2})^2
,
\end{align}
which as calculated above is always an identity.
Thus, there is only one trivial $Z_4^f$ SPT fermionic phase in 2+1D.  This
agrees with the result obtained in \Ref{LW160205946}.

The $Z_4^f$ symmetry was denoted by $G_-(C)$ symmetry in \Ref{W11116341}.
There, it was found that for non-interacting fermion systems with $Z_4^f=G_-(C)$ symmetry,
the SPT phases in 2+1D is classified by $\Z$.  
The result from \Ref{LW160205946} indicates that all of those non-interacting fermion
$Z_4^f$-SPT states actually correspond to trivial SPT states in the presence of
interactions.

\subsection{3+1D}
\label{Z4f31D}

\subsubsection{Without extension of $SO_\infty$}

\noindent\textbf{Calculate $\bar n_3$}:  
$\bar n_3 \in H^3(\cB Z_2;\Z_2)$ has two
choices:
\begin{align}
\bar n_3 = \al_n (\bar a^{\Z_2})^3, 
\end{align}
$\al_n=0,1$, since $H^3(\cB Z_2;\Z_2) = \Z_2$.  

\noindent\textbf{Calculate $\bar \nu_4$}:  
Next we want to solve
\begin{align}
\label{tom4}
 -\dd \bar \nu_4 & \se{1}  \frac12 (\Sq^2\bar n_3 + \bar n_3 \bar e_2) \se{1} 0 .
\end{align}
where we have used \eqn{a3a5}.  Since $H^4(\cB Z_2;\RZ)=0$, the solution of
\eqn{tom4} is unique $\bar \nu_4\se{1}0$.

\noindent\textbf{SPT invariant}:
This leads to the SPT invariant 
\begin{align}
\label{Z4finv}
Z(\cM^3,A^{\Z_4^f}) 
& = \ee^{\ii \pi \int_{\cM^4}  
\al_n (A^{\Z_2})^3 A^{\Z_2^f}
} 
\ee^{
\ii \pi \int_{\cN^{5}}  \Sq^2 f_2+f_2  \w_2 
} ,
\nonumber\\ &
f_3 \big|_{\prt \cN^{4}} \se{2} \al_n (A^{\Z_2})^3
,
\end{align}
where we have written $A^{\Z_4^f}$ as
\begin{align}
 A^{\Z_4^f}\se{4} A^{\Z_2}+2A^{\Z_2^f},
\end{align}
which satisfies
\begin{align}
 \dd A^{\Z_4^f} &\se{4} 2\w_2,  
\\
 \dd A^{\Z_2} & \se{2} 0,\ \ \ \
 \dd A^{\Z_2^f} \se{2} \w_2+\Bs_2 A^{\Z_2} \se{2} \w_2+(A^{\Z_2})^2 .
\nonumber 
\end{align}

\subsubsection{With extension of $SO_\infty$}

\noindent\textbf{Calculate $\bar n_3$}:  
In general, $\bar n_3(\bar a^{G_{fSO}}) \in H^3(\cB G_{fSO};\Z_2)$ can be written as
\begin{align}
& \bar n_3(\bar a^{G_{fSO}}) \se{2} 
    \al_{n}\bar \w_3(\bar a^{SO}) 
+ \al_{n}'(\bar a^{Z_2})^3
+ \al_{n}'' \bar \w_2(\bar a^{SO}) \bar a^{Z_2} ,
\nonumber\\
&\ \ \ \ \ \ \ \al_{n},\al_{n}',\al_{n}''=0,1 .
\end{align}
However, from $\bar \w_2 \se{2,\dd}(\bar a^{Z_2})^2$, we find that $\bar \w_2
\bar a^{Z_2} \se{2,\dd}(\bar a^{Z_2})^3$ and $\Sq^1 (\bar \w_2+ (\bar
a^{Z_2})^2) \se{2,\dd} \bar \w_1 \bar \w_2+\bar \w_3 \se{2,\dd}\bar \w_3
\se{2,\dd} 0$ (see Appendix \ref{Rswc} and notice $\bar \w_1\se{2}0$).  Thus the
above expression for $n_3(\bar a^{G_{fSO}})$ is reduced to
\begin{align}
 n_3(\bar a^{G_{fSO}}) &\se{2} 
 \al_{n}(\bar a^{Z_2})^3 ,
\nonumber\\
 \al_{n}&=0,1 .
\end{align}
There are two choices of $n_3$.

\noindent\textbf{Calculate $\bar \nu_4$}:  
Next, we consider $\bar \nu_4$ in \eqn{fSPTclSO} which becomes (see \eqn{a2w2}
and \eqn{a3a5})
\begin{align}
 -\dd \bar \nu_4(\bar a^{G_{fSO}}) 
& \se{1}  
\frac12 \Sq^2 \bar n_3(\bar a^{G_{fSO}})+\frac12 \bar n_3(\bar a^{G_{fSO}}) \bar \w_2(\bar a^{SO})
\nonumber\\
& \se{1}  
\frac{\al_n}2 (\bar a^{\Z_2})^5 +\frac{\al_n}2 (\bar a^{\Z_2})^3 \bar \w_2(\bar a^{SO})
\nonumber\\
& \se{1} \frac{\al_n}2 (\bar a^{\Z_2})^3 \dd \bar a^{Z_2^f}
.
\end{align}

From $\frac12 \w_2 \se{1,\dd} \frac12 (\bar a^{Z_2})^2 $, we obtain that
$\frac12 \w_2 (\bar a^{Z_2})^2 \se{1,\dd} \frac12 (\bar a^{Z_2})^4 $ and
$\frac12 (\w_2)^2 \se{1,\dd} \frac12 (\bar a^{Z_2})^2 \w_2 $.  Thus  $\frac12
(\bar \w_2)^2 \se{1,\dd} \frac12 (\bar a^{Z_2})^4$.  Since $H^4(\cB
Z_2;\R/\Z)=0$, we find that $ \frac12 (\bar a^{Z_2})^4 \se{1,\dd} 0 $.  We also
find (see Appendix \ref{Rswc}) $0\se{2,\dd} \Sq^1 (\bar \w_2+ (\bar a^{Z_2})^2)
\se{2,\dd} \bar \w_1 \bar \w_2+\bar \w_3 \se{2,\dd}\bar \w_3 $ and $0\se{2,\dd}
\gSq^2 (\bar \w_2+ (\bar a^{Z_2})^2)= \bar \w_2^2+(\bar a^{Z_2})^4 $.  To
summarize, we have
\begin{align}
 \frac12 (\bar a^{Z_2})^4 \se{1,\dd} 
 \frac12 \bar \w_2^2  \se{1,\dd} 
 \frac12 \bar a^{Z_2} \bar \w_3  \se{1,\dd} 
 \frac12 (\bar a^{Z_2})^2 \w_2 \se{1,\dd} 0  .
\end{align}
This allows us to conclude that $\bar \nu_4$ must have a form
\begin{align}
\bar  \nu_4 &\se{1} \frac{\al_n}2 (\bar a^{\Z_2})^3  \bar a^{Z_2^f} 
+\frac{\al_\nu}2 \bar p_1.
\end{align} 
But on $\cM^4$, we have $p_1 \se{2,\dd} \w_2^2 \se{2,\dd}0$.
Thus the pullback of $\bar  \nu_4$ to $\cM^4$ reduces to
\begin{align}
\nu_4 &\se{1} \frac{\al_n}2 ( a^{\Z_2})^3  a^{Z_2^f} .
\end{align} 

\noindent\textbf{SPT invariant}:
The corresponding fermionic model (in the bosonized form) is given by
\begin{align}
\label{Z4fZ}
&\ \ \ \
 Z(\cM^4,A^{G_{fSO}}) 
\nonumber\\
& = \sum_{g_i \in G_{fO}} \ee^{\ii \pi \int_{\cM^4}  
\al_n (a^{\Z_2})^3 a^{\Z_2^f}
} 
\ee^{
\ii \pi \int_{\cN^{5}}  \Sq^2 f_2+f_2  (\w_2+\w_1^2) 
} ,
\nonumber\\ & \ \ \ \
f_3 \big|_{\prt \cN^{5}} \se{2} \al_n (a^{\Z_2})^3,
\end{align}
which lead to a SPT invariant given by
\begin{align}
\label{Z4finvSO}
&\ \ \ \
 Z^\text{top}(\cM^4,A^{G_{fSO}}) 
\nonumber\\
& = \ee^{\ii \pi \int_{\cM^4}  
\al_n (A^{\Z_2})^3 A^{\Z_2^f}
} 
\ee^{
\ii \pi \int_{\cN^{5}}  \Sq^2 f_2+f_2  \w_2 
} ,
\nonumber\\ & \ \ \ \
f_3 \big|_{\prt \cN^{5}} \se{2} \al_n (A^{\Z_2})^3 ,\ \ \
\dd A^{\Z_2^f} \se{2} \w_2+(A^{\Z_2})^2 ,
\end{align}
where $A^{G_{fSO}}$ is labeled by $(A^{Z_2},A^{Z_2^f},A^{SO})$ (see \eqn{fSOZ4}
and \eqn{Z4Z2Z2}).  This agrees with \eqn{Z4finv}

Is the above SPT invariant trivial or not trivial?  One way to show the
non-trivialness is to change $A^{Z_2^f}$ by a $\Z_2$-valued 1-cocycle
$A_0^{Z_2^f}$. In this case $ Z^\text{top}(\cM^4,A^{G_{fSO}})$ changes by a
factor
\begin{align}
 \t  Z^\text{top} 
= \ee^{\ii \pi \int_{\cM^4}  \al_n (A^{\Z_2})^3 A_0^{\Z_2^f} }
= \ee^{\ii \pi \int_{\cM^4}  \al_n \w_2 A^{\Z_2} A_0^{\Z_2^f} } 
\end{align}
where we have used $\w_2 \se{2,\dd} (A^{\Z_2})^2$.  Since 
\begin{align}
\w_2 A^{\Z_2} A_0^{\Z_2^f} 
&\se{2,\dd} \Sq^2 (A^{\Z_2} A_0^{\Z_2^f}) 
\se{2,\dd} (\Sq^1 A^{\Z_2})(\Sq^1 A_0^{\Z_2^f})
\nonumber\\
& \se{2,\dd} (\frac12 \dd A^{\Z_2})(\frac12 \dd A_0^{\Z_2^f}).
\end{align}
We find
\begin{align}
\t  Z^\text{top} 
= \ee^{\ii \pi \int_{\cM^4}  \al_n (\frac12 \dd A^{\Z_2})(\frac12 \dd A_0^{\Z_2^f}) }=1, 
\end{align}
which is independent of $\al_n$. This suggests that $\al_n=0,1$ describes the
same SPT phases.

\subsection{$Z_4^f\times Z_2^T$-SPT model}

We have seen that the model \eq{Z4fZ} describes a trivial $Z_4^f$-SPT phase
even when $\al_n=1$.  However, we note that the model \eq{Z4fZ} actually has a
$Z_4^f\times Z_2^T$ symmetry.  It turns out that the model \eq{Z4fZ} realizes a
non-trivial $Z_4^f\times Z_2^T$-SPT phase when $\al_n=1$.

To physically detect that non-trivial $Z_4^f\times Z_2^T$-SPT phase, we note
that $n_3=\al_n (A^{\Z_2})^3$ is the 3-cocycle fermion current.  Let us put the
$Z_4^f$-SPT state on space-time $S^1_t\times M^3$, where $A^{G_{fSO}}$ on
$S^1_t\times M^3$ is the pullback of a $A^{G_{fSO}}$ on the space $M^3$.  In
this case, using $(A^{\Z_2})^2\se{2,\dd} \w_2 $ and $\w_1^2\se{2,\dd} \w_2 $ on
$M^3$, we can rewrite $(A^{\Z_2})^3$ on $M^3$ as $A^{\Z_2}\w_2\se{2,\dd}
A^{\Z_2} \w_1^2$.  Now we can choose $M^3=\R P^2\times S^1$. Then we have
\begin{align}
\int_{M^3} n_3 \se{2}
\al_n \int_{\R P^2\times S^1} A^{\Z_2} \w_1^2
\se{2} \al_n\int_{S^1_z} A^{\Z_2}.
\end{align}
We find that, if we put the $Z_4^f$-SPT state on space $\R P^2\times S^1$,
the fermion number in the ground state will be given by
$N_f\se{2}\al_n\int_{S^1} A^{\Z_2}$.  Thus adding a $Z_4^f$ symmetry twist
around $S^1_z$ will change the fermion number in the ground state by $\al_n$
mod 2.  When $\al_n=1$, the non-trivial change in the fermion number indicates
the non-trivialness of the SPT phase.

To realize such a fermionic $Z_4^f\times Z_2^T$-SPT phase in 3+1D, we start
with a symmetry breaking state that break the $Z_4^f$ to $Z_2^f$. The
order-parameter has a $Z_2$-value.  We then consider a random configuration of
order-parameter in 3d space.  A fermion decoration that gives rise to the
$\al_n=1$ phase is realized by binding a fermion worldline to $*(*W_3)^3$.
Here $W_3$ is the 3-dimensional domain wall of the $Z_2$-order-parameter.  $*$
is the Poincar\'e dual.  Thus $*W_3$ is a 1-cocycle. $*(*W_3)^3$ is the
Poincar\'e dual of a 3-cocycle which is a closed loop.  The fermion worldline is
attached to such a loop.

\section{Fermionic $Z_2^f\times Z_2^T$-SPT state}

The fermion symmetry $Z_2^f\times Z_2^T$ is realized by electron
superconductors with coplanar spin polarization.  The time reversal $Z_2^T$ is
generated by the standard time reversal followed by a 180$^\circ$ spin
rotation.

For $G_f=Z_2^T\times Z_2^f$, $G_{fO}$ is a $Z_2^f$ extension of $O_\infty$:
$G_{fO}=Z_2^f\gext O_\infty$.  Such kind of extensions are classified by
$H^2(\cB O_\infty;\Z_2)=\Z_2^2$ which is generated by $\w_1^2$ and $\w_2$.  For
fermion symmetry $Z_2^f\times Z_2^T$, we should choose the extension
$G_{fO}=Z_2^f\gext_{\w_2+\w_1^2} O_\infty=Pin^-_\infty$.  According to Appendix
\ref{cenext}, this implies that, on $\cB Pin^-_\infty$, the canonical 1-cochain
$\bar a^{G_{fO}}$ satisfy a relation
\begin{align}
\label{w1w2}
\bar \w_2(\bar a^{G_{fO}})+\bar \w_1^2(\bar a^{G_{fO}})\se{2} \dd \bar a^{Z_2^f}.  
\end{align}

\subsection{2+1D}
\label{Z2TZ2f2d}

\subsubsection{Without extension of $O_\infty$}

\noindent\textbf{Calculate $\bar n_2$}:  
From $\bar n_2 \in H^2(\cB Z_2^T;\Z_2)=\Z_2$, we obtain
\begin{align}
 \bar n_2\se{2,\dd}\al_n (\bar a^{Z_2^T})^2 \se{2,\dd} \al_n \bar \w_1^2 ,
\ \ \ \  \al_n=0,1
.
\end{align}  

\noindent\textbf{Calculate $\bar \nu_3$}:  
Next, we consider $\bar \nu_3$ in \eqn{fSPTcl}.  Since $\bar e_2=0$, only the
term $\frac12 \Sq^2 \bar n_2 = \frac12 \bar n_2^2$ is non-zero.  The term
$\frac12 \bar n_2^2=\frac{\al_n}2 (\bar a^{\Z_2^T})^4$ is a cocycle in
$\cZ^4(Z_2^T;(\RZ)_T)$.  Note that now $Z_2^T$ has a non-trivial action on the
value $\RZ$ and the differential operator $\dd$ is modified by this non-trivial
action [which corresponds to the cases of non-trivial $\al_i$ in \Ref{ZLW} and
Appendix \ref{hgroup}, see \eqn{cn4}].  This modifies the group cohomology.
Since $H^4(\cB Z_2^T;(\RZ)_T)=\Z_2$, and $\frac12 (a^{\Z_2^T})^4$ is the
non-trivial cocycle in $H^4(\cB Z_2^T;(\RZ)_T)$, \eqn{fSPTcl} has solution only
when $\al_n=0$.  In case $\al_n=0$, there is only one solution $\nu_3\se{1}0$,
since $H^3(\cB Z_2^T;(\RZ)_T)=0$.

From \eqn{ZfSPTZ2Z2}, we see that when $\al_n=1$, the action amplitude is not
real, and breaks the time reversal symmetry.  This is why  $\al_n=1$ is not a
solution for time-reversal symmetric cases.

\subsubsection{With extension of $O_\infty$}

\noindent\textbf{Calculate $\bar n_2$}:  Due to the relation \eqn{w1w2} on $\cB
G_{fO}=\cB Pin^-_\infty$, $H^2(\cB Pin^-_\infty)=\Z_2$ which is generated by
$\bar \w_2\se{2,\dd}\bar \w_1^2$.  Therefore, $\bar n_2\in H^2(\cB
Pin^-_\infty;\Z_2)=\Z_2$ has two choices 
\begin{align}
n_2 \se{2} \al_n \bar \w_1^2,\ \ \ \
\al_n=0,1.  
\end{align}

\noindent\textbf{Calculate $\bar \nu_3$}:  
Next, we consider $\bar \nu_3$ that satisfy \eqn{fSPTclO}
which becomes, in the present case,
\begin{align}
 -\dd \bar \nu_{3} 
& \se{1} \frac{\al_n}2 [\bar \w_1^4 +\bar \w_1^2 (\bar \w_2+\bar \w_1^2)] .
\nonumber\\
& \se{1} \frac{\al_n}2 [\bar \w_1^4 +\bar \w_1^2 \dd \bar a^{\Z_2^f}] .
\end{align}
where we have used
\begin{align}
 \dd \bar a^{\Z_2^f} \se{2} \bar \w_2+\bar \w_1^2.
\end{align}
$\frac12 \bar \w_1^4$ is a non-trivial cocycle in $H^4(\cB
Pin^-_\infty;(\RZ)_T)$.  Thus \eqn{fSPTclO} has solution only when $\al_n=0$.
In this case, there are four solutions 
\begin{align}
\bar  \nu_{3} \se{1}  \frac{\al_\nu}2 \bar \w_3 + \frac{\t\al_\nu}2 \bar \w_1^3,\ \ \ 
\al_\nu,\t\al_\nu=0,1.
\end{align}

\noindent\textbf{SPT invariant}:
The corresponding SPT invariant is given by
\begin{align}
& \ \ \ \
Z^\text{top}(\cM^{3},A^{G_{fO}})
\nonumber \\
& = 
\ee^{\ii 2\pi \int_{\cM^{3}} 
 \frac{\al_\nu}2 \w_3(A^{O}) + \frac{\t\al_\nu}2 \w_1^3(A^{O})
}
\ee^{ \ii \pi \int_{\cN^{4}} \Sq^2 f_2 + f_2 \w_2},
\nonumber \\
& f_2\big|_{\prt \cN^4} \se{2} 0\ \to \
\ee^{ \ii \pi \int_{\cN^{4}} \Sq^2 f_2 + f_2 \w_2} =1
,
\end{align}
where the background connection $A^{G_{fO}}$ is labeled by
$(A^{\Z_2^f},A^{O})$.  On 2+1D space-time $\cM^3$, we always have
$\w_2+\w_1^2\se{2,\dd}0$ and $\w_1^3 \se{2,\dd} \w_2\w_1 \se{2,\dd}\w_3
\se{2,\dd}0$ (see Appendix \ref{Rswc3D}).  The above four solutions give rise
to the same SPT invariant and the same fermionic $Z_2^f\times Z_2^T$ SPT phase.

For non-interacting fermions, there is no non-trivial fermionic $Z_2^f\times
Z_2^T$-SPT phase.  The above result implies that, for interacting fermions,
fermion decoration also fail to produce any non-trivial fermionic $Z_2^f\times
Z_2^T$-SPT phase.  The spin cobordism consideration\cite{KTT1429,FH160406527}
tells us that there is no non-trivial fermionic $Z_2^f\times Z_2^T$-SPT phase,
even beyond the fermion decoration construction.

\subsection{3+1D}
\label{Z2TZ2f3d}
 
\subsubsection{Without extension of $O_\infty$}

\noindent\textbf{Calculate $\bar n_3$}:  
From $H^3(\cB Z_2^T;\Z_2)=\Z_2$, we find that:
\begin{align}
\bar n_3 = \al_n (\bar a^{\Z_2^T})^3=
\al_n \bar \w_1^3 ,
\end{align}
$\al_n=0,1$.  

\noindent\textbf{Calculate $\bar \nu_4$}:  
Next, we consider $\bar \nu_4$ in \eqn{fSPTcl}, which has a form  
\begin{align}
 -\dd \bar \nu_4 \se{1}
\frac{\al_n}2 \Sq^2 \bar \w_1^3 \se{1} \frac{\al_n}2 \bar \w_1^5 
\end{align}
where \eqn{a3a5} is used.  It turns out that
$\frac12 \bar \w_1^5$ is a trivial element in $H^5(\cB Z_2^T;(\RZ)_T)=0$:  
\begin{align}
\label{eta4}
 \frac12 \bar \w_1^5 \se{1} \dd_{\w_1} \bar \eta_4.
\end{align}
To calculate $\bar \eta_4$,
we first note that (see \eqn{eq:differentialw1})
\begin{align}
(\dd_{\bar\w_1}\frac14)_{01} = \frac14 (-)^{(\bar\w_1)_{01}} -\frac14=-\frac12 (\bar\w_1)_{01},
\end{align}
or
\begin{align}
\label{dd14}
 \dd_{\bar\w_1}\frac14 \se{1} \frac12 \bar\w_1.
\end{align}
We also note that
\begin{align}
 \bar\w_1^4 \se{2} (\Sq^1 \bar\w_1)^2 \se{2} (\Bs_2 \bar\w_1)^2 ,
\end{align}
where $\Bs_2 \bar\w_1$ is a $\Z$-valued cocycle.
Thus
\begin{align}
\label{ddw5}
 \frac12 \bar\w_1^5 \se{1} \frac12 \bar\w_1 (\Bs_2 \bar\w_1)^2
\se{1} (\dd_{\bar\w_1}\frac14) (\Bs_2 \bar\w_1)^2
\se{1} \dd_{\bar\w_1} [\frac14 (\Bs_2 \bar\w_1)^2].
\end{align}
or
\begin{align}
 \bar \nu_4 = \frac14 (\Bs_2 \bar\w_1)^2
\end{align}
We see that $\bar \nu_4$ has a form
\begin{align}
\bar \nu_4 \se{1,\dd} \frac{\al_n}4  (\Bs_2 \bar\w_1)^2 +
\frac{\al_\nu}2 \bar\w_1^4 \ \  \ \al_\nu=0,1.
\end{align}

\subsubsection{With extension of $O_\infty$}

\noindent\textbf{Calculate $\bar n_3$}:  
$\bar n_3\in
H^3(\cB Pin^-_\infty;\Z_2)$ may have a form 
\begin{align}
n_3 \se{2} \al_n \bar \w_1^3 + \al_n' \bar \w_1\bar \w_2+\al_n'' \bar \w_3, \ \ \ \
\al_n, \al_n, \al_n'' =0,1.  
\end{align}
From \eqn{w1w2}, we have $\bar \w_1\bar \w_2 \se{2} \bar \w_1^3$ and $0 \se{2}
\Sq^1(\bar \w_1^2+\bar \w_2) \se{2} \bar \w_1\bar \w_2+\bar \w_3$.  Thus $\bar
n_3$ has two choices
\begin{align}
\bar n_3 \se{2} \al_n \bar \w_1^3 , \ \ \ \
\al_n  =0,1.  
\end{align}

\noindent\textbf{Calculate $\bar \nu_4$}:  
Next, we consider $\bar \nu_4$ that satisfy \eqn{fSPTclO}
which becomes, after using \eqn{a3a5}
\begin{align}
 -\dd \bar \nu_{4} 
& \se{1} \frac{\al_n}2 [\bar \w_1^5 +\bar \w_1^3 (\bar \w_2+\bar \w_1^2)] .
\nonumber\\
& \se{1} \al_n (\dd \bar \eta_4 +\frac12 \bar \w_1^3 \dd \bar a^{Z_2^f} ).
\end{align}
where we have used \eqn{w1w2} and \eqn{eta4}.

From $\frac12 \bar \w_2 \se{1,\dd} \frac12 \bar \w_1^2 $, we obtain that
$\frac12 \bar \w_2 \bar \w_1^2 \se{1,\dd} \frac12 \bar \w_1^4 $ and $\frac12
\bar \w_2^2 \se{1,\dd} \frac12 \bar \w_1^2 \bar \w_2 $.  We also find (see
Appendix \ref{Rswc}) $0\se{2,\dd} \Sq^1 (\bar \w_2+ \bar \w_1^2) \se{2,\dd}
\bar \w_1 \bar \w_2+\bar \w_3 $ and $0\se{2,\dd} \gSq^2 (\bar \w_2+ \bar
\w_1^2)= \bar \w_2^2+\bar \w_1^4 $.  To summarize, we have
\begin{align}
\label{w4rel}
 \frac12 \bar \w_1^4 \se{1,\dd} 
 \frac12 \bar \w_2^2  \se{1,\dd} 
 \frac12 \bar \w_1 ^2 \bar \w_2 \se{1,\dd}
 \frac12 \bar \w_1 \bar \w_3 \se{1,\dd} \frac12 \bar p_1. 
\end{align}
where $\bar p_1(a^{O})$ is the first Pontryagin class (see \eqn{piw2i}).

Thus \eqn{fSPTclO} has a solution of form
\begin{align}
\bar \nu_{4} &\se{1}  
\al_n (\frac14 (\Bs_2 \bar\w_1)^2 +\frac12 \bar \w_1^3  \bar a^{Z_2^f} )
+ \frac{\al_\nu}2 \bar \w_1^4 
+ \frac{\t \al_\nu}2 \bar \w_4 ,
\nonumber\\
&\ \ \ \ \
\al_\nu,\t\al_\nu=0,1 ,
\end{align}

\noindent\textbf{SPT invariant}:
The corresponding SPT invariant is given by
\begin{align}
& 
Z^\text{top}(\cM^4,A^{G_{fO}})
= \ee^{ \ii \pi \int_{\cN^5} \Sq^2 f_3 + f_3 (\w_2+\w_1^2)}
\nonumber \\
& \ \ \ \ \ \ \ \ \ \ \ \
\ee^{\ii 2\pi \int_{\cM^4} 
\al_n (\frac14 (\Bs_2 \w_1)^2 +\frac12 \w_1^3  A^{Z_2^f})
+\frac{\al_\nu}2 \w_1^4 
+ \frac{\t\al_\nu}2 \w_4
}
,
\nonumber \\
& f_3\big|_{\prt \cN^5} \se{2} \al_n \w_1^3, \ \ \ 
\dd A^{\Z_2^f} \se{2} \w_2+\w_1^2
,
\end{align}
where the background connection $A^{G_{fO}}$ is labeled by
$(A^{\Z_2^f},A^{O})$.  In 3+1D space-time, we have some additional relations
\eqn{wrel4}. When combined with \eqn{w4rel}, we find
\begin{align}
\frac12 \w_1^4 &\se{1,\dd} 
\frac12 \w_2^2  \se{1,\dd} 
\frac12 \w_1 ^2  \w_2 \se{1,\dd}
\frac12 \w_1  \w_3 \se{1,\dd} \frac12  p_1 \se{1,\dd} 0, 
\nonumber\\
\frac12 \w_4 &\se{1,\dd} 0 .
\end{align}
Therefore,  the SPT invariant is independent of $\al_\nu$ and $\t\al_\nu$, and
they fail to label different $Z_2^f\times Z_2^T$-SPT phases.  Also from
$\w_1^2+\w_2 \se{2,\dd}0$ and $\w_1\w_2\se{2,\dd}0$ on the 3+1D space-time, we
see that $\w_1^3\se{2,\dd}0$.  Thus $n_3$ is always a $\Z_2$-valued coboundary
when pulled back on $\cM^4$.  As a result, the SPT invariant is independent of
$\al_n$ and it fails to label different $Z_2^f\times Z_2^T$-SPT phases (see
\eqn{gaugendSOM}).  Therefore, there is only one trivial $Z_2^f\times Z_2^T$
SPT phase.  

The fermionic $Z_2^f\times Z_2^T$ symmetry is denoted by $G_+^T$ in
\Ref{W11116341}. It was found that for non-interacting fermions, there is no
non-trivial fermionic $Z_2^f\times Z_2^T$-SPT phase in
3+1D.\cite{K0986,SRF0825,W11116341}  The spin cobordism
consideration\cite{KTT1429,FH160406527} also tells us that there is no non-trivial
fermionic $Z_2^f\times Z_2^T$-SPT phase, even beyond the fermion decoration
construction.

\section{Fermionic $Z_4^{T,f}$-SPT state}

In this section, we consider the fermionic SPT states with $G_b=Z_2^T$ and a
non-trivial $\bar e_2=(\bar a^{\Z_2^T})^2=\bar \w_1^2$ in $H^2(\cB
Z_2^T;\Z_2)=\Z_2$.  In this case, the full fermionic symmetry is
$G_f=Z_4^{T,f}$.  For $G_f=Z_4^{T,f}$, $G_{fO}$ is a $Z_2^f$ extension of
$O_\infty$: $G_{fO}=Z_2^f\gext_{\w_2} O_\infty=Pin^+_\infty$.  This implies
that on $\cB G_{fO}=\cB Pin^+_\infty$, we have a relation (see Appendix
\ref{cenext})
\begin{align}
\label{w2}
\bar \w_2(\bar a^{G_{fO}})\se{2} \dd \bar a^{Z_2^f} .
\end{align}
The fermion symmetry $Z_4^{T,f}$ is realized by charge $2e$ electron
superconductors with spin-orbital couplings.

\subsection{2+1D}

\subsubsection{Without extension of $O_\infty$}

\noindent\textbf{Calculate $\bar n_2$}:  
First, the possible $\bar n_2$'s have a form
\begin{align}
\bar n_2= \al_n  (\bar a^{\Z_2^T})^2 \se{2} \al_n \bar \w_1^2 ,
\end{align}
where $\al_n=0,1$.  

However, after $\bar n_2$ is pulled back on $\cM^3$, it becomes $ n_2\se{2}
\al_n \w_1^2$.  On 2+1D manifold, we have $\w_2+ \w_1^2\se{2,\dd}0$, and on a
Pin$^+$ manifold we have $ \w_2\se{2,\dd}0$.  Thus, on a 2+1D Pin$^+$ manifold,
$ \w_1^2$ is always a coboundary. The two solutions of $n_2$ are equivalent
(see \eqn{gaugendSOM}) and we can choose $\al_n=0$.

\noindent\textbf{Calculate $\bar \nu_3$}:  
Now $\bar \nu_3$ is obtained from $\dd \bar \nu_3  \se{1} 0 $.  $\bar \nu_3$ has only
one solution $\bar \nu_3 \se{1}0$, since $H^3(\cB Z_2^T;(\RZ)_T)=0$.  

\subsubsection{With extension of $O_\infty$}

\noindent\textbf{Calculate $\bar n_2$}:  
Due to the relation \eqn{w2}, $H^2(\cB Pin^+_\infty)=\Z_2$ which is generated
by $\bar \w_1^2$.  Thus $\bar n_2\in H^2(\cB Pin^+_\infty;\Z_2)$ has two
choices $\bar n_2 \se{2} \al_n \bar \w_1^2$, $\al_n=0,1$.  

On a 2+1D space-time $\cM^3$ with $G_{fO}= Z_2^f\gext_{\bar \w_2}
O_\infty=Pin^+_\infty$ connection $A^{G_{fO}}$, the connection can be viewed as
a pullback from the canonical 1-cochain $\bar a^{G_{fO}}$ on $\cB G_{fO}$.
Such a 2+1D space-time satisfies $\w_2\se{2,\dd}0$ and is a Pin$^+$ manifold.
Since any 3-manifold $\cM^3$ satisfies $\w_2+ \w_1^2\se{2,\dd}0$ (see Appendix
\ref{Rswc}), therefore, $\cM^3$ with $Z_2^f\gext_{\bar \w_2}
O_\infty=Pin^+_\infty$ connection satisfies $ \w_2\se{2,\dd}
\w_1^2\se{2,\dd}0$.  The pullback of $\bar n_2$ on $\cM^3$ is given by $ n_2
\se{2} \al_n  \w_1^2$.  We see that, after pulled back to $\cM^3$, $\al_n=1$
and $\al_n=1$ are equivalent. We may choose $\al_n=0$.

\noindent\textbf{Calculate $\bar \nu_3$}:  
Next, we consider $\bar \nu_3$ that satisfy $ -\dd \bar \nu_{3} \se{1} 0 $.
There are four solutions 
\begin{align}
\bar  \nu_{3} \se{1}  \frac{\al_\nu}2 \bar \w_3 + \frac{\t\al_\nu}2 \bar \w_1^3,\ \ \ 
\al_\nu,\t\al_\nu=0,1.
\end{align}

After pulled back to $\cM^3$, $\bar \nu_3$ becomes $\nu_3 \se{1}
\frac{\al_\nu}2  \w_3 + \frac{\t\al_\nu}2 \w_1^3$.  But on 2+1D space-time
$\cM^3$, we have a relation $\w_3 \se{2,\dd}0$ (see \eqn{wrel3}).  Combined the
$\w_2\se{2,\dd} \w_1^2\se{2,\dd}0$ obtained above, we find that the above four
solutions differ only by coboundaries, which give rise to the same fermionic
$Z_4^{T,f}$ SPT phase.  Therefore the fermion decoration construction fails to
produce a non-trivial fermionic $Z_4^{T,f}$ SPT phase.  Thus, the fermion
decoration construction fail to produces any non-trivial  fermionic
$Z_4^{T,f}$-SPT state.

In fact, 2+1D $Z_4^{T,f}$ fermionic SPT phases is classified by $\Z_2$
\cite{KTT1429,FH160406527}.  The non-trivial $Z_4^{T,f}$ SPT phase can be realized as a
$p+\ii p$ superconductor for spin-up fermions stacking with a $p-\ii p$
superconductor for spin-down fermions \cite{R0664,QHR0901}.  It has the
following special property: After we gauge the $Z_2^f$ symmetry, we obtain a
$Z_2^f$ gauge theory with $G_b=Z_2^T$ symmetry.  In 2+1D, $Z_2^f$ charge $e$,
$Z_2^f$-flux, and their bound state $em$ are all point-like topological
excitations. The time-reversal symmetry in this case exchanges the bosonic $e$
and $m$, and $em$ is a Kramers doublet \cite{VS1306,WS1334,CG150101313}.

\subsection{3+1D}

\subsubsection{Without extension of $O_\infty$}

\noindent\textbf{Calculate $\bar n_3$}:  
$\bar n_3 \in H^3(\cB Z_2^T;\Z_2)=\Z_2$ is given
\begin{align}
\bar n_3 \se{2,\dd} \al_n (\bar a^{\Z_2^T})^3 \se{2,\dd}  \al_n \bar \w_1^3. 
\end{align}

For the 3+1D space-time with a Pin$^+$ structure (\ie $\w_2\se{2,\dd}0$),  it
turns out that, even when $\w_2\se{2,\dd}0$,  $\w_1^3$ is still non-trivial.
So $n_3\se{1} \al_n  \w_1^3$ indeed has two choices.

\noindent\textbf{Calculate $\bar \nu_4$}:  
Next we want to solve
\begin{align}
 -\dd \bar \nu_4 & \se{1}  
\frac12 \Sq^2 \bar n_3 
+ \frac12  \bar n_3 \bar e_2 
\nonumber\\
& \se{1}  \frac{\al_n} 2  \gSq^2 (\bar a^{\Z_2^T})^3 + \frac{\al_n} 2   
(\bar a^{\Z_2^T})^5
\se{1} 0
,
\end{align}
where we have used \eqn{a3a5}).  Since $H^4(\cB Z_2^T;(\RZ)_T)=\Z_2$, $\nu_4$
has two solutions
\begin{align}
\bar   \nu_4 & \se{1} \frac{\al_\nu}2 (\bar a^{\Z_2^T})^4
\se{1} \frac{\al_\nu}2 \bar \w_1^4
.
\nonumber 
\end{align}

On 3+1D space-time with $\w_2\se{2,\dd}0$, $\frac12 \w_1^4$ is still a
non-trivial $\R/\Z$-valued cocycle.  The pullback on $\cM^4$, $\nu_4 \se{1}
\frac{\al_\nu}2 \w_1^4$, is still non-trivial.

\subsubsection{With extension of $O_\infty$}

\noindent\textbf{Calculate $\bar n_3$}:  
Due to the relation \eqn{w2}, and $0 \se{2,\dd} \Sq^1 \bar \w_2 \se{2,\dd} \bar
\w_1\bar \w_2+\bar \w_3 \se{2,\dd} \bar \w_3$, we find that $H^2(\cB
Pin^+_\infty;\Z_2)=\Z_2$ is generated by $\bar \w_1^3$.  Thus $\bar n_3\in
H^2(\cB Pin^+_\infty;\Z_2)$ has two choices $\bar n_3 \se{2} \al_n \bar
\w_1^3$, $\al_n=0,1$.  

\noindent\textbf{Calculate $\bar \nu_4$}:  
Next, we consider $\bar \nu_4$ that satisfy 
\begin{align}
 -\dd \bar \nu_{4} & \se{1} 
\frac{\al_n}2 [\bar \w_1^5 + \bar \w_1^3 (\bar \w_2+\bar \w_1^2)]
\nonumber\\
& \se{1} \frac{\al_n}2 \bar \w_1^3 \dd \bar a^{Z_2^f}
.  
\end{align}
Since $\bar \w_2 \se{2,\dd} \bar \w_3 \se{2,\dd} 0$, there are eight solutions 
\begin{align}
\bar  \nu_{4} &\se{1}  
\frac{\al_n}2 \bar \w_1^3 \bar a^{Z_2^f} 
+ \frac{\al_{\nu,1}}2 \bar \w_1^4 
+ \frac{\al_{\nu,2}}2 \bar \w_4
+ \frac{\al_{\nu,3}}2 \bar p_1
,
\nonumber\\
\al_{\nu,i}& =0,1.
\end{align}

But on 3+1D space-time $\cM^3$ with $\w_2\se{2,\dd}0$, we have relations $\w_3
\se{2,\dd}0$, $\w_4 \se{2,\dd}\w_1^4$, and $p_1\se{2} \w_2^2\se{2}0$ (see
\eqn{wrel4}).  So the pullback of $\bar \nu_{4}$ on $\cM^4$ becomes
\begin{align}
\nu_{4} \se{1}  
\frac{\al_n}2 \w_1^3 \bar a^{Z_2^f} +
\frac{\al_\nu}2  \w_1^4 ,\ \ \ 
\al_\nu=0,1.
\end{align}

\noindent\textbf{SPT invariant}:
The corresponding SPT invariant is given by
\begin{align}
& 
Z^\text{top}(\cM^4,A^{G_{fO}})
= \ee^{ \ii \pi \int_{\cN^5} \Sq^2 f_3 + f_3 (\w_2+\w_1^2)}
\nonumber \\
& \ \ \ \ \ \ \ \ \ \ \ \
\ee^{\ii 2\pi \int_{\cM^4} 
\frac{\al_n}2 \w_1^3 \bar a^{Z_2^f} +
\frac{\al_\nu}2  \w_1^4
}
,
\nonumber \\
& f_3\big|_{\prt \cN^5} \se{2} \al_n \w_1^3, \ \ \ 
\dd A^{\Z_2^f} \se{2} \w_2
,
\end{align}
where the background connection $A^{G_{fO}}$ is labeled by
$(A^{\Z_2^f},A^{O})$.  We see that fermion decoration construction produces
four different fermionic $Z_4^{T,f}$-SPT phases.

For non-interacting fermion systems, the $Z_4^{T,f}$-SPT phases are classified
by $\Z$.\cite{K0986,SRF0825,W11116341} However, after include interaction,
\Ref{KTT1429,FH160406527} found that the $Z_4^{T,f}$-SPT phases are classified by
$\Z_{16}$. Thus fermion decoration construction does not produce all the
$Z_4^{T,f}$-SPT phases.

\section{Fermionic $(U_1^f \rtimes_\phi Z_4^{T,f})/Z_2$-SPT state
-- interacting topological insulators}

The symmetry group 
\begin{align}
 G_f&=(U_1^f \rtimes_\phi Z_4^{T,f})/Z_2 
\end{align} 
is the symmetry group for topological insulator, \ie for electron systems with
time reversal $Z_4^{T,f}$ and charge conservation $ U_1^f$ symmetries.  Such a
symmetry can be realized by electron systems with spin-orbital coupling.  

In the above expression, $\phi$ is a homomorphism $\phi: Z_4^{T,f} \to
\text{Aut}(U_1^f)$.  Let $T$ be the generator of $Z_4^{T,f}$, then $\phi(T)$
changes an element in $U_1^f$ to its inverse.  The semi-direct product $U_1^f
\rtimes_\phi Z_4^{T,f}$ is defined using such an automorphism $\phi(T)$.  $Z_2$
in $(U_1^f \rtimes_\phi Z_4^{T,f})/Z_2$  is generated by the product of the
$\pi$-rotation in $U_1^f$ and $T^2$. It is in the center of $U_1^f \rtimes_\phi
Z_4^{T,f}$.

The symmetry group can be written as
\begin{align}
 G_f&=
U_1^f \gext_{\bar \veps_2,\phi} Z_2^T.
\end{align} 
In other words, the elements in $G_f$ can be labeled by $(\bar a^{(\RZ)^f},\bar
a^{\Z_2^T})$, $\bar a^{(\RZ)^f} \in (\RZ)^f$ and $\bar a^{\Z_2^T}\in
\Z_2^T=\{0,1\}$, such that
\begin{align}
&
 (\bar a^{(\RZ)^f}_1,\bar a^{\Z_2^T}_1)
 (\bar a^{(\RZ)^f}_2,\bar a^{\Z_2^T}_2) =
\\
&\Big(\bar a^{(\RZ)^f}_1+\phi(\bar a^{\Z_2^T}_1)\circ \bar a^{(\RZ)^f}_2
+\bar\veps_2(\bar a^{\Z_2^T}_1,\bar a^{\Z_2^T}_2)
,\bar a^{\Z_2^T}_1+\bar a^{\Z_2^T}_2\Big)
\nonumber 
\end{align}
where 
\begin{align}
 \bar\veps_2(\bar a^{\Z_2^T}_1,\bar a^{\Z_2^T}_2) &= \frac12 \bar a^{\Z_2^T}_1\bar a^{\Z_2^T}_2,\ \ \ \
\phi(0)=1,\ \phi(1)=-1.
\end{align}
In terms of cochains, the above can be rewritten as
\begin{align}
\bar\veps_2 = \frac12 (\bar a^{\Z_2^T})^2.
\end{align}

We may write $U_1^f$ as $\Z_2^f\gext_{\bar c_1} U_1$, \ie write
\begin{align}
 \bar a^{(\RZ)^f} =\frac12 \bar a^{\RZ}+\frac12 \bar a^{\Z_2^f}
\end{align}
where $\bar a^{\RZ} \in \RZ=(-\frac12,\frac12]$ and $\bar a^{\Z_2^f} \in
\Z_2^f=\{0,1\}$.  From
\begin{align}
 \bar a^{(\RZ)^f}_1 & +\bar a^{(\RZ)^f}_2
 =
\frac12 \bar a^{\RZ}_1+\frac12 \bar a^{\RZ}_2
+\frac12 \bar a^{\Z_2^f}_1 +\frac12 \bar a^{\Z_2^f}_2
\nonumber\\
&= \frac12 (\bar a^{\RZ}_1+\bar a^{\RZ}_2-\toZ{\bar a^{\RZ}_1+\bar a^{\RZ}_2})
\nonumber\\
&\ \ \ \ \
+\frac12 (\bar a^{\Z_2^f}_1+\frac12 \bar a^{\Z_2^f}_2+\toZ{\bar a^{\RZ}_1+\bar a^{\RZ}_2})
\nonumber\\
&= \frac12 (\bar a^{\RZ}_1+\bar a^{\RZ}_2-\toZ{\bar a^{\RZ}_1+\bar a^{\RZ}_2})
\nonumber\\
&\ \ \ \ \
+\frac12 (\bar a^{\Z_2^f}_1+\frac12 \bar a^{\Z_2^f}_2
+ \bar c_1(\bar a^{\RZ}_1,\bar a^{\RZ}_2) ),
\end{align}
we see that
\begin{align}
 \bar c_1(\bar a^{\RZ}_1,\bar a^{\RZ}_2) =\toZ{\bar a^{\RZ}_1+\bar a^{\RZ}_2}
\end{align}
which is the first Chern class of $U_1$.  Note that $\bar c_1(\bar
a^{\RZ}_1,\bar a^{\RZ}_2)$ is a smooth function of $\bar a^{\RZ}_1,\bar
a^{\RZ}_2$ near $\bar a^{\RZ}_1,\bar a^{\RZ}_2=0$.  But it has dicontinuities
away from $\bar a^{\RZ}_1,\bar a^{\RZ}_2=0$.

Now we can label elements in $G_f$ by triples
$(\bar a^{\Z_2^f},\bar a^{\RZ}, \bar a^{\Z_2^T})$.
The group multiplication is
\begin{align}
&\ \ \ \
 (\bar a^{\Z_2^f}_1,\bar a^{\RZ}_1, \bar a^{\Z_2^T}_1)
 (\bar a^{\Z_2^f}_2,\bar a^{\RZ}_2, \bar a^{\Z_2^T}_2)
\nonumber\\
&=
 \Big(\bar a^{\Z_2^f}_1+\bar a^{\Z_2^f}_2
+2\bar\veps_2(\bar a^{\Z_2^T}_1,\bar a^{\Z_2^T}_2)
+\bar c_1(\bar a^{\RZ}_1, \bar a^{\RZ}_2)\ , 
\nonumber\\
&\ \ \  \ \ \ \ \
\bar a^{\RZ}_1+\phi(\bar a^{\Z_2^T}_1)\circ \bar a^{\RZ}_2\ , \
\bar a^{\Z_2^T}_1+\bar a^{\Z_2^T}_2\Big)
\end{align}
We see that $G_f$ can also be written as
\begin{align}
 G_f & =Z_2^f \gext_{\bar e_2} (U_1\rtimes_\phi Z_2^T),
\end{align} 
where
\begin{align}
 \bar e_2 \se{2,\dd} \bar c_1 
+ (\bar a^{\Z_2^T})^2\se{2,\dd} \bar c_1 + \bar \w_1^2.
\end{align}

For fermion symmetry $G_f=(U_1^f \rtimes_\phi Z_4^{T,f})/Z_2$, the
corresponding extended group $G_{fO}$ can be written as
\begin{align}
\label{GfO2exp}
 G_{fO} &=G_f^0\gext_{\bar \veps_2'} O_\infty 
= U_1^f \gext_{\bar \veps_2'} O_\infty
= Z_2^f \gext_{\bar e_2'} (U_1\rtimes_\phi O_\infty )
\nonumber\\
\bar \veps_2' &\se{1} \frac12 \bar \w_2,\ \ \ 
\bar e_2' \se{2} \bar \w_2+\bar c_1 
.
\end{align}
where $G_f^0$ is the fermion symmetry with time reversal removed:
$G_f^0=U_1^f$.  We like to mention that $\bar \veps_2' \in H^2(\cB O_\infty;
\R/\Z)$ which describes how we extend $O_\infty$ by $U_1^f$ and $\bar e_2' \in
H^2(\cB (U_1\rtimes_\phi O_\infty); \Z_2^f)$ which describes how we extend
$U_1\rtimes_\phi O_\infty$ by $Z_2^f$.

We may view $\bar \veps_2'(\bar a^{O})$ as $\bar \veps_2'(\pi^{U_1}(\bar
a^{G_{fO}}))$ an element in $H^2(\cB G_{fO}; (\R/\Z)_T)$, where $\pi^{U_1}$ is
the projection $G_{fO} \xrightarrow{\pi^{U_1}} O_\infty$. Then $\bar
\veps_2'(\pi^{U_1}(\bar a^{G_{fO}}))$ is a trivial element in $H^2(\cB G_{fO};
(\R/\Z)_T)$ (\ie is a $(\RZ)_T$-valued coboundary, see Appendix \ref{cenext}):
\begin{align}
\label{e2ptri}
& \frac 12 \bar \w_2(\pi^{U_1}(\bar a^{G_{fO}})) 
\se{1} \dd \bar \eta_1(\bar a^{G_{fO}}),
\nonumber\\ &\ \ \ \
\bar \eta_1 \in C^1(\cB G_{fO}, (\R/\Z)_T).
\end{align}
Similarly, we may view $\bar e_2'$ as a trivial element in $H^2(\cB G_{fO}; \Z_2)$:
\begin{align}
\label{e2pptri}
& \bar \w_2(\pi^{Z_2^f}(\bar a^{G_{fO}})) +\bar c_1(\pi^{Z_2^f}(\bar a^{G_{fO}})) \se{2} \dd \bar u_1(\bar a^{G_{fO}}),
\nonumber\\ & \ \ \ \
\bar u_1 \in C^1(\cB G_{fO}, \Z_2),
\end{align}
where $\pi^{Z_2^f}$ is the projection $G_{fO} \xrightarrow{\pi^{Z_2^f}}
U_1\rtimes O_\infty$.  Eqns, \eq{e2ptri} and \eq{e2pptri} imply that although
$\bar c_1$ mod 2 can be a non-trivial $\Z_2$-valued cocycle, $\frac12 \bar c_1$
is always a $(\RZ)_T$-valued coboundary on $\cB G_{fO}$.

In other words, on space-time $\cM^{d+1}$, we have a $G_{fO}$ connection
$a^{G_{fO}}$. 
The $G_{fO}$ connection can be labeled in two ways
\begin{align}
 a^{G_{fO}} = (a^{(\RZ)^f}, a^{O}) =
(a^{Z_2^f}, a^{\RZ}, a^{O})
\end{align}
using the two expressions of $G_{fO}$ \eq{GfO2exp}.  In the above $a^{O}$ is
the connection of the tangent bundle of the space-time.  The above results
implies that, if the
$a^{O}$ can be lifted to a $a^{G_{fO}}$ connection, then
 $\frac12 \w_2(a^O)$ on $\cM^{d+1}$ is a $(\RZ)_T$-valued coboundary and
$\w_2 + c_1$ is a $\Z_2$-valued coboundary.
Here $O$ has a non-trivial action on the $(\RZ)_T$ coefficient:
$\R/\Z \xrightarrow{T} -\R/\Z$, as indicated by the subscript $T$.

\subsection{2+1D}

\subsubsection{Without extension of $O_\infty$}

\noindent\textbf{Calculate $\bar n_2$}:
To construct fermionic $(U_1^f\rtimes Z_4^{T,f}) /Z_2$ SPT states using fermion
decoration,  we need to find $\bar n_2 \in H^2(\cB G_b;\Z_2) = H^2(\cB
(U_1\rtimes Z_2^T);\Z_2)$.
Notice that
\begin{align}
& 
 H^2[\cB (U_1\rtimes Z_2^T);\Z_2] 
\rightarrowtail 
H^2(\cB Z_2^T;\Z_2)\oplus 
\nonumber\\
& \ \ \ \ \ \
H^1[\cB Z_2^T;H^1(\cB U_1,\Z_2)]\oplus
H^2(\cB U_1;\Z_2)
\end{align}
We can construct $\bar n_2(\bar a^{G_b})$ using flat $Z_2^T$-connection $a^{Z_2^T}$ and nearly flat $U_1$ connection $a^{\RZ}$:  
\begin{align}
 n_2(a^{G_b}) 
& \se{2} \al_n (\bar a^{\Z_2^T})^2 +\t \al_n  \bar c_1
\se{2} \al_n \bar \w_1^2 +\t \al_n  \bar c_1 ,
\nonumber\\
& \al_n,\t \al_n=0,1 .
\end{align}
When $\al_n=1$, we decorate the intersection of $Z_2^T$ symmetry-breaking
domain walls by fermion.  When $\t \al_n=1$, we decorate the $2\pi$-flux of
bosonic $U_1$ (which is the $\pi$-flux of bosonic $U_1^f$) by fermion.

We note that the space-time $\cM^{d+1}$ has a twisted spin
structure
\begin{align}
 \dd a^{Z_2^f} \se{2} \w_2+\w_1^2 +e_2 \se{2} \w_2+c_1.
\end{align}
So an electron system with time reversal and charge conservation symmetry
$G_f=(U_1^f \rtimes_\phi Z_4^{T,f})/Z_2$ can only lives on space-time with
trivial $\w_2+c_1$.  A 2+1D space-time $\cM^3$ also satisfies
$\w_2+\w_1^1\se{2,\dd}0$. Thus decorating the intersection of $Z_2^T$
symmetry-breaking domain walls and decorating the $2\pi$-flux of bosonic
$U_1$ give rise to the same SPT phase. The inequivalent pullback of $\bar n_2$
to $\cM^3$ has a form
\begin{align}
 n_2 \se{2} \al_n \w_1^2,\ \ \ \ \ 
\al_n=0,1.
\end{align}
So $n_2$ has two choices.

\noindent\textbf{Calculate $\bar \nu_3$}:
Next, we calculate $\bar \nu_3(\bar a^{G_b})$ from \eqn{fSPTcl}, which becomes
\begin{align}
 -\dd \bar \nu_3
& \se{1} \frac{\al_n}2 [\bar \w_1^4 + \bar \w_1^2
(\bar \w_1^2 + \bar c_1)] .
\nonumber\\
&\se{1} \frac {\al_n}{2} \bar c_1\bar \w_1^2 
\end{align}

$\frac12 \bar c_1\bar \w_1^2 $ is a coboundary in $H^4(\cB
(U_1\rtimes Z_2^T);(\RZ)_T)$:
\begin{align}
 \frac 12 \bar c_1 \bar \w_1^2  \se{1} \dd \frac14 \bar c_1\bar \w_1 .
\end{align}
We first note that $\frac14 \bar c_1\bar \w_1 $ is a product of three values: a
$\R/\Z$-value $\frac14$, a $\Z$-value in $\bar c_1$, and a $\Z_2$-value in
$\bar \w_1$.  The time reversal $Z_2^T$ has a non-trivial action on the
$\R/\Z$-value $\frac14 \to -\frac14$ and a non-trivial action on the $\Z$-value
$\Z \to -\Z$.  Thus time reversal $Z_2^T$ acts on the Chern class $\bar c_1\to
-\bar c_1$. Therefore, $Z_2^T$ acts trivially on the combination $\frac14 \bar
\w_1 \bar c_1$, and $\dd$ in $\dd \frac14 \bar \w_1 \bar c_1$ is the ordinary
differentiation operator, not the $\dd_{\w_1}$ in \eqn{eq:differentialw1}.  Now
using \eqn{Sq1Bs2}, we find
\begin{align}
 \dd \frac14 \bar \w_1 \bar c_1=\frac12 (\Bs_2 \bar \w_1) c_1 
\se{1} \frac12 \bar \w_1^2 \bar c_1 .
\end{align}
This means that $\bar \nu_3$ has solution for all  two cases
$\al_n=0,1$:  
\begin{align}
\bar \nu_3 \se{1} \frac{\al_n}4 \bar \w_1 \bar c_1 .
\end{align}
We note that
\begin{align}
&H^3[\cB (U_1\rtimes Z_2^T);(\R/\Z)_T]
\rightarrowtail
H^3[\cB Z_2^T ;H^0(\cB U_1; \R/\Z)]\oplus
\nonumber\\ &
H^2[\cB Z_2^T ;H^1(\cB U_1; \R/\Z)]\oplus
H^1[\cB Z_2^T ;H^2(\cB U_1; \R/\Z)]\oplus
\nonumber\\ &
H^0[\cB Z_2^T ;H^3(\cB U_1; \R/\Z)]
\nonumber\\ &
=
H^3[\cB Z_2^T; (\R/\Z)_T]
\oplus 
H^2[\cB Z_2^T;\Z] \oplus
H^0[\cB Z_2^T;(\Z)_T] 
\nonumber\\
& = 0 
\end{align}
Thus, for each $\bar n_2$, there is only one solution of $\bar \nu_3$ since $H^3[\cB
(U_1\rtimes Z_2^T);(\R/\Z)_T]=0$.

\subsubsection{With extension of $O_\infty$}

\noindent\textbf{Calculate $\bar n_2$}:
To construct fermionic $G_f=(U_1^f \rtimes_\phi Z_4^{T,f})/Z_2$ SPT states
using fermion decoration and extension of $O_\infty$ , we first calculate $\bar
n_2 \in H^2(\cB G_{fO};\Z_2)= H^2(\cB (U_1^f \gext_{\bar \veps_2'}
O_\infty);\Z_2)$.  $\bar n_2$ has a form
\begin{align}
 \bar n_2 \se{2,\dd} 
\al_n \bar \w_1^2 +
\t\al_n \bar \w_2 , \ \ \ \
\al_n,\t\al_n=0,1.
\end{align}
We do not have the $\bar c_1$ term due to the relation $\bar c_1\se{2,\dd} \bar
\w_2$.

However, on 2+1D manifold, $\w_2\se{2,\dd}\w_1^2$. Thus the pullback of 
$\bar n_2$ on space-time $\cM^3$ has a simpler form
\begin{align}
  n_2 \se{2,\dd} \al_n \w_1^2 , \ \ \ \ \ \al_n=0,1.
\end{align}

\noindent\textbf{Calculate $\bar \nu_3$}:
Next, we calculate $\bar \nu_3(\bar a^{G_{fO}})$ from \eqn{fSPTclO}, which
becomes
\begin{align}
 -\dd \bar \nu_3 
& \se{1} \frac{\al_n}2 [\bar \w_1^4 + \bar \w_1^2
(\bar \w_2 + \bar \w_1^2 )] .
\nonumber\\
&\se{1} \frac {\al_n}{2} \bar \w_1^2 \bar c_1
\nonumber 
\end{align}
Similarly, we find $\bar \nu_3 \in H^3(\cB(U_1^f \gext_{\bar \veps_2'}
O_\infty);(\RZ)_T)$ to have a form
\begin{align}
\bar \nu_3 \se{1} \frac{\al_n}4 \bar \w_1 \bar c_1 +
\frac{\al_{\nu}}{2} \bar \w_1^3 +
\frac{\t\al_{\nu}}{2} \bar \w_3 .
\end{align}
The term $\frac12 \bar \w_1 \bar c_1 \se{1,\dd} \frac12 \bar \w_1 \bar \w_2$ is
not included since $\frac12 \bar \w_2$ is a coboundary (see \eqn{e2ptri}).

In 2+1D space-time, we have $\w_1^2 +\w_2 \se{2,\dd} \w_3 \se{2,\dd}0 $ (see
\eqn{wrel3}).  This also implies that $ \w_1^3 $ is a coboundary.
So the pullback of $\bar \nu_3$ on $\cM^3$ is reduced to
\begin{align}
\nu_3 \se{1} \frac{\al_n}4 \w_1 c_1 .
\end{align}

\noindent\textbf{SPT invariant}: The above $\nu_3$
gives rise to two fermionic $ (U_1^f \rtimes_\phi Z_4^{T,f})/Z_2 $-SPT states,
whose SPT invariant is given by
\begin{align}
&\ \ \ \
Z^\text{top}(\cM^4,A^{G_{fO}})
\nonumber\\ &
= 
\ee^{\ii 2\pi \int_{\cM^4}   \frac{\al_n}4 \w_1 c_1 }
\ee^{
\ii \pi \int_{\cN^4} \Sq^2 f_2 + f_2(\w_2+\w_1^2) } ,
\nonumber\\ &  \ \ \ \ \
\dd A^{\Z_2^f} \se{2} \w_2+c_1(A^{U_1}) , \ \ \ 
f_2 \big|_{\prt \cN^4} \se{2} \al_n \w_1^2 ,
\nonumber\\ &  \ \ \ \ \
\al_n=0,1.
\end{align}
Here we label $A^{G_{fO}}\in U_1^f \gext_{\bar \veps_2'} O_\infty$ by $(A^{U_1^f},
A^{O})$, and label $A^{U_1^f}$ by $A^{U_1^f}_{ij}=\frac12 A^{U_1}_{ij}+\frac12
A^{Z_2^f}_{ij}$, where $A^{U_1^f}_{ij},A^{U_1}_{ij} \in [0,1)$ and
$A^{Z_2^f}_{ij} = 0,1$.  The 2+1D space-time and its $G_{fO}$ connection
satisfies \eqn{e2ptri} and \eqn{e2pptri}.

For non-interacting fermion systems, the $(U_1^f \rtimes_\phi
Z_4^{T,f})/Z_2$-SPT phases (the topological insulators) are classified by
$\Z_2$.\cite{K0986,SRF0825,W11116341} In the above, we show that, after
including interaction and via fermion decoration, the resulting interacting
topological insulators are still described by $\Z_2$.

\subsection{3+1D}

\subsubsection{Without extension of $O_\infty$}

\noindent\textbf{Calculate $\bar n_3$}:
$\bar n_3(\bar a^{G_b}) \in H^2(\cB G_b;\Z_2) = H^2(\cB (U_1\rtimes
Z_2^T);\Z_2)$ has a form
\begin{align}
 \bar n_3 \se{2} 
\al_n \bar \w_1^3 +
\t \al_n \bar \w_1 \bar c_1 
\end{align}

On a 3+1D space-time $\cM^4$ with $\w_2\se{2,\dd} c_1$, we have $\w_1 c_1
\se{2,\dd} \w_1 \w_2 \se{2,\dd} 0$ (see \eqn{wrel4}).  Thus after pulled back
on $\cM^4$, $\bar n_3$ reduces to
\begin{align}
n_3 \se{2} \al_n \w_1^3 , \ \ \ \ \al_n =0,1.
\end{align}

\noindent\textbf{Calculate $\bar \nu_4$}:
$\bar \nu_4(\bar a^{G_b})$ is calculated from \eqn{fSPTcl}, which becomes
\begin{align}
 -\dd \bar \nu_4
& \se{1} \frac{\al_n}2 [\bar \w_1^5 + \bar \w_1^3 (\bar \w_1^2 + \bar c_1)] .
\nonumber\\
&\se{1} \frac {\al_n}{2} \bar c_1\bar \w_1^3 
\nonumber 
\end{align}
To see if $\frac {1}{2} \bar c_1\bar \w_1^3 $ is a non-trivial cocycle, we note
that
\begin{align}
&H^5[\cB (U_1\rtimes Z_2^T);(\R/\Z)_T]
\rightarrowtail
H^5[\cB Z_2^T ;H^0(\cB U_1; \R/\Z)]\oplus
\nonumber\\ &
H^4[\cB Z_2^T ;H^1(\cB U_1; \R/\Z)]\oplus
H^3[\cB Z_2^T ;H^2(\cB U_1; \R/\Z)]\oplus
\nonumber\\ &
H^2[\cB Z_2^T ;H^3(\cB U_1; \R/\Z)]\oplus
H^1[\cB Z_2^T ;H^4(\cB U_1; \R/\Z)]\oplus
\nonumber\\
& \ \ \ \
H^0[\cB Z_2^T ;H^5(\cB U_1; \R/\Z)]
\nonumber\\ &
=
H^5[\cB Z_2^T; (\R/\Z)_T]
\oplus 
H^4[\cB Z_2^T;\Z] \oplus
H^2[\cB Z_2^T;(\Z)_T] \oplus
\nonumber\\ & 
\ \ \ \ H^5(\cB U_1; \R/\Z)
= \Z_2\oplus \Z
\end{align}

The $\Z$ in $H^4[\cB Z_2^T;\Z]$ comes from $H^1(\cB U_1; \R/\Z)=\Z$.  The unit
in $\Z$ correspond to the generator of $H^1(\cB U_1; \R/\Z)$: the
$\R/\Z$-valued nearly-flat 1-cochain $\bar a^{\RZ}$.  The time-reversal $Z_2^T$
has a non-trivial action on the $\R/\Z$-value: $\R/\Z \to - \R/\Z$.  It also
has a non-trivial action on $U_1$: $\bar a^{\RZ} \to -\bar a^{\RZ}$.  So the
total action of $Z_2^T$ is given by $ \bar a^{\RZ} \to \bar a^{\RZ}$.  Thus the
action of $Z_2^T$ on the coefficient $\Z$ is trivial. In this case $H^4[\cB
Z_2^T;\Z]=\Z_2$.
It is generated by $\bar \eta_5$ that satisfy
\begin{align}
 \dd \bar \eta_5 
\se{1} \bar c_1 \Bs_2 (\bar \w_1^3) 
\se{1} \frac12 \bar c_1 \dd_{\w_1} (\bar \w_1^3) .
\end{align}
Thus
\begin{align}
 \bar \eta_5 = \frac12 \bar c_1 \bar \w_1^3.
\end{align}

The $\Z$ in $H^2[\cB Z_2^T;\Z_T]$ comes from $H^3(\cB U_1; \R/\Z)=\Z$.  The
unit in $\Z$ correspond to the generator of $H^3(\cB U_1; \R/\Z)$: the
$\R/\Z$-valued nearly-flat 3-cochain $\bar a^{\RZ}c_1$, where $c_1$ is the
first Chern class of $\bar a^{\RZ}$.  The total action of $Z_2^T$ is given by $
\bar a^{\RZ}c_1 \to -\bar a^{\RZ}c_1$.  Thus the action of $Z_2^T$ on the
coefficient $\Z$ is non-trivial $\Z \to -\Z$. This is why we denote $\Z$ as
$\Z_T$.  In this case $H^2[\cB Z_2^T;\Z_T]=0$.

So, $H^5[\cB (U_1\rtimes Z_2^T);(\R/\Z)_T]$ is generated by $\frac12 \bar
\w_1^3 \bar c_1 $ and $\bar a^{\RZ} \bar c_1^2$.  We see that $ \frac12 \bar
\w_1^3 \bar c_1$ is a non-trivial cocycle in $H^5[\cB (Z_2^T\rtimes
U_1);(\R/\Z)_T]$ This means that $\bar \nu_4$ has solution only when $\al_n=0$.
In this case, $\bar \nu_4 $ is given by the cocycles in $H^4[\cB (U_1\rtimes
Z_2^T);(\R/\Z)_T]$.  We note that 
\begin{align}
&H^4[\cB (U_1\rtimes Z_2^T);(\RZ)_T]
\rightarrowtail
H^4[\cB Z_2^T ;H^0(\cB U_1; \R/\Z)]\oplus
\nonumber\\ &
H^3[\cB Z_2^T ;H^1(\cB U_1; \R/\Z)]\oplus
H^2[\cB Z_2^T ;H^2(\cB U_1; \R/\Z)]\oplus
\nonumber\\ &
H^1[\cB Z_2^T ;H^3(\cB U_1; \R/\Z)]\oplus
H^0[\cB Z_2^T ;H^4(\cB U_1; \R/\Z)]
\nonumber\\ &
=
H^4[\cB Z_2^T; (\RZ)_T]
\oplus 
H^3[\cB Z_2^T;\Z] \oplus
H^1[\cB Z_2^T;\Z_T] 
\nonumber\\
& = \Z_2\oplus \Z_2 
\end{align}

The $\Z$ in $H^3[\cB Z_2^T;\Z]$ comes from $H^1(\cB U_1; \R/\Z)=\Z$.  The unit
in $\Z$ correspond to the generator of $H^1(\cB U_1; \R/\Z)$: the
$\R/\Z$-valued nearly-flat 1-cochain $\bar a^{\RZ}$.  The time-reversal $Z_2^T$ has
a non-trivial action on the $\R/\Z$-value: $\R/\Z \to - \R/\Z$.  It also has a
non-trivial action on $U_1$: $\bar a^{\RZ} \to -\bar a^{\RZ}$.  So the total action of
$Z_2^T$ is given by $ \bar a^{\RZ} \to \bar a^{\RZ}$.  Thus the action of $Z_2^T$ on the
coefficient $\Z$ is trivial. In this case $H^3[\cB Z_2^T;\Z]=0$.

The $\Z$ in $H^1[\cB Z_2^T;\Z_T]$ comes from $H^3(\cB U_1; \R/\Z)=\Z$.  The
unit in $\Z$ correspond to the generator of $H^3(\cB U_1; \R/\Z)$: the
$\R/\Z$-valued nearly-flat 3-cochain $\bar a^{\RZ}c_1$, where $c_1$ is the first
Chern class of $\bar a^{\RZ}$.  The total action of $Z_2^T$ is given by $ \bar a^{\RZ}c_1
\to -\bar a^{\RZ}c_1$.  Thus the action of $Z_2^T$ on the coefficient $\Z$ is
non-trivial $\Z \to -\Z$. This is why we denote $\Z$ as $\Z_T$.  In this case
$H^1[\cB Z_2^T;\Z_T]=\Z_2$.

So, $H^4[\cB (U_1\rtimes Z_2^T);(\R/\Z)_T]$ is generated by $\frac12 \bar
c_1^2$ and $\frac12 (\bar a^{Z_2^T})^4$.
$\bar \nu_4$ may have a form
\begin{align}
 \bar \nu_4 \se{1} \frac{\al_\nu}2 \bar \w_1^4 +  \frac{\t\al_\nu}2 \bar c_1^2 
,\ \ \ \ \
\al_\nu,\t\al_\nu =0,1 .
\end{align}

\subsubsection{With extension of $O_\infty$}

\noindent\textbf{Calculate $\bar n_3$}:
Let us first calculate $\bar n_3 \in H^2(\cB G_{fO};\Z_2)= H^3(\cB (U_1^f
\gext_{\bar \veps_2'} O_\infty);\Z_2)$.  $\bar n_3$ has a form
\begin{align}
 \bar n_3 \se{2,\dd} \al_n \bar \w_1^3 + \t\al_n \bar \w_1 \bar \w_2 , \ \ \ \
\al_n,\t\al_n=0,1.
\end{align}
We do not have the $\bar \w_1 \bar c_1$ term due to the relation $\bar
c_1\se{2,\dd} \bar \w_2$.

However, on 3+1D manifold, $\w_1 \w_2\se{2,\dd}0$. Thus the pullback of 
$\bar n_3$ on space-time $\cM^3$ has a simpler form
\begin{align}
  n_3 \se{2,\dd} \al_n \w_1^3 , \ \ \ \ \ \al_n=0,1.
\end{align}

\noindent\textbf{Calculate $\bar \nu_4$}:
Next, we calculate $\bar \nu_4(\bar a^{G_{fO}})$ from \eqn{fSPTclO}, which
becomes
\begin{align}
 -\dd \bar \nu_4 
& \se{1} \frac{\al_n}2 [\bar \w_1^5 + \bar \w_1^3
(\bar \w_2 + \bar \w_1^2 )] 
\nonumber\\
&
\se{1} \frac {\al_n}{2} \bar \w_1^3 \bar \w_2
\se{1} \frac {\al_n}{2} \bar \w_1^3 \bar c_1 +   \frac {\al_n}{2} \bar \w_1^3 \dd \bar u_1,
\end{align}
where we have used \eqn{e2pptri}.
We note that $\frac 12 \bar \w_1^3$ is a $(\RZ)_T$-valued coboundary:
\begin{align}
 \frac 12 \bar \w_1^3 \se{1}\dd \eta_2,
\end{align}
where $Z_2^T$ has a non-trivial action on the value $(\RZ)_T$.
Thus $\frac 12 \bar \w_1^3\bar \w_2$ is also a coboundary
\begin{align}
 \frac 12 \bar \w_1^3\bar \w_2
\se{1}  \dd (\bar \eta_2 \bar c_1 +  \frac {1}{2} \bar \w_1^3 \bar u_1).
\end{align}
We find $\bar \nu_4$
to have a form
\begin{align}
 \bar \nu_4  & \se{1} 
\al_n (\bar \eta_2 \bar c_1 +\frac12 \bar \w_1^3 \bar u_1) 
  + \frac{\al_{\nu,1}}{2} \bar \w_1^4 
  + \frac{\al_{\nu,2}}{2} \bar \w_2^2 
  + \frac{\al_{\nu,3}}{2} \bar \w_4 
\nonumber\\
& + \frac{\al_{\nu,4}}{2} \bar \w_1^2 \bar \w_2
  + \frac{\al_{\nu,5}}{2} \bar \w_1 \bar \w_3 
,\ \ \ \ \al_{\nu,i} =0,1.
\end{align}
The terms  $\bar p_1$, $\bar \w_2\bar c_1$, and $\bar \w_1^2 \bar c_1$ are
not included since $\bar \w_2^2 \se{2} \bar p_1$ and $\bar \w_2 \se{2} \bar
c_1$.  

In 3+1D space-time, we have $\w_1\w_2 \se{2,\dd} \w_1\w_3
\se{2,\dd}  \w_1^4+  \w_2^2+  \w_4 \se{2,\dd}
 0 $ (see \eqn{wrel4}).  Thus after pulled back
to space-time $\cM^4$, $\bar \nu_4$ reduces to
\begin{align}
  \nu_4 & \se{1} \al_n  ( \eta_2  c_1 +\frac12  \w_1^3  u_1) 
+\frac{\al_{\nu}}{2}  \w_1^4 
+\frac{\t\al_{\nu}}{2}  \w_2 c_1
\nonumber\\
& \al_{\nu}, \t\al_{\nu} =0,1.
\end{align}
We note that $\w_2 c_1 \se{2,\dd} \w_2^2 \se{2,\dd} p_1$.  

From \eqn{e2ptri}, we see that $\frac12  \w_2$ is a coboundary.  So one might
expect the term $\frac12 \w_2 c_1 $ to be a coboundary, and can be dropped.
$\frac12 \w_2 c_1$ is indeed a coboundary if we view $\frac12 \w_2$ as a
$(\RZ)_T$-valued cocycle, where subscript $T$ indicate the non-trivial action
by time-reversal.  (Note that $\frac12 \w_2$ can also be viewed  as a
$\RZ$-valued cocycle, and in this case it may not be a coboundary.) Since $c_1$
is a $\Z_T$-valued cocycle, this implies that $\frac12 \w_2 c_1$ is a
coboundary if we view it as $\RZ$-valued cocycle, where the time-reversal acts
trivially.  But $\frac{\t\al_{\nu}}{2}  \w_2 c_1$ in $\nu_4$ is viewed as a
$(\RZ)_T$ valued cocycle, which may be non-trivial.

\noindent\textbf{SPT invariant}: The above $\nu_4$
gives rise to two fermionic $ (U_1^f \rtimes_\phi Z_4^{T,f})/Z_2 $-SPT states,
whose SPT invariant is given by
\begin{align}
Z^\text{top}(\cM^4,&A^{G_{fO}})
= 
\ee^{\ii 2\pi \int_{\cM^4}   
\al_n  ( \eta_2  c_1 +  \frac {1}{2} \w_1^3  u_1) +
\frac{\al_{\nu}}{2}  \w_1^4
+\frac{\t\al_{\nu}}{2}  p_1
}
\nonumber\\
&
\ \ \ \ \
\ \ \ \ \
\ \ \
\ee^{
\ii \pi \int_{\cN^5} \Sq^2 f_3 + f_3(\w_2+\w_1^2) } ,
\nonumber\\ &  \ \ \ \ \
\dd A^{\Z_2^f} \se{2} \w_2+c_1(A^{U_1}) , \ \ \ 
f_3 \big|_{\prt \cN^4} \se{2} \al_n \w_1^3 ,
\nonumber\\ &  \ \ \ \ \
\al_n,\al_\nu =0,1.
\end{align}
Here we label $A^{G_{fO}}\in U_1^f \gext_{\bar \veps_2'} O_\infty$ by
$(A^{U_1^f}, A^{O})$, and label $A^{U_1^f}$ by $A^{U_1^f}_{ij}=\frac12
A^{U_1}_{ij}+\frac12 A^{Z_2^f}_{ij}$, where $A^{U_1^f}_{ij},A^{U_1}_{ij} \in
[0,1)$ and $A^{Z_2^f}_{ij} = 0,1$.  The 3+1D space-time and its $G_{fO}$
connection satisfies \eqn{e2ptri} and \eqn{e2pptri}.

For non-interacting fermion systems, the $(U_1^f \rtimes_\phi
Z_4^{T,f})/Z_2$-SPT phases (the topological insulators) in 3+1D are classified
by $\Z_2$.\cite{K0986,SRF0825,W11116341}

In the above, we show that, after
including interaction and via fermion decoration, we obtain 4 types of
interacting topological insulators (including the trivial type).
The charge $0$ bosonic
electron-hole pairs can give rise to 

We know that bosonic $Z_2^T$-SPT phases are classified by
$\Z_2^2$,\cite{CGL1314,VS1306,W1477,K1459} which are generated by bosonic
$Z_2^T$-SPT states with SPT invariants $\frac12 \w_1^4,\  \frac12
p_1$.\cite{W1477,K1459} Our above result indicates that the 4 bosonic
$Z_2^T$-SPT phases correspond to 4 different fermionic SPT phases when the
bosons are formed by electron-hole pairs.  This gives rise to a total of 8
fermionic $(U_1^f \rtimes_\phi Z_4^{T,f})/Z_2$-SPT phases in 3+1D, obtained via
fermion decoration construction.  This is consistent with the previous physical
argument \cite{WS13063238} and the cobordism
calculation.\cite{KTT1429,FH160406527}

\section{Fermionic $SU_2^f$-SPT state}

In this section, we are going to study fermionic SPT phases with $G_f=SU_2^f$
symmetry in 2+1D and in 3+1D.  Such a symmetry can be realized by a charge-$2e$
spin-singlet superconductor of electrons.  For non-interacting electron, there
is no non-trivial $SU_2^f$-SPT phase in 3+1D.

For fermion systems with bosonic symmetry $G_b=SO_3$, the full fermionic
symmetry $G_f$ is an extension of $G_b$ by $Z_2^f$.  The extension
$G_f=SU_2^f=Z_2^f \gext_{\w_2^{SO_3}} SO_3$ is characterized by
\begin{align}
 \bar e_2\se{2} \bar \w_2^{SO_3} \in H^2(\cB SO_3;\Z_2).
\end{align}

For $G_f=SU_2^f$, the group $G_{fSO}$ is an extension of $SO_\infty$ by
$SU_2^f$:
\begin{align} 
G_{fSO}=SU_2^f\gext_{e_2} SO_\infty=Z_2^f\gext_{\bar e_2'} (SO_3 \times SO_\infty) .  
\end{align} 
where $\bar e_2' \in H^2(\cB (SO_3\times SO_\infty); \Z_2)$ is given by
\begin{align}
\bar e_2' \se{2} \bar \w_2(\bar a^{SO})+ \bar \w_2^{SO_3}(\bar a^{SO_3}).
\end{align}
On $\cB G_{fSO}$, $\bar e_2'$ is trivialized
\begin{align}
\label{a2w2SU2}
\bar \w_2(\bar a^{SO})+ \bar \w_2^{SO_3}(\bar a^{SO_3}) 
\se{2} \dd \bar a^{\Z_2^f} ,
\end{align}
where a $G_{fSO}$ connection is labeled by
\begin{align}
 a^{G_{fSO}} = (a^{\Z_2^f},  a^{SO_3}, a^{SO}).
\end{align}

\subsection{2+1D}

\subsubsection{Without extension of $SO_\infty$}

\noindent\textbf{Calculate $\bar n_2$}:  
First, $\bar n_2\in H^2(\cB SO_3;\Z_2) =\Z_2$. It has two choices: 
\begin{align}
\bar n_2 =\al_n \bar \w_2^{SO_3}, \ \ \ \ \ \al_n=0,1.  
\end{align}

\noindent\textbf{Calculate $\bar \nu_3$}:  
Similar to the last section, $\bar \nu_3$ satisfies
\begin{align}
-\dd \bar \nu_3 & \se{1}  
 \frac{\al_n^2}2 \Sq^2 \bar \w_2^{SO_3}
+ \frac{\al_n}2 \bar \w_2^{SO_3} \bar e_2 
\se{1} 0 .
\end{align}
Thus, $\bar \nu_3$ has solutions classified by $\Z$: 
\begin{align}
\bar  \nu_3 \se{1} \al_\nu \om_3^{SO_3},\ \ \ \al_\nu \in \Z,
\end{align}
since $H^3(\cB SO_3;\RZ) =\Z$ (see \eqn{HBSOU1}).

\noindent\textbf{SPT invariant}:
The $SU_2^f$ fermionic SPT states labeled by $\al_n,\al_\nu$ 
have the following SPT invariant
\begin{align}
\label{SPTinvSU2f}
& \ \ \ \
Z^\text{top}(\cM^3,A^{Z_2},A^{\Z_2^f}) 
\nonumber\\
& = 
 \ee^{\ii 2\pi \int_{\cM^3}  \al_\nu  \om_3^{SO_3}
+\frac{\al_n}2   \w_2^{SO_3} A^{\Z_2^f} } 
\ee^{ \ii \pi \int_{\cN^{4}}  \Sq^2 f_2+f_2\w_2   }, 
\nonumber\\
& \ \ \ \
\dd A^{\Z_2^f} \se{2} \w_2+\w_2^{SO_3}
, \ \ \ \ f_2\big|_{\prt \cN^4} \se{2} \al_n \w_2^{SO_3}
\end{align}
where the space-time $\cM^3$ is orientable and $\w_1\se{2}0$.  However, as we
will see below, the $SU_2^f$ fermionic SPT phase are only labeled by $\al_\nu
\in \Z$. 

On 2+1D space-time manifold, $\w_2+\w_1^2 \se{2,\dd}0$ (see Appendix
\ref{Rswc3D}).  The $SU_2^f$ fermionic symmetry requires the space-time $\cM^3$
to be a orientable manifold with $\w_2+\w_2^{SO_3} \se{2,\dd}0$ and $\w_1
\se{2,\dd}0$. Thus $n_2\se{2,\dd}\w_2^{SO_3}$ is always a coboundary, and
$\al_n=0,1$ describe the same SPT phase.

\subsubsection{With extension of $SO_\infty$}
\label{SU2f2d}

\noindent\textbf{Calculate $\bar n_2$}:  
With extension of $SO_\infty$,
in general, $\bar n_2(\bar a^{G_{fSO}}) \in H^2(\cB
G_{fSO};\Z_2)$ is given by [using the triple $(\bar a^{Z_2^f},\bar a^{SO_3},\bar a^{SO})$ to
label $\bar a^{G_{fSO}}$]
\begin{align}
\bar n_2(\bar a^{G_{fSO}}) 
&\se{2} \al_{n,1}\bar \w_2(\bar a^{SO}) 
+ \al_{n,2} \bar \w_2^{SO_3}(\bar a^{SO_3}),
\end{align}
$\al_{n,1},\al_{n,2}=0,1$.  
The above can be reduced to
\begin{align}
\bar n_2(\bar a^{G_{fSO}}) \se{2} 
    \al_{n}  \bar \w_2 
\end{align}
$\al_n=0,1$,  due to the relation \eq{a2w2SU2}.  So $\bar n_2(\bar
a^{G_{fSO}})$ has two choices.  But on 2+1D orientable space-time $\cM^3$,
$\w_2\se{2,\dd}0$.  Thus after pulled back to $\cM^3$, $ \w_2$ is a coboundary.
Thus $n_2$ is always trivial.

\noindent\textbf{Calculate $\bar \nu_3$}:  
Next, we consider $\nu_3$ in \eqn{fSPTclSO} which becomes $\dd \bar \nu_3
\se{1} 0$.  We find that $\bar \nu_3$ is given by
\begin{align}
\bar  \nu_3 
\se{1} \al_\nu \bar \om_3^{SO_3} + \frac{\t\al_\nu}2 \bar \w_3
\ \ \ \ \ \al_\nu \in \Z,\ \ \  \t\al_\nu=0,1.
\end{align} 
However, $\bar \w_3 \se{2,\dd} \Bs_2 \bar \w_2$ (see \eqn{Sq1Bs2} and
\eqn{WuF})
So the term $\frac{\t\al_\nu}2 \bar \w_3$ can be rewritten as
\begin{align}
 \frac{\t\al_\nu}2 \bar \w_3 \se{1,\dd} \frac{\t\al_\nu}2 \Bs_2 \bar \w_2 .
\end{align} 
In this form, since $\Bs_2 \bar \w_2$ is a $\Z$-valued cocycle, $\t\al_\nu$ do
not have to be quantized as $\t\al_\nu=0,1$.  $ \t\al_\nu$ can take any real
values and $\frac{\t\al_\nu}2 \bar \w_3$ is still a $\RZ$-valued cocycle.  Thus
$\t\al_\nu$ is not quantized and can be tuned to zero.  Thus we drop the
$\frac{\t\al_\nu}2 \bar \w_3$ term.
Thus, The fermionic $SU_2^f$-SPT phase obtained via fermion decoration is
classified by $\al_n\in \Z$.

\subsection{3+1D}
\label{SU2f3d}

\subsubsection{Without extension of $SO_\infty$}

\noindent\textbf{Calculate $\bar n_3$}:  
$\bar n_3 \in H^3(\cB SO_3;\Z_2)$ has two
choices:
\begin{align}
\bar n_3 = \al_n \bar \w_3^{SO_3}
\end{align}
$\al_n=0,1$, since $H^3(\cB SO_3;\Z_2) = \Z_2$.  

\noindent\textbf{Calculate $\bar \nu_4$}:  
Next we want to solve
\begin{align}
 -\dd \bar \nu_4 & \se{1}  \frac12 (\Sq^2\bar n_3 + \bar n_3 \bar e_2)  .
\nonumber\\
 & \se{1}  \frac{\al_n}2 (\Sq^2\bar \w_3^{SO_3} + \bar \w_3^{SO_3} \bar \w_2^{SO_3})
\se{1} \dd \bar s_4^{SO_3} ,
\end{align}
where we have used (see \eqn{WuF})
\begin{align}
\Sq^2\bar \w_3^{SO_3} \se{2}
\bar \w_2^{SO_3}\bar \w_3^{SO_3} + \dd \bar s_4^{SO_3}(\bar a^{SO_3})
\end{align}
  Since $H^4(\cB SO_3;\RZ)= \Z_2$, the solution of
$\bar \nu_4$ has a form
\begin{align}
 \bar \nu_4 \se{1} \frac{\al_\nu}2 \bar \w_4^{SO_3} + \frac{\al_n}2 \bar s_4^{SO_3},\ \ \ \
\al_\nu=0,1.
\end{align}

\subsubsection{With extension of $SO_\infty$}

\noindent\textbf{Calculate $\bar n_3$}:  
In general, $\bar n_3(\bar a^{G_{fSO}}) \in H^3(\cB G_{fSO};\Z_2)$ can be written as
\begin{align}
& \bar n_3(\bar a^{G_{fSO}}) \se{2} 
    \al_{n}\bar \w_3 + \al_{n}' \bar \w_3^{SO_3} ,
\nonumber\\
&\ \ \ \ \ \ \ \al_{n},\al_{n}'=0,1 .
\end{align}
However, from $\bar \w_2 \se{2,\dd} \bar \w_2^{SO_3}$, we find that $\Sq^1
(\bar \w_2+ \bar \w_2^{SO_3}) \se{2,\dd} \bar \w_3 + \bar \w_2^{SO_3}
\se{2,\dd} 0$ (see \eqn{WuF} and notice $\bar \w_1\se{2}\bar
\w_1^{SO_3}\se{2}0$).  
Furthermore, on a orientable 4-manifold $\cM^4$, $\w_3^{SO_3}\se{2,\dd} \w_3$
is always a coboundary.  From Appendix \ref{Rswc4D}, we have
\begin{align}
a^{Z_2}\w_3 & \se{2,\dd} (a^{Z_2})^2 \w_2 \se{2,\dd} \Sq^2 (a^{Z_2})^2  
\nonumber\\
& \se{2,\dd}  (a^{Z_2})^4 \se{2,\dd}  \w_1 (a^{Z_2})^3 \se{2,\dd} 0.
\end{align}
Since  $a^{\Z_2}$ can be 
an arbitrary $\Z_2$-valued 1-cocycle, we find $\w_3\se{2,\dd} 0$. 
Thus the above expression for $n_3(\bar a^{G_{fSO}})$ is reduced to
\begin{align}
 n_3(\bar a^{G_{fSO}}) &\se{2} 0.
\end{align}

\noindent\textbf{Calculate $\bar \nu_4$}:  Next, we consider $\bar \nu_4$ in
\eqn{fSPTclSO} which becomes $ \dd \bar \nu_4(\bar a^{G_{fSO}})  \se{1} 0 $.
We find that 
\begin{align}
\bar  \nu_4 &\se{1} 0. 
\end{align} 
Note that we do not have the $\bar p_1$ term since $\bar p_1$ is $\Z$-valued
and its coefficient is not quantized.  Also on $\cM^4$, we have $\w_4\se{2,\dd}
\w_2^2 \se{2,\dd} p_1$.  So we do not have the $\w_4$ and $\w_2^2$ terms.  Due
to \eqn{a2w2SU2}, we also do not have the $\w_2\w_2^{SO_3}$ and
$(\w_2^{SO_3})^2$ terms.  Thus there is no non-trivial fermionic $SU_2^f$-SPT
phases in 3+1D from fermion decoration.

\section{Fermionic $Z_2\times Z_4\times Z_2^f$-SPT state}

In this section, we are going to study fermionic SPT phases, where the fermion
symmetry is given by $G_f=Z_2\times Z_4\times Z_2^f$ symmetry.  For
non-interaction fermions, there is no non-trivial $Z_2\times Z_4\times
Z_2^f$-SPT phases in 3+1D.  For $G_f=Z_2\times Z_4\times Z_2^f$, the
corresponding $G_{fSO}=G_f\gext SO_\infty$ is given by $G_{fSO}=Z_2\times
Z_4\times Spin_\infty$.  $\bar \w_2$ is trivialized on $\cB G_{fSO}$:
\begin{align}
 \bar \w_2(a^{SO}) \se{2} \dd \bar a^{\Z_2^f} ,
\end{align}
where we have labeled the $G_{fSO}$ connection as
\begin{align}
a^{G_{fSO}}=( a^{\Z_2^f}, a^{\Z_2}, a^{\Z_4},a^{SO}).
\end{align}
where $a^{\Z_4}$ is a $\Z_4$-valued 1-cocycle.

\subsection{2+1D}

\noindent\textbf{Calculate $\bar n_2$}:  
First $\bar n_2 \in H^2(\cB G_{fSO};\Z_2)$ is given by
\begin{align}
\bar n_2 & \se{2} 
 \al_{n,1} (\bar a^{\Z_2})^2 
+\al_{n,2} (\bar a^{\Z_4})^2 
+\al_{n,3} \bar a^{\Z_2}\bar a^{\Z_4},
\nonumber\\
&\ \ \ \ \ \  \ \ \
\al_{n,i}=0,1
.
\end{align}

\noindent\textbf{Calculate $\bar \nu_3$}:  
Next, we consider $\bar \nu_3(\bar a^{G_{fSO}})$ in \eqn{fSPTclSO} which becomes
\begin{align}
 -\dd \bar \nu_3 &\se{1} \frac12 \Sq^2 \bar n_2 +\frac12 \bar n_2 \dd \bar a^{\Z_2^f}
\nonumber\\ &
 \se{1} 
\frac{\al_{n,1}}2 (\bar a^{\Z_2})^4 
+\frac{\al_{n,2}}2 (\bar a^{\Z_4})^4 
+\frac{\al_{n,3}}2 (\bar a^{\Z_2})^2(\bar a^{\Z_4})^2
\nonumber\\ &\ \
+ \frac{\al_{n,3}}2 \bar a^{\Z_2} [\dd (\bar a^{\Z_2}\hcup{1}\bar a^{\Z_4})] \bar a^{\Z_4}
\nonumber\\ &\ \
+ \frac{\al_{n,1}\al_{n,2}}2 \dd [(\bar a^{\Z_2})^2 \hcup{1}(\bar a^{\Z_4})^2 ]
\nonumber\\ &\ \ 
+ \frac{\al_{n,1}\al_{n,3}}2 \dd [(\bar a^{\Z_2})^2 \hcup{1}\bar a^{\Z_2}\bar a^{\Z_4} ]
\\ &\ \ 
+ \frac{\al_{n,2}\al_{n,3}}2 \dd [(\bar a^{\Z_4})^2 \hcup{1}\bar a^{\Z_2}\bar a^{\Z_4} ]
\nonumber\\ &\ \ 
+ [ \frac{\al_{n,1}}2 (\bar a^{\Z_2})^2 
+\frac{\al_{n,2}}2 (\bar a^{\Z_4})^2 
+\frac{\al_{n,3}}2 \bar a^{\Z_2}\bar a^{\Z_4} ] \dd \bar a^{\Z_2^f}
.
\nonumber 
\end{align}
The solution of the above equation has a form
\begin{align}
\bar  \nu_3 &\se{1} 
\frac{\al_{n,1}}4 \bar a^{\Z_2} \Bs_2 \bar a^{\Z_2}
+\frac{\al_{n,2}}4 \bar a^{\Z_4} \Bs_2 \bar a^{\Z_2}
+\frac{\al_{n,3}}4 \bar a^{\Z_4} \Bs_2 \bar a^{\Z_2}
\nonumber\\ &\ \
+ \frac{\al_{n,3}}2 \bar a^{\Z_2} (\bar a^{\Z_2}\hcup{1}\bar a^{\Z_4}) \bar a^{\Z_4}
\nonumber\\ &\ \
+ \frac{\al_{n,1}\al_{n,2}}2 (\bar a^{\Z_2})^2 \hcup{1}(\bar a^{\Z_4})^2 
\nonumber\\ &\ \ 
+ \frac{\al_{n,1}\al_{n,3}}2 (\bar a^{\Z_2})^2 \hcup{1}\bar a^{\Z_2}\bar a^{\Z_4}
\\ &\ \ 
+ \frac{\al_{n,2}\al_{n,3}}2 (\bar a^{\Z_4})^2 \hcup{1}\bar a^{\Z_2}\bar a^{\Z_4}
\nonumber\\ &\ \ 
+ [ \frac{\al_{n,1}}2 (\bar a^{\Z_2})^2 
+\frac{\al_{n,2}}2 (\bar a^{\Z_4})^2 
+\frac{\al_{n,3}}2 \bar a^{\Z_2}\bar a^{\Z_4} ] \bar a^{\Z_2^f}
\nonumber\\ &\ \ 
+\frac{\al_{\nu,1}}2 (\bar a^{\Z_2})^3
+\frac{\al_{\nu,2}}2 \bar a^{\Z_4}\Bs_2 \bar a^{\Z_2}
+\frac{\al_{\nu,3}}4 (\bar a^{\Z_4})^3
.
\nonumber 
\end{align}
where $\al_{\nu,i}|_{i=1,2} =0,1$ and $\al_{\nu,3} =0,1,2,3$.
$\al_{n,i},\al_{\nu,i}$ label 256 different SPT phases, which can be
divided into 8 classes with 16 SPT phases in each class.  The 8 classes are
labeled by $\al_{n,i}$.  The 16 SPT phases in each class only differ by
stacking the bosonic $Z_2\times Z_4$-SPT phases realized by fermion pairs.

\subsection{3+1D}

\noindent\textbf{Calculate $\bar n_3$}:  
Since $G_{fSO}=\Z_2\times Z_4 \times Spin_\infty$, $\bar n_3 \in
H^3(G_{fSO};\Z_2)$ is generated by $\bar a^{\Z_2}$, $\bar a^{\Z_4}$, and
Stiefel-Whitney class $\bar \w_n$.  For $Spin_\infty$, $\bar \w_1=\bar \w_2=0$.
Also, $\Sq^1\bar \w_2 \se{2,\dd} \bar \w_1\bar \w_2 +\bar \w_3 \se{2,\dd} \bar
\w_3$.  Since $\bar \w_2$ is a coboundary for $Spin_\infty$, $\bar \w_3$ is
also a coboundary.  Thus, $\bar n_3$ is given by
\begin{align}
 \bar n_3 &\se{2} 
 \al_{n,1} (\bar a^{\Z_2})^3 
+\al_{n,2} (\bar a^{\Z_4})^3 
+\al_{n,3} \bar a^{\Z_4} (\bar a^{\Z_2})^2 
\nonumber\\
&\ \ \ \
+\al_{n,4} \bar a^{\Z_2} (\bar a^{\Z_4})^2 ,
\ \ \ \ \ \al_{n,i}=0,1.
\end{align}

\noindent\textbf{Calculate $\bar \nu_4$}:  
$\bar \nu_4$ is calculated from \eqn{fSPTclSO}
\begin{align}
 \dd \bar \nu_4 
&\se{1,\dd}
\frac{\al_{n,1}}2 \Sq^2 (\bar a^{\Z_2})^3
+\frac{\al_{n,2}}2 \Sq^2 (\bar a^{\Z_4})^3
\nonumber\\ &\ \ \ \ \ 
+\frac{\al_{n,3}}2 \Sq^2 \bar a^{\Z_4}(\bar a^{\Z_2})^2
+\frac{\al_{n,4}}2 \Sq^2 \bar a^{\Z_2}(\bar a^{\Z_4})^2
\nonumber\\
&\se{1,\dd}
\frac{\al_{n,1}}2  (\bar a^{\Z_2})^5
+\frac{\al_{n,2}}2  (\bar a^{\Z_4})^5
\nonumber\\ &\ \ \ \ \ 
+\frac{\al_{n,3}}2 \bar a^{\Z_4}(\bar a^{\Z_2})^4
+\frac{\al_{n,3}}2 \bar a^{\Z_2}(\bar a^{\Z_4})^4
,
\nonumber 
\end{align}
where we have used \eqn{Sq2xy}.  
$\frac12 (\bar a^{\Z_2})^5$, $\frac12
(\bar a^{\Z_4})^5$, and $\frac{\al_{n,3}}2 \bar a^{\Z_2}(\bar a^{\Z_4})^4$ are
non-trivial cocyles in $H^5(\cB G_{fSO};\RZ)$. But
$\frac12 \bar a^{\Z_4}(\bar a^{\Z_2})^4$
is a coboundary
\begin{align}
 \frac12 \bar a^{\Z_4} (\Bs_2 \bar a^{\Z_2}) (\Bs_2 \bar a^{\Z_2})
\se{1} \frac14 \bar a^{\Z_4} (\dd \bar a^{\Z_2}) \Bs_2 \bar a^{\Z_2}
\se{1} \dd (\frac14 \bar a^{\Z_4}  \bar a^{\Z_2} \Bs_2 \bar a^{\Z_2})
\end{align}
Thus $\nu_4 $ has a form
\begin{align}
 \bar \nu_4(\bar a^{G_{fSO}}) &\se{1} 
\frac{\al_n}4 \bar a^{\Z_4}  \bar a^{\Z_2} \Bs_2 \bar a^{\Z_2}
+ \frac{\al_{\nu,1}}2 \bar a^{\Z_2} (\bar a^{\Z_4})^3
\nonumber\\ &\ \ \ \ \
+\frac{\al_{\nu,2}}2 \bar a^{\Z_4} (\bar a^{\Z_2})^3,
\ \ \ \ \ \ \ \ \al_{\nu,i}=0,1 
.
\end{align}
We find that there are eight fermionic $Z_2\times Z_4\times Z_2^f$-SPT phases
in 3+1D from fermion decoration.  Those phases can be divided into two groups
of four, and the four phases in each group differ by bosonic $Z_2\times
Z_4$-SPT phases from fermion pairs.

\Ref{WY180105416} has calculated $Z_2\times Z_4\times Z_2^f$-SPT phases using
cobordism approach and found $\Z_2\times \Z_4$ phases.  On the other hand,
non-interacting fermions can only realize the trivial  $Z_2\times Z_4\times
Z_2^f$-SPT phase.\cite{K0986,SRF0825,W11116341} So the intrinsic fermionic
$Z_2\times Z_4\times Z_2^f$-SPT phase (the phases that cannot be realized by
bosonic fermion pairs) cannot come from non-interacting
fermions.\cite{CW170508911}  In the above, we see that the intrinsic fermionic
$Z_2\times Z_4\times Z_2^f$-SPT phases can all come from fermion decoration.

We thank Zheng-Cheng Gu and Juven  Wang for many very helpful discussions.  XGW
is supported by NSF Grant No.  DMR-1506475 and DMS-1664412.

\appendix

\section{Space-time complex, cochains, and cocycles} 

\label{cochain}

In this paper, we consider models defined on a space-time lattice.  A
space-time lattice is a triangulation of the $D$-dimensional space-time $M^D$, which
is denoted by $\cM^D$.  We will also call the triangulation
$\cM^D$ as a space-time complex, which is formed by simplices -- the
vertices, links, triangles, \etc.  We will use $i,j,\cdots$ to label vertices
of the space-time complex.  The links of the complex (the 1-simplices) will be
labeled by $(i,j),(j,k),\cdots$.  Similarly, the triangles of the complex  (the
2-simplices)  will be labeled by $(i,j,k),(j,k,l),\cdots$.

\begin{figure}[t]
\begin{center}
\includegraphics[scale=0.5]{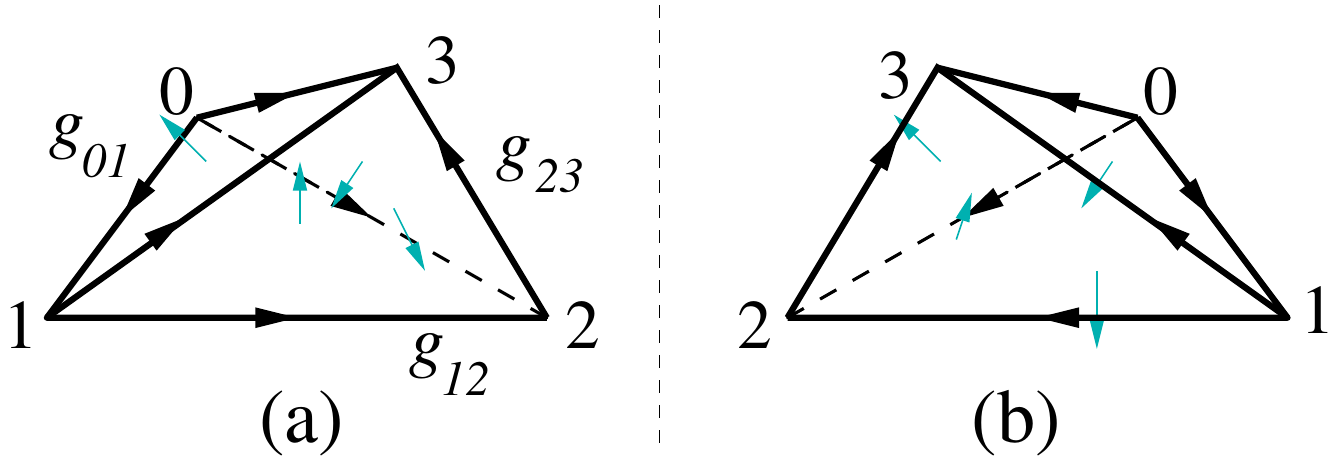} \end{center}
%4
\caption{ (Color online) Two branched simplices with opposite orientations.
(a) A branched simplex with positive orientation and (b) a branched simplex
with negative orientation.  }
\label{mir}
\end{figure}

In order to define a generic lattice theory on the space-time complex
$\cM^D$ using local Lagrangian term on each simplex, it is important
to give the vertices of each simplex a local order.  A nice local scheme to
order  the vertices is given by a branching
structure.\cite{C0527,CGL1314,CGL1204} A branching structure is a choice of
orientation of each link in the $d$-dimensional complex so that there is no
oriented loop on any triangle (see Fig. \ref{mir}).

The branching structure induces a \emph{local order} of the vertices on each
simplex.  The first vertex of a simplex is the vertex with no incoming links,
and the second vertex is the vertex with only one incoming link, \etc.  So the
simplex in  Fig. \ref{mir}a has the following vertex ordering: $0,1,2,3$.

The branching structure also gives the simplex (and its sub-simplices) a
canonical orientation.  Fig. \ref{mir} illustrates two $3$-simplices with
opposite canonical orientations compared with the 3-dimension space in which
they are embedded.  The blue arrows indicate the canonical orientations of the
$2$-simplices.  The black arrows indicate the canonical orientations of the
$1$-simplices.

Given an Abelian group $(\M, +)$, an $n$-cochain $f_n$ is an assignment of
values in $\M$ to each $n$-simplex, for example a value $f_{n;i,j,\cdots,k}\in
\M$ is assigned to $n$-simplex $(i,j,\cdots,k)$.  So \emph{a cochain $f_n$ can
be viewed as a bosonic field on the space-time lattice}. 

$\M$ can also be viewed a $\Z$-module (\ie a vector space with
integer coefficient) that also allows scaling by an integer:  
\begin{align}
	 x+y &= z,\ \ \ \ x*y=z, \ \ \ \ mx=y,
	\nonumber\\
	x,y,z & \in \M,\ \ \ m \in \Z.
\end{align}
The direct sum of two modules
$\M_1\oplus \M_2$ (as vector spaces) is equal to the direct product of the two
modules (as sets):
\begin{align}
 \M_1\oplus \M_2 \stackrel{\text{as set}}{=} \M_1\times \M_2
\end{align}

\begin{figure}[tb]
\begin{center}
\includegraphics[scale=0.5]{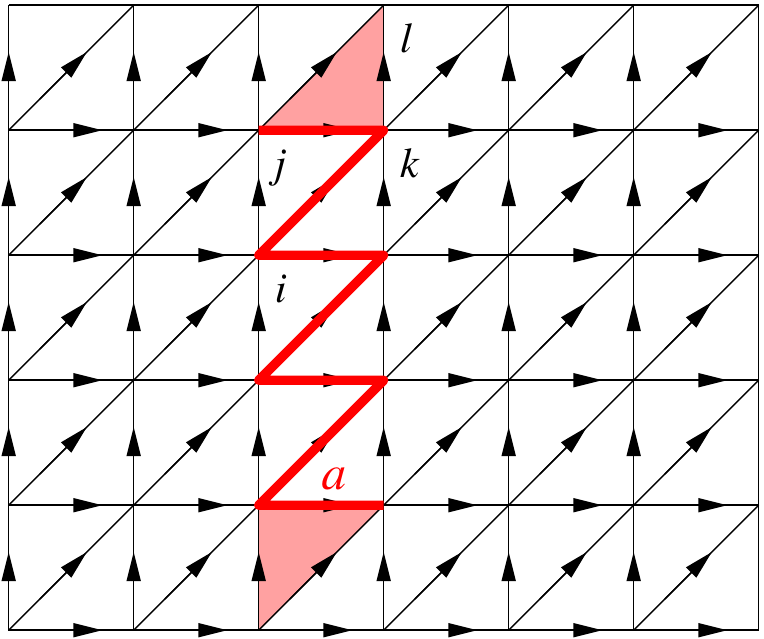} \end{center}
%5
\caption{ (Color online)
A 1-cochain $a$ has a value $1$ on the red links: $ a_{ik}=a_{jk}= 1$ and a
value $0$ on other links: $ a_{ij}=a_{kl}=0 $.  $\dd a$ is non-zero on the
shaded triangles: $(\dd a)_{jkl} = a_{jk} + a_{kl} - a_{jl}$.  For such
1-cohain, we also have $a\smile a=0$.  So when viewed as a $\Z_2$-valued cochain,
$\Bs_2 a \neq a\smile a$ mod 2.
}
\label{dcochain}
\end{figure}

We like to remark that a simplex $(i,j,\cdots,k)$ can have two different
orientations. We can use $(i,j,\cdots,k)$ and $(j,i,\cdots,k)=-(i,j,\cdots,k)$
to denote the same simplex with opposite orientations.  The value
$f_{n;i,j,\cdots,k}$ assigned to the simplex with opposite  orientations should
differ by a sign: $f_{n;i,j,\cdots,k}=-f_{n;j,i,\cdots,k}$.  So to be more
precise $f_n$ is a linear map $f_n: n\text{-simplex} \to \M$. We can denote the
linear map as $\<f_n, n\text{-simplex}\>$, or
\begin{align}
 \<f_n, (i,j,\cdots,k)\> = f_{n;i,j,\cdots,k} \in \M.
\end{align}
More generally, a \emph{cochain} $f_n$ is a linear map
of $n$-chains:
\begin{align}
	f_n:  n\text{-chains} \to \M,
\end{align}
or (see Fig. \ref{dcochain})
\begin{align}
 \<f_n, n\text{-chain}\> \in \M,
\end{align}
where a \emph{chain} is a composition of simplices. For example, a 2-chain can
be a 2-simplex: $(i,j,k)$, a sum of two 2-simplices: $(i,j,k)+(j,k,l)$, a more
general composition of 2-simplices: $(i,j,k)-2(j,k,l)$, \etc.  The map $f_n$ is
linear respect to such a composition.  For example, if a chain is $m$ copies of
a simplex, then its assigned value will be $m$ times that of the simplex.
$m=-1$ correspond to an opposite orientation.  

We will use $C^n(\cM^D;\M)$ to denote the set of all
$n$-cochains on $\cM^D$.  $C^n(\cM^D;\M)$ can also be
viewed as a set all $\M$-valued fields (or paths) on  $\cM^D$.  Note
that $C^n(\cM^D;\M)$ is an Abelian group under the $+$-operation.

The total space-time lattice $\cM^D$ correspond to a $D$-chain.  We
will use the same $\cM^D$ to denote it.  Viewing $f_D$ as a linear
map of $D$-chains, we can define an ``integral'' over $\cM^D$:
\begin{align}
 \int_{\cM^D} f_D &\equiv \<f_D,\cM^D\>
\\
&=\sum_{(i_0,i_1,\cdots,i_D)}
s_{i_0i_1\cdots i_D} (f_D)_{i_0,i_1,\cdots,i_D}
.
\nonumber 
\end{align}
Here $s_{i_0i_1\cdots i_D}=\pm 1$, such that a $D$-simplex in the $D$-chain
$\cM^D$ is given by $s_{i_0i_1\cdots i_D} (i_0,i_1,\cdots,i_D)$.

\begin{figure}[tb]
\begin{center}
\includegraphics[scale=0.5]{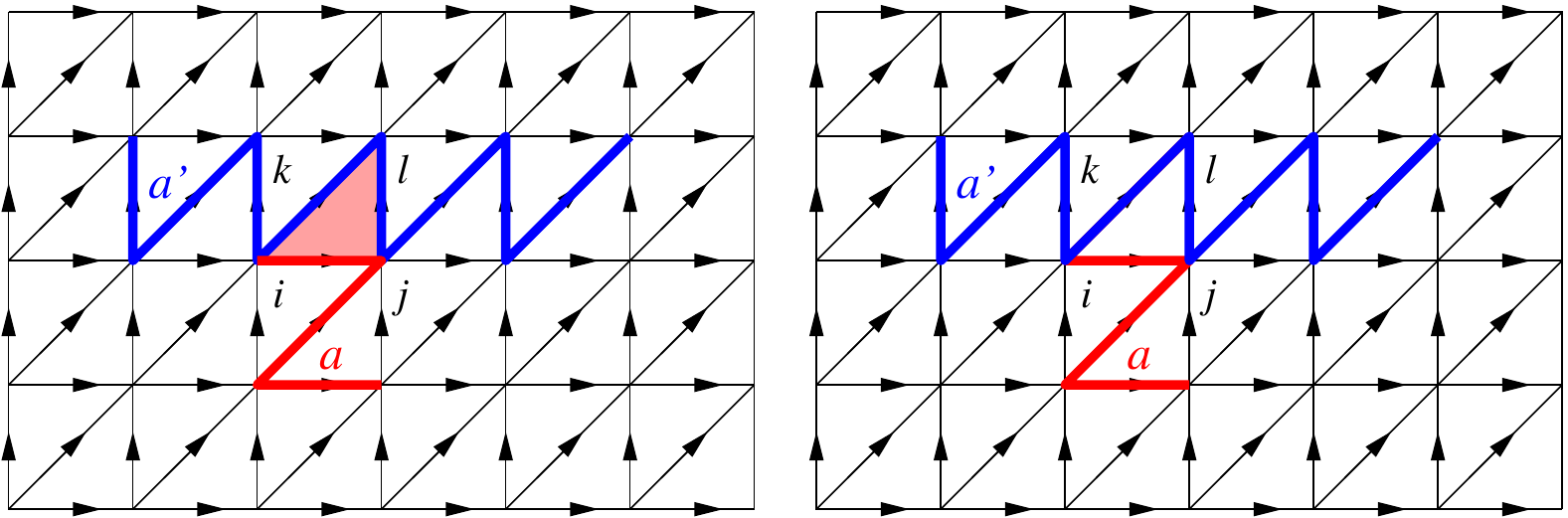} \end{center}
%6
\caption{ (Color online)
A 1-cochain $a$ has a value $1$ on the red links, Another
1-cochain $a'$ has a value $1$ on the blue links.
On the left, $a\smile a'$ is non-zero on the shade triangles:
$(a\smile a')_{ijl}=a_{ij}a'_{jl}=1$.
On the right, $a'\smile a$ is zero on every triangle.
Thus $a\smile a'+a'\smile a$ is not a coboundary.
}
\label{cupcom}
\end{figure}

We can define a derivative operator $\dd$ acting on an $n$-cochain $f_n$, which
give us an $n+1$-cochain (see Fig. \ref{dcochain}):
\begin{align} 
\label{eq:differential}
&\ \ \ \ \<\dd f_n, (i_0i_1i_2\cdots i_{n+1})\>
\nonumber\\
&=\sum_{m=0}^{n+1} (-)^m 
\<f_n, (i_0i_1i_2\cdots\hat i_m\cdots i_{n+1})\>
\end{align}
where $i_0i_1i_2\cdots \hat i_m \cdots i_{n+1}$ is the sequence
$i_0 i_1 i_2 \cdots i_{n+1}$ with $i_m$ removed, and
$i_0, i_1,i_2 \cdots i_{n+1}$ are the ordered vertices of the $(n+1)$-simplex
$(i_0 i_1 i_2 \cdots i_{n+1})$.

A cochain $f_n \in C^n(\cM^D;\M)$ is called a \emph{cocycle} if $\dd
f_n=0$.   The set of cocycles is denoted by $Z^n(\cM^D;\M)$.  A
cochain $f_n$ is called a \emph{coboundary} if there exist a cochain $f_{n-1}$
such that $\dd f_{n-1}=f_n$.  The set of coboundaries is denoted by
$B^n(\cM^D;\M)$.  Both $Z^n(\cM^D;\M)$ and
$B^n(\cM^D;\M)$ are Abelian groups as well.  Since $\dd^2=0$, a
coboundary is always a cocycle: $B^n(\cM^D;\M) \subset
Z^n(\cM^D;\M)$.  We may view two  cocycles differ by a coboundary as
equivalent.  The equivalence classes of cocycles, $[f_n]$, form the so called
cohomology group denoted by \begin{align} H^n(\cM^D;\M)=
Z^n(\cM^D;\M)/ B^n(\cM^D;\M), \end{align}
$H^n(\cM^D;\M)$, as a group quotient of $Z^n(\cM^D;\M)$ by
$B^n(\cM^D;\M)$, is also an Abelian group.

When we study systems with time reversal symmetry, we need to consider cochains
with \emph{local value}.  To define  cochains with local value, we first note
that a manifold $\cM^D$ has a Stiefel-Whitney class $\w_1$ to describe
its orientation structure.  On each link $(ij)$ of $\cM^D$, $\w_1$ has
two values $(\w_1)_{ij}=0,1$.  For a cochain $f$ with local values, we cannot
directive compare their values on different simplexes, say $f_{ijk\cdots}$ and
$f_{lmn\cdots}$.  To compare $f_{ijk\cdots}$ and $f_{lmn\cdots}$, we need to
use $\w_1$ to parallel transport the value $f_{lmn\cdots}$ with base point $l$
to the value with base point $i$: $f_{lmn\cdots} \to (\w_1)_{il} f_{lmn\cdots}$
then we can compare $f_{ijk\cdots}$ and $(\w_1)_{il} f_{lmn\cdots}$.

Such a interpretation of the value of a cochain will modify
our definition of derivative operator
\begin{align} 
\label{eq:differentialw1}
&\ \ \ \ \<\dd_{\w_1} f_n, (i_0i_1i_2\cdots i_{n+1})\>
\nonumber\\
&=
(-)^{(\w_1)_{01}} \<f_n, (i_1i_2\cdots i_{n+1})\>
\nonumber\\
&\ \ \ \
 +\sum_{m=1}^{n+1} (-)^m 
\<f_n, (i_0i_1i_2\cdots\hat i_m\cdots i_{n+1})\>
\end{align}
we note that on the right-hand-side of the above equation, all the terms have
the base point $0$, except the first term which has a base point $1$.

In this paper, on orientable manifold $M^D$, we will choose $\w_1\se{2}0$, and
the cochains will all be the cochains with global values and the derivative
operator is defined by \eqn{eq:differential}.  On the other hand, on
unorientable manifold $M^D$, the cochains will all be the cochains with local
values and the derivative operator is defined by \eqn{eq:differentialw1}.  One
can show that $\dd$ and $\dd_{\w_1}$ have the same local properties.  We will
drop the subscript $\w_1$ in $\dd_{\w_1}$ and write $\dd_{\w_1}$ as $\dd$.

For the $\Z_N$-valued cocycle $x_n$, $\dd x_n \se{N} 0$. Thus 
\begin{align}
 \Bs_N x_n \equiv \frac1N \dd x_n 
\end{align}
is a $\Z$-valued cocycle. Here $\Bs_N$ is Bockstrin homomorphism.

We notice the above definition for cochains still makes sense if we
have a non-Abelian group $(G, \cdot)$ instead of an Abelian group $(\M,
+)$, however the differential $\dd$ defined by \eqn{eq:differential}
will not satisfy $\dd \circ \dd = 1$, except for the first two
$\dd$'s. That is, one may still make sense of 0-cocycle and 1-cocycle,
but no more further naively by formula
\eqn{eq:differential}. For us, we only use non-Abelian 1-cocycle in
this article. Thus it is ok.  Non-Abelian cohomology is then thoroughly
studied in mathematics motivating concepts such as gerbes to enter.

From two cochains $f_m$  and $h_n$, we can construct a third cochain
$p_{m+n}$ via the cup product (see Fig. \ref{cupcom}):
\begin{align}
p_{m+n} &= f_m \smile h_n ,
\nonumber\\
\<p_{m+n}, (0 \to {m+n})\> 
&= 
\<f_m, (0 \to m)\> \times
\nonumber\\
&\ \ \ \ 
\<h_n,(m \to {m+n}) \>,
\end{align}
where $i\to j$ is the consecutive sequence from $i$ to $j$: 
\begin{align}
i\to j\equiv i,i+1,\cdots,j-1,j. 
\end{align}
Note that the above definition applies to cochains with global.
If $h_n$ has a local value, we then have
\begin{align}
p_{m+n} &= f_m \smile h_n ,
\nonumber\\
\<p_{m+n}, (0 \to {m+n})\> 
&= 
(-)^{(\w_1)_{0m}}\<f_m, (0 \to m)\> \times
\nonumber\\
&\ \ \ \ 
\<h_n,(m \to {m+n}) \>,
\end{align}

The cup product has the following property 
\begin{align}
\label{cupprop}
 \dd(f_m \smile h_n) &= 
(\dd h_n) \smile f_m 
+ (-)^n h_n \smile (\dd f_m) 
\end{align}
for  cochains with global or local values.  We note that the above is a local
relation. Locally, we can always choose a gauge to make $\w_1\se{2}0$.  Thus,
the local relations are valid for cochains with both global and local values.

We see that $f_m \smile h_n$ is a
cocycle if both $f_m$ and $h_n$ are cocycles.  If both $f_m$ and $h_n$ are
cocycles, then $f_m \smile h_n$ is a coboundary if one of $f_m$ and $h_n$ is a
coboundary.  So the cup product is also an operation on cohomology groups
$\hcup{} : H^m(M^D;\M)\times H^n(M^D;\M) \to H^{m+n}(M^D;\M)$.  The cup product
of two \emph{cocycles} has the following property (see Fig. \ref{cupcom}) 
\begin{align}
 f_m \smile h_n &= (-)^{mn} h_n \smile f_m + \text{coboundary}
\end{align}

We can also define higher cup product $f_m \hcup{k}
h_n$ which gives rise to a $(m+n-k)$-cochain \cite{S4790}:
\begin{align}
&\ \ \ \
 \<f_m \hcup{k} h_n, (0,1,\cdots,m+n-k)\> 
\nonumber\\
&
 = 
\hskip -1em 
\sum_{0\leq i_0<\cdots< i_k \leq n+m-k} 
\hskip -3em  
(-)^p
\<f_m,(0 \to i_0, i_1\to i_2, \cdots)\>\times
\nonumber\\
&
\ \ \ \ \ \ \ \ \ \
\ \ \ \ \ \ \ \ \ \
\<h_n,(i_0\to i_1, i_2\to i_3, \cdots)\>,
\end{align} 
and $f_m \hcup{k} h_n =0$ for  $k<0$ or for $k>m \text{ or } n$.  Here $i\to j$
is the sequence $i,i+1,\cdots,j-1,j$, and $p$ is the number of permutations to
bring the sequence
\begin{align}
 0 \to i_0, i_1\to i_2, \cdots; i_0+1\to i_1-1, i_2+1\to i_3-1,\cdots
\end{align}
to the sequence
\begin{align}
 0 \to m+n-k.
\end{align}
For example
\begin{align}
&
 \<f_m \hcup1 h_n, (0\to m+n-1)\> 
 = \sum_{i=0}^{m-1} (-)^{(m-i)(n+1)}\times
\nonumber\\
&
\<f_m,(0 \to i, i+n\to m+n-1)\>
\<h_n,(i\to i+n)\>.
\end{align} 
We can see that $\hcup0 =\smile$.  
Unlike cup product at $k=0$, the higher cup product of two
cocycles may not be a cocycle. For cochains $f_m, h_n$, we have
\begin{align}
\label{cupkrel}
& \dd( f_m \hcup{k} h_n)=
\dd f_m \hcup{k} h_n +
(-)^{m} f_m \hcup{k} \dd h_n+
\\
& \ \ \
(-)^{m+n-k} f_m \hcup{k-1} h_n +
(-)^{mn+m+n} h_n \hcup{k-1} f_m 
\nonumber 
\end{align}
If $h_n$ has a local value, we then have
\begin{align}
&\ \ \ \
 \<f_m \hcup{k} h_n, (0,1,\cdots,m+n-k)\> 
\nonumber\\
&
 = 
\hskip -1em 
\sum_{0\leq i_0<\cdots< i_k \leq n+m-k} 
\hskip -3em  
(-)^p
(-)^{(\w_1)_{0i_0}}
\<f_m,(0 \to i_0, i_1\to i_2, \cdots)\>\times
\nonumber\\
&
\ \ \ \ \ \ \ \ \ \
\ \ \ \ \ \ \ \ \ \
\<h_n,(i_0\to i_1, i_2\to i_3, \cdots)\>,
\end{align}

Let $f_m$ and $h_n$ be cocycles and $c_l$ be a chain, from \eqn{cupkrel} we
can obtain
\begin{align}
\label{cupkrel1}
 & \dd (f_m \hcup{k} h_n) = (-)^{m+n-k} f_m \hcup{k-1} h_n 
\nonumber\\
&
\ \ \ \ \ \ \ \ \ \
 \ \ \ \ \ \ \
+ (-)^{mn+m+n}  h_n \hcup{k-1} f_m,
\nonumber\\
 & \dd (f_m \hcup{k} f_m) = [(-)^k+(-)^m] f_m \hcup{k-1} f_m,
\nonumber\\
& \dd (c_l\hcup{k-1} c_l + c_l\hcup{k} \dd c_l)
= \dd c_l\hcup{k} \dd c_l 
\nonumber\\
&\ \ \ -[(-)^k-(-)^l]
(c_l\hcup{k-2} c_l + c_l\hcup{k-1} \dd c_l) .
\end{align}

From \eqn{cupkrel1}, we see that, for $\Z_2$-valued cocycles $z_n$,
\begin{align}
 \Sq^{n-k}(z_n) \equiv z_n\hcup{k} z_n
\end{align}
is always a cocycle.  Here $\Sq$ is called the Steenrod square.  More generally
$h_n \hcup{k} h_n$ is a cocycle if $n+k =$ odd and $h_n$ is a cocycle.
Usually, the Steenrod square is defined only for $\Z_2$-valued cocycles or
cohomology classes.  Here, we like to define a generalized
Steenrod square for $\M$-valued
cochains $c_n$:
\begin{align}
\label{Sqdef}
 \gSq^{n-k} c_n \equiv c_n\hcup{k} c_n +  c_n\hcup{k+1} \dd c_n .
\end{align}
From \eqn{cupkrel1}, we see that
\begin{align}
\label{Sqd1}
 \dd \gSq^{k} c_n &= \dd(
c_n\hcup{n-k} c_n +  c_n\hcup{n-k+1} \dd c_n )
\\
&= \gSq^k \dd c_n +(-)^{n}
\begin{cases}
0, & k=\text{odd} \\ 
2  \gSq^{k+1} c_n  & k=\text{even} \\ 
\end{cases}
.
\nonumber 
\end{align}
In particular, when $c_n$ is a $\Z_2$-valued cochain, we have
\begin{align}
\label{Sqd}
  \dd \gSq^{k} c_n \se{2} \gSq^k \dd c_n.
\end{align}

Next, let us consider the action of $\gSq^k$ on the sum of two
 $\M$-valued cochains $c_n$ and $c_n'$:
\begin{align}
\label{Sqplus1}
& \gSq^{k} (c_n+c_n')
 = \gSq^{k} c_n + \gSq^k c_n' +
\nonumber\\
&\ \ \
 c_n \hcup{n-k} c_n' + c_n' \hcup{n-k} c_n 
+ c_n \hcup{n-k+1} \dd c_n' + c_n' \hcup{n-k+1} \dd c_n 
\nonumber\\
&=\gSq^{k} c_n + \gSq^k c_n' 
+[1 + (-)^k]c_n \hcup{n-k} c_n'
\nonumber\\
&\ \ \
-(-)^{n-k} [ - (-)^{n-k} c_n' \hcup{n-k} c_n + (-)^n c_n \hcup{n-k} c_n']
\nonumber\\
&\ \ \
+ c_n \hcup{n-k+1} \dd c_n' + c_n' \hcup{n-k+1} \dd c_n
\nonumber\\
& = 
\gSq^{k} c_n + \gSq^k c_n' 
+[1 + (-)^k]c_n \hcup{n-k} c_n'
\nonumber\\
&
+(-)^{n-k} [ \dd c_n' \hcup{n-k+1} c_n +(-)^n c_n' \hcup{n-k+1} \dd c_n
]
\nonumber\\
&
-(-)^{n-k} 
\dd (c_n'\hcup{n-k+1}c_n) 
+c_n \hcup{n-k+1} \dd c_n'+ c_n' \hcup{n-k+1} \dd c_n
\nonumber\\
&=
\gSq^{k} c_n + \gSq^k c_n'  
+[1 + (-)^k]c_n \hcup{n-k} c_n'
\nonumber \\
&\  \ \
+[1+(-)^{k}]c_n' \hcup{n-k+1} \dd c_n 
-(-)^{n-k} \dd (c_n'\hcup{n-k+1}c_n)
\nonumber\\
&\ \ \
-[(-)^{n-k+1}\dd c_n' \hcup{n-k+1} c_n
- c_n \hcup{n-k+1} \dd c_n']
\nonumber\\
&=
\gSq^{k} c_n + \gSq^k c_n'  
+[1 + (-)^k]c_n \hcup{n-k} c_n'
\nonumber \\
&\  \ \
+[1+(-)^{k}]c_n' \hcup{n-k+1} \dd c_n 
-(-)^{n-k} \dd (c_n'\hcup{n-k+1}c_n)
\nonumber\\
&\ \ \
-\dd (\dd c_n'\hcup{n-k+2} c_n )
+ \dd c_n'\hcup{n-k+2} \dd c_n 
\nonumber\\
&=
\gSq^{k} c_n + \gSq^k c_n'  
+ \dd c_n'\hcup{n-k+2} \dd c_n 
\nonumber \\
&\ \ \
+[1+(-)^{k}][c_n \hcup{n-k} c_n'+ c_n' \hcup{n-k+1} \dd c_n] 
\nonumber\\
&\ \ \
-(-)^{n-k} \dd (c_n'\hcup{n-k+1}c_n)
-\dd (\dd c_n'\hcup{n-k+2} c_n )
.
\end{align}
We see that, if one of the $c_n$ and $c_n'$ is a cocycle,
\begin{align}
\label{Sqplus}
  \gSq^{k} (c_n+c_n') \se{2,\dd} \gSq^{k} c_n + \gSq^k c_n' .
\end{align}
We also see that
\begin{align}
\label{Sqgauge}
&\ \ \ \
 \gSq^{k} (c_n+\dd f_{n-1})
\\
& = \gSq^{k} c_n + \gSq^k \dd f_{n-1} +
[1+(-)^k] \dd f_{n-1}\hcup{n-k} c_n
\nonumber\\
&\ \ \
-(-)^{n-k} \dd (c_n\hcup{n-k+1}\dd f_{n-1})
-\dd (\dd c_n\hcup{n-k+2} \dd f_{n-1} )
\nonumber\\
& = \gSq^{k} c_n 
+ [1+(-)^k] [\dd f_{n-1}\hcup{n-k} c_n +(-)^n \gSq^{k+1}f_{n-1}]
\nonumber\\
&
+\dd [\gSq^k  f_{n-1}
-(-)^{n-k} c_n \hskip -0.5em \hcup{n-k+1} \hskip -0.5em \dd f_{n-1}
-\dd c_n \hskip -0.5em \hcup{n-k+2}  \hskip -0.5em \dd f_{n-1} ]
\nonumber\\
& = \gSq^{k} c_n 
+ [1+(-)^k] [c_n\hcup{n-k}  \dd f_{n-1} +(-)^n \gSq^{k+1}f_{n-1}]
\nonumber\\
&
+\dd [\gSq^k  f_{n-1} -(-)^{n-k} \dd f_{n-1} \hcup{n-k+1} c_n ]
.
\nonumber 
\end{align}
Using \eqn{Sqplus2}, we can also obtain the following result
if $\dd c_n = $ even
\begin{align}
\label{Sqplus2}
& \ \ \ \
 \gSq^k (c_n+2c_n')
\nonumber\\
& \se{4} \gSq^k c_n+2 \dd (c_n\hcup{n-k+1} c_n') +2 \dd c_n\hcup{n-k+1} c_n'
\nonumber\\
& \se{4} \gSq^k c_n+2 \dd (c_n\hcup{n-k+1} c_n') 
\end{align}

As another application, we note that, for a $\M$-valued cochain $m_d$ and using
\eqn{cupkrel},
\begin{align}
\label{Sq1Bs}
& \gSq^1(m_{d}) = m_{d}\hcup{d-1} m_{d} + m_{d}\hcup{d} \dd m_{d}
\nonumber\\
&=\frac12 (-)^{d} 
[\dd (m_{d}\hcup{d} m_{d}) 
-\dd m_{d} \hcup{d} m_{d}] 
+\frac12  m_{d} \hcup{d} \dd m_{d} 
\nonumber\\
&=
(-)^{d} \Bs_2 (m_{d}\hcup{d} m_{d}) -(-)^d \Bs_2 m_{d} \hcup{d} m_{d}
+  m_{d} \hcup{d} \Bs_2 m_{d}
\nonumber\\
&=
(-)^{d} \Bs_2  \gSq^0 m_{d} 
-2 (-)^d \Bs_2 m_{d} \hcup{d+1} \Bs_2 m_{d}
\nonumber\\
&=
(-)^{d} \Bs_2 \gSq^0 m_{d} 
-2 (-)^d \gSq^0 \Bs_2 m_{d} 
\end{align}
This way, we obtain a relation between Steenrod square and Bockstein
homomorphism, when $m_d$ is a $\Z_2$-valued cochain
\begin{align}
\label{Sq1Bs2}
  \gSq^1(m_{d}) \se{2} \Bs_2 m_{d} ,
\end{align}
where we have used $\gSq^0 m_{d}= m_d$ for $\Z_2$-valued cochain.

\section{Almost cocycle and almost coboundary}
\label{almco}

A $\RZ$-valued cocycle $c$ satisfies
\begin{align}
 \dd c \se{1} 0.
\end{align}
A $\RZ$-valued almost-cocycle $\t c$ satisfies
\begin{align}
 \dd \t c \ase{1} 0.
\end{align}
Physically, it is impossible to constrain an $\RZ$-value to exactly zero.  So
almost-cocycle is more relevant for our model construction.  In this paper,
when we refer $\RZ$-valued cocycle, we really mean $\RZ$-valued almost-cocycle.

Similarly, a $\RZ$-valued almost-coboundary is given by
\begin{align}
 \dd c + \eps,
\end{align}
where $\eps$ is an almost-zero cochain.

The cohomology class $H^*_\text{a}(M;\RZ)$ for $\RZ$-valued almost-cocycles is
different from the cohomology class $H^*(M;\RZ)$ for $\RZ$-valued cocycles.
For example, let us consider $H^1_\text{a}(S^2;\RZ)$. We parametrize $S^2$ by
the polar angle $\th$, and the azimuthal angle $\phi$.  The 1-cochain
\begin{align}
\t a^{\RZ}=\frac{1-\cos (\th)}{4\pi} \dd \phi
\end{align}
is not a cocycle since $\dd \t a^{\RZ} \neq 0$ mod 1.  But it is an
1-almost-cocycle since
\begin{align}
 \dd \t a^{\RZ}  \approx c_1,
\end{align}
and $c_1$ is a $\Z$-valued.  $\t a^{\RZ}$ is not an 1-almost-coboundary, since
$c_1$ generates $H^2(S^2;\Z)$.  In fact, $\t a^{\RZ}$ is the generator of
$H^1_\text{a}(S^2;\RZ)$ and we have
\begin{align}
 H^1_\text{a}(S^2;\RZ)\cong H^2(S^2;\Z)=\Z.
\end{align}
In contrast $H^1(S^2;\RZ)=0$.

In general, for an almost-cocycle $\t c$,
\begin{align}
 \toZ{\dd \t c} = C 
\end{align}
is a $\Z$-valued cocycle.  The non-trivialness of $\t c$ is given by the
non-trivialness of $C$. Thus we have
\begin{align}
\label{HalmH}
 H^n_\text{a}(M;\RZ)\cong H^{n+1}(M;\Z) .
\end{align}
In this paper, we will drop the subscript and write $H^n_\text{a}(M;\RZ)$ as
$H^n(M;\RZ)$.

\section{Some additional discussion of fermion decoration}

\subsection{Another form of exactly soluble local fermionic models}

We can also write the path integral \eqn{ZfSPT} as one on $M^{d+1}$.  To do
so, we introduce a new $\Z_2$-valued $(d-1)$-cochain field $b_{d-1}$ that
satisfy
\begin{align}
\label{dbf}
 \dd b_{d-1} \se{2} f_d
\end{align}
to write the integrant of $\int_{N^{d+2}}$ as a total derivative (using
\eqn{Sqd})
\begin{align}
&\ \ \ \
 \dd [\gSq^2 b_{d-1} + b_{d-1} (\w_2+\w_1^2)]
\nonumber\\
&
\se{2}  \Sq^2 n_d + n_d (\w_2+\w_1^2) .
\end{align}
This way, we can change the path integral \eqn{ZfSPT} to one on $M^{d+1}$
only:
\begin{align}
\label{ZfSPT1}
Z(M^{d+1},A^{G_b}) 
&= 
\hskip -4.7em
\sum_{ g \in C^0(M^{d+1};G_b); f_d\se{2} n_d(a^{G_b})} 
\hskip -4.7em
\ee^{\ii 2\pi \int_{M^{d+1}} \nu_{d+1}(a^{G_b})
 +\frac12 \gSq^2 b_{d-1} + \frac12 b_{d-1} e_2  
}
,
\end{align}
where $b_{d-1}$ is a function of $f_d$ as determined from \eqn{dbf}.
The summation  $\sum_{ g \in C^0(M^{d+1};G_b); f_d\se{2} n_d(a^{G_b})}$ (\ie
the path integral) is over a $G_b$-valued 0-cochain field $g$ and $\Z_2$-valued
$d$-cochain field $f_d$.  But $f_d$ subject to a constrain $f_d\se{2}
n_d(a^{G_b})=n_d(g_iA^{G_b}_{ij}g_j^{-1})$, which can be imposed as an energy
penalty.  The term $\frac12  \gSq^2 b_{d-1}$ makes the current $f_d$ a fermion
current (\ie makes the field $f_d$ to describe a fermion).  

In order for \eqn{ZfSPT1} to be well defined, the action amplitude $\ee^{\ii
2\pi \int_{M^{d+1}} \nu_{d+1}(a^{G_b}) +\frac12 [ \gSq^2 b_{d-1} + b_{d-1}
e_2(a^{G_b}) ] }$ should be a function of $f_d$ and does not depend on which
solution $b_{d-1}$ of $\dd b_{d-1} \se{2} f_d$ that we choose.  Different
solutions can differ by a cocycle $\bar b_{d-1}$. Using \eqn{Sqplus1}, we
find that
\begin{align}
\label{barbw2}
&
 \gSq^2 (b_{d-1}+\bar b_{d-1}) + (b_{d-1}+\bar b_{d-1}) e_2 -\gSq^2 b_{d-1} - b_{d-1} e_2
\nonumber\\
&\se{2,\dd}
\gSq^2(\bar b_{d-1}) + \bar b_{d-1} e_2(a^{G_b})
\nonumber\\
&\se{2,\dd}
 \bar b_{d-1}[\w_2+\w_1^2+ e_2(A^{G_b})].
\end{align}
Thus the path integral is well defined only on $M^{d+1}$ with a symmetry twist
$A$ such that $\w_2+\w_1^2+ e_2(A^{G_b}) \se{2,\dd} 0$.  This implies that $f_d$
describes a fermion.  This also implies that the fermion is described by a
representation of $G_f = Z_2\gext_{e_2} G_b$ (see \Ref{ZLW}). 

\subsection{Exactly soluble local bosonic models with emergent fermions}
\label{exc}

If we treat the field $b_{d-1}$ in \eqn{ZfSPT1}
as an independent dynamical field (instead of as a function of $f_d$),
then we will get a very different theory:
\begin{align}
\label{ZEF}
&Z(M^{d+1},A^{G_b}) 
= 
\hskip -5.5em
\sum_{ g \in C^0(M^{d+1};G_b); \dd b_{d-1}\se{2} n_d(a^{G_b})} 
\hskip -5.5em
\ee^{\ii \pi \int_{M^{d+1}} 2 \nu_{d+1}(a^{G_b})
+ \gSq^2 b_{d-1} + b_{d-1} e_2(a^{G_b})  }
,
\end{align}
The new path integral \eqn{ZEF} sums over the 0-cochains $g_i$ and
$(d-1)$-cochains $b_{d-1}$ satisfying $\dd b_{d-1} =n_d(a^{G_b})$.  Such a model
is actually a local bosonic model.  The local bosonic model has emergent
fermions whose current is given by $f_d = \dd b_{d-1} - n_d(a^{G_b})$.

The above  local bosonic model has a $G_b$ symmetry. Thus the model describes
symmetry $G_b$ enriched topological order with emergent fermions.  If we break
the $G_b$ symmetry, the model describes a $Z_2$-topological order with emergent
fermion. Such a $Z_2$-topological is described by a $Z_2$ gauge theory where
the $Z_2$ charge is fermionic. In the presence of $G_b$ symmetry, the emergent
fermions carry fractionalized symmetry quantum number.  Since the partition
function of the local bosonic model vanishes when $\w_2+\w_1^2 + e_2( A^{G_b}) \neq 0
$, we conclude that the emergent fermions carry representations of $G_f = Z_2
\gext_{e_2} G_b$.  The partition function of the local bosonic model vanishes
when $\w_2+\w_1^2 + e_2( A^{G_b}) \neq 0 $.
We may view \eqn{ZfSPT} as a fermionic model with $G_f$ symmetry, and the
bosonic model \eq{ZEF} as the $Z_2^f$ gauged fermionic model.  If we gauge
all the symmetry $G_f$ in the fermionic model \eq{ZfSPT}, we will obtain a
higher gauge theory as described in \Ref{ZLW}:
\begin{align}
\label{Zhg}
&Z(M^{d+1}) 
= 
\hskip -9em
\sum_{ 
\ \ \ \ \ \ \ \ \ \ \ \ \ \ \ \ \ \ \ \ 
a^{G_b} \in C^1(M^{d+1};G_b); \dd b_{d-1} =n_d(a^{G_b})} 
\hskip -9em
\ee^{\ii \pi \int_{M^{d+1}} 2\nu_{d+1}(a^{G_b})
+ \gSq^2 b_{d-1} + b_{d-1} e_2(a^{G_b})  }
.
\end{align}
Despite their similarity, the two local bosonic models \eq{ZEF} and \eq{Zhg}
are very different.  In  \eqn{ZEF}, the dynamical fields are $g, b_{d-1}$ (with
$a^{G_b}_{ij}=g_iA^{G_b}_{ij}g_j^{-1}$), where $g$ is a $G_b$-valued 0-cochain
living on vertices.  In contrast, in  \eqn{ZEF}, the dynamical fields are
$a^{G_b}, b_{d-1}$ where $a^{G_b}$ is a $G_b$-valued 1-cochain living on links.

\subsection{Another connection to higher gauge theory}

There is another connection to higher gauge theory.  After gauging all the
$G_f$ symmetry in the fermionic model \eqn{ZfSPT}, we obtain a local bosonic
model \eqn{Zhg}.  Such a local bosonic model is a higher gauge theory with
emergent fermion.  Such a higher gauge theory is characterized by a higher
group and its cocycle. Thus the data that characterizes a fermionic SPT phase
is closely related to a higher group and its cocycle.

In fact the local bosonic theory \eqn{Zhg} is characterized by a higher group
$\cG_{n_d}(G_b,1;Z_2^f,d-1)$ and its higher-group cocycle $\om_{d+1}$.  The
field content $a^{G_b}, b_{d-1}$ and their conditions
\begin{align}
(\del a^{G_b})_{ijk} &\equiv
a^{G_b}_{ij} a^{G_b}_{jk} a^{G_b}_{ki} = 1 
\nonumber\\
\dd b_{d-1}   & \se{2} n_d(a^{G_b}) ,
\end{align}
determine the  higher group $\cG_{n_d}(G_b,1;Z_2^f,d-1)$, which can be viewed
as  a space with homotopy groups $\pi_1 =G_b$, $\pi_{d-1} =Z_2^f$, and other
$\pi_n=0$.  

The local bosonic model \eqn{Zhg} and \eqn{ZEF} are exactly soluble if the
Lagrangian is given by a higher-group cocycle satisfying $\dd  \om_{d+1} \se{1}
0$ (\ie $ \om_{d+1} \in Z^{d+1}[\cG_{n_d}(G_b,1;Z_2^f,d-1);\R/\Z]$):
\begin{align}
&\ \ \ \
 \om_{d+1} (a^{G_b}, b_{d-1})
\nonumber\\
&=\nu_{d+1}(a^{G_b}) + \frac12 [\gSq^2 b_{d-1} + n_d(a^{G_b}) e_2(a^{G_b})] ,
\end{align}
(See \Ref{ZLW} and Appendix \ref{hgroup} for an introduction on higher groups
and higher-group cocycles.)

\section{Operations on modules}
\label{tentor}

The tensor-product operation $\otimes_R$ and the torsion-product operation
$\Tor_1^R$ act on $R$-modules $\M,\M',\M''$.  Here $R$ is a ring and a
$R$-module is like a vector space over $R$ (\ie we can ``multiply'' a
``vector'' in $\M$ by an element of $R$, and two ``vectors'' in $\M$ can add.)
The tensor-product operation $\otimes_R$ has the following properties:
\begin{align}
\label{tnprd}
& \M \otimes_\Z \M' \simeq \M' \otimes_\Z \M ,
\nonumber\\
&  (\M'\oplus \M'')\otimes_R \M = (\M' \otimes_R \M)\oplus (\M'' \otimes_R \M)   ,
\nonumber\\
& \M \otimes_R (\M'\oplus \M'') = (\M \otimes_R \M')\oplus (\M \otimes_R \M'')   ;
\nonumber\\
& \Z \otimes_\Z \M \simeq \M \otimes_\Z \Z =\M ,
\nonumber\\
& \Z_n \otimes_\Z \M \simeq \M \otimes_\Z \Z_n = \M/n\M ,
\nonumber\\
& \Z_n \otimes_\Z \RZ \simeq \RZ \otimes_\Z \Z_n = 0,
\nonumber\\
& \Z_m \otimes_\Z \Z_n  =\Z_{\<m,n\>} ,
\nonumber\\
%& \RZ \otimes_\Z \RZ = 0 (???),
%\nonumber\\
& \R \otimes_\Z \RZ = 0,
\nonumber\\
& \R \otimes_\Z \R = \R.
\end{align}
The torsion-product operation $\Tor_1^R$ has the following properties:
\begin{align}
\label{trprd}
& \Tor^1_R(\M , \M') \simeq \Tor^1_R (\M', \M)  ,
\nonumber\\
& \Tor^1_R(\M'\oplus \M'' , \M) = \Tor^1_R(\M', \M) \oplus \Tor^1_R(\M'',\M),
\nonumber\\
& \Tor^1_R(\M,\M'\oplus \M'') = \Tor^1_R(\M,\M')\oplus \Tor^1_R(\M,  \M'')
\nonumber\\
& \Tor^1_\Z(\Z,  \M) = \Tor^1_\Z(\M,  \Z) = 0,
\nonumber\\
& \Tor^1_\Z(\Z_n, \M) = \{m\in \M| nm=0\},
\nonumber\\
& \Tor^1_\Z(\Z_n, \RZ) = \Z_n,
\nonumber\\
& \Tor^1_\Z(\Z_m, \Z_n) = \Z_{\<m,n\>} ,
\nonumber\\
& \text{Tor}^1_\Z(\R/\Z, \R/\Z) = 0 .
\end{align}
These expressions allow us to compute the tensor-product
$\otimes_R$ and  the torsion-product $\Tor^1_R$.  We will use abbreviated
$\Tor$ to denote $\Tor^1_\Z$.

In addition to $\otimes_{\Z}$ and $\Tor$, we also have
Ext and Hom operations on modules.
 Ext operation is given by
\begin{align}
\label{extprd}
& \Ext^1_R(\M'\oplus \M'' , \M) = \Ext^1_R(\M', \M) \oplus \Ext^1_R(\M'',\M),
\nonumber\\
& \Ext^1_R(\M,\M'\oplus \M'') = \Ext^1_R(\M,\M')\oplus \Ext^1_R(\M,  \M'')
\nonumber\\
& \Ext^1_\Z(\Z,  \M) = 0,
\nonumber\\
& \Ext^1_\Z(\Z_n, \M) = \M/n\M,
\nonumber\\
& \Ext^1_\Z(\Z_n, \Z) = \Z_n,
\nonumber\\
& \Ext^1_\Z(\Z_n, \RZ) = 0,
\nonumber\\
& \Ext^1_\Z(\Z_m, \Z_n) = \Z_{\<m,n\>} ,
.
\end{align}
The Hom operation on modules is given by
\begin{align}
\label{homprd}
& \Hom_R(\M'\oplus \M'' , \M) = \Hom_R(\M', \M) \oplus \Hom_R(\M'',\M),
\nonumber\\
& \Hom_R(\M,\M'\oplus \M'') = \Hom_R(\M,\M')\oplus \Hom_R(\M,  \M'') ,
\nonumber\\
& \Hom_\Z(\Z,  \M) = \M,
\nonumber\\
& \Hom_\Z(\Z_n, \M) = \{m\in \M| nm=0\},
\nonumber\\
& \Hom_\Z(\Z_n, \Z) = 0,
\nonumber\\
& \Hom_\Z(\Z_n, \RZ) = \Z_n,
\nonumber\\
& \Hom_\Z(\Z_m, \Z_n) = \Z_{\<m,n\>} 
.
\end{align}
We will use abbreviated $\Ext$ and $\Hom$ to
denote $\Ext^1_\Z$ and $\Hom_\Z$.

\section{K\"unneth formula and
universal coefficient theorem
} \label{KunnUCT}

The K\"unneth formula is a very helpful formula that allows us to calculate the
cohomology of chain complex $X\times X'$ in terms of  the cohomology of chain
complex $X$ and chain complex $X'$.
The K\"unneth formula is expressed in terms of
the tensor-product operation $\otimes_R$ and the torsion-product operation
$\text{Tor}_1^R$ described in the last section
(see \Ref{Spa66} page 247):
\begin{align}
\label{kunn}
0&\to
\bigoplus_{k=0}^d H^k(X,\M)\otimes_R H^{d-k}(X',\M')
\nonumber\\
&\to
H^d(X\times X',\M\otimes_R \M')
\nonumber\\
& \to
\bigoplus_{k=0}^{d+1}
\Tor_R^1 (H^k(X,\M), H^{d-k+1}(X',\M')) \to 0,
\end{align}
where the exact sequence is split.  Here $R$ is a principle ideal domain and
$\M,\M'$ are $R$-modules such that $\Tor_R^1(\M, \M')=0$.  We also require
either\\
(1) $H_d(X,\Z)$ and  $H_d(X',\Z)$ are finitely generated, or\\
(2) $\M'$ and $H_d(X',\Z)$ are finitely generated.\\
(For example, $\M'=\Z\oplus \cdots \oplus \Z\oplus \Z_n\oplus\cdots\oplus\Z_m$
is finitely generated, with $R=\Z$.)  

For more details on principal ideal domain and $R$-module, see the
corresponding Wiki articles.  Note that ring $\Z$ and $\R$ are principal ideal
domains.  Also, $\R$ and $\R/\Z$ are not finitely generate $R$-modules if
$R=\Z$.  

The K\"unneth formula works for topological cohomology where $X$ and $X'$ are
treated as topological spaces.  

As the first application of K\"unneth formula, we like to use it to calculate
$H^*(X',\M)$ from $H^*(X',\Z)$,  by choosing $R=\M'=\Z$. In this case, the
condition $\Tor_R^1(\M,\M')=\Tor_{\Z}^1 (\M, \Z)=0$ is always
satisfied. $\M$ can be $\R/\Z$, $\Z$, $\Z_n$ \etc. So we have
\begin{align}
\label{kunnZ}
0 &\to
\bigoplus_{k=0}^d H^k(X,\M)\otimes_{\Z} H^{d-k}(X',\Z)
\nonumber\\
& \to
H^d(X\times X',\M)
\nonumber\\
& \to
\bigoplus_{k=0}^{d} \Tor (H^k(X,{\M}), H^{d-k+1}(X',\Z)) \to 0
\end{align}
Again, the exact sequence is split.

We can further choose $X$ to be the space of one point in \eqn{kunnZ},
and use
\begin{align}
H^{d}(X,\M))=
\begin{cases}
\M, & \text{ if } d=0,\\
0, & \text{ if } d>0,
\end{cases}
\end{align}
to reduce \eqn{kunnZ} to
\begin{align}
\label{ucf}
 H^d(X,\M)
&\simeq  \M \otimes_{\Z} H^d(X,\Z)
\oplus
\Tor( \M, H^{d+1}(X,\Z) )  .
\end{align}
where $X'$ is renamed as $X$.  The above is a form of the universal coefficient
theorem which can be used to calculate $H^*(X,\M)$ from $H^*(X,\Z)$ and the
module $\M$.  

Using the universal
coefficient theorem, we can rewrite \eqn{kunnZ} as
\begin{align}
\label{kunnHX}
H^d(X\times X',\M) \simeq \bigoplus_{k=0}^d H^k[X, H^{d-k}(X',\M)] .
\end{align}
The above is valid for topological cohomology.

%The $\Z$-modules $\R,\RZ,\Z$ can be viewed as abelian groups.
%They satisfy
%\begin{align}
% 0\to \Z \to \R \to \RZ \to 0.
%\end{align}
%It induces the following exact sequence
%\begin{align}
%& \to H^n(X;\Z) 
%\to H^n(X;\R) 
%\to H^n(X;\RZ) 
%\\
%%& \to H^{n+1}(X;\Z) \to H^{n+1}(X;\R) 
%\to H^{n+1}(X;\RZ) \to 
%\nonumber 
%\end{align}
%It has the following meaning: Let $a_{n+1}\in H^{n+1}(X;\Z)$ be a $\Z$-valued
%$n$-cocycle. Assume when viewed as a $\R$-valued $n+1$-cocycle, it is a
%coboundary of a $\R$-valued $n$-cochain: $\dd \om_n = a_{n+1}$.  Since $\dd
%\om_n \se{1} 0$, thus $\om_n$ is a $\RZ$-valued $n$-cocycle.
%If $H^n(X;\R)=H^{n+1}(X;\R)=0$, we have
%\begin{align}
% H^n(X;\RZ) \cong H^{n+1}(X;\Z) .
%\end{align}
%

\section{Lyndon-Hochschild-Serre spectral sequence}
\label{LHS}

The Lyndon-Hochschild-Serre spectral sequence (see \Ref{L4871} page 280,291,
and \Ref{HS5310}) allows us to understand the structure of of the cohomology of
a fiber bundle $F\to X\to B$, $H^*(X;\M)$, from $H^*(F;\M)$ and
$H^*(B;\M)$.  In general, $H^d(X;\M)$, when viewed as an Abelian group,
contains a chain of subgroups
\begin{align}
\label{Lyndon}
\{0\}=H_{d+1}
\subset H_d
\subset \cdots
\subset H_0
=
 H^d(X;\M)
\end{align}
such that $H_l/H_{l+1}$ is a subgroup of a factor
group of $H^l[B,H^{d-l}(F;\M)_{B}]$,
\ie $H^l[B,H^{d-l}(F;\M)_{B}]$
contains a   subgroup $\Ga^k$, such that
\begin{align}
 H_l/H_{l+1} &\subset H^l[B,H^{d-l}(F;\M)_{B}]/\Ga^l,
\nonumber\\
l&=0,\cdots,d.
\end{align}
Note that $\pi_1(B)$  may have a non-trivial action on $\M$ and $\pi_1(B)$ may
have a non-trivial action on $H^{d-l}(F;\M)$ as determined by the structure
$F \to X \to B$.  We add the subscript $B$ to $H^{d-l}(F;\M)$ to
indicate this action.  We also have
\begin{align}
 H_0/H_{1} &\subset H^0[B,H^{d}(F;\M)_{B}],
\nonumber\\
 H_d/H_{d+1}&=H_d = H^d(B;\M)/\Ga^d.
\end{align}
In other words, all the elements in $H^d(X;\M)$ can be one-to-one
labeled by $(x_0,x_1,\cdots,x_d)$ with
\begin{align}
 x_l\in H_l/H_{l+1} \subset H^l[B,H^{d-l}(F;\M)_{B}]/\Ga^l.
\end{align}

Let $x_{l,\al}$,
$\al=1,2,\cdots$, be the generators of $H^l/H^{l+1}$. Then we say $x_{i,\al}$
for all $l,\al$ are the generators of $H^d(X;\M)$.  We also call
$H_l/H_{l+1}$, $l=0,\cdots,d$, the generating sub-factor groups of
$H^d(X;\M)$.

The above result implies that we can use $(k_0,k_1,\cdots,k_d)$ with $ k_l\in
H^l[B,H^{d-l}(F;\M)_{B}] $ to  label all the elements in
$H^d(X;\M)$. However, such a labeling scheme may not be one-to-one, and it
may happen that only some of $(k_0,k_1,\cdots,k_d)$ correspond to  the
elements in $H^d(X;\M)$.  But, on the other hand, for every element in
$H^d(X;\M)$, we can find a $(k_0,k_1,\cdots,k_d)$ that corresponds to it.
Such a relation can be described by an injective map
\begin{align}
\label{HtoH}
  H^d(F\gext B;\M) \rightarrowtail \bigoplus_{l=0}^dH^l[B,H^{d-l}(F;\M)_{B}]
\end{align}

For the special case $X=B \times F$, $(k_0,k_1,\cdots,k_d)$ will give us
an one-to-one labeling of the elements in $H^d(B\times F;\M)$. In fact
\begin{align}
\label{Kunn}
H^d(B & \times F;\M) = \bigoplus_{l=0}^{d} H^{l}[B,H^{d-l}(F;\M)]
.
\end{align}

\section{The ring of $H^*(\cB SO_\infty;\Z)$}
\label{HBSO}

The ring $H^{*}(\cB SO_n;\Z_2)$ has a simple structure:
\begin{align}
 H^{*}(\cB SO_n;\Z_2)=\Z_2[\w_2,\w_3,\cdots,\w_n].
\end{align}
According to \Ref{B8283}, the ring $H^{*}(\cB SO_n;\Z)$ is given by
\begin{align}
 H^{*}(\cB SO_\infty;\Z)=\Z[\{p_i\}, \{\Bs_2 (\w_{2i_1}\w_{2i_2}\cdots)\},X_n]/ \sim ,
\end{align}
where $\Z[\{p_i\}, \{\Bs_2 (\w_{2i_1}\w_{2i_2}\cdots)\},X_n]$
is a polynomial ring
generated by $p_i$ and $\Bs_2(\w_{2i_1}\w_{2i_2}\cdots)$, 
$0<i\leq \toZ{\frac{n-1}2}$,
$0<i_1<i_2<\cdots\leq \toZ{\frac{n-1}2}$,
with integer coefficients. 
Here $p_i\in H^{4i}(\cB SO_\infty;\Z)$ is the
Pontryagin class of dimension $4i$ and $\w_i\in H^{i}(\cB SO_\infty;\Z_2)$ is the
Stiefel-Whitney class of dimension $i$.  Since $\text{Tor}H^d(\cB
G,\RZ)=\text{Tor}H^{d+1}(\cB G;\Z)$ (see, for example, \Ref{W1313}), the
natural map $H^d(\cB G;\Z_2)\to \text{Tor}H^d(\cB G,\RZ)$ induces the Bockstein
homomorphism $H^d(\cB G;\Z_2)\to H^{d+1}(\cB G;\Z)$: $\Bs_2:
H^{i}(\cB SO_\infty;\Z_2)\to H^{i+1}(\cB SO_\infty;\Z)$.  Note that $f \in
H^{i}(\cB SO_\infty;\Z_2)$ satisfies $\dd f \se{2}0$. Thus $\frac12 \dd f
\se{1}0$, or $\frac12 \dd f$ is an integral cocycle. This allows us to write
the Bockstein homomorphism as
\begin{align}
 \Bs_2 f = \frac12 \dd f \in H^{i+1}(\cB SO_\infty;\Z).
\end{align}

To obtain the ring $H^{*}(\cB SO_\infty;\Z)$ from a polynomial ring generated by
$p_i$ and $\Bs_2(\w_{2i_1}\w_{2i_2}\cdots)$, we need to quotient out certain
equivalence relations $\sim$.  The  equivalence relations $\sim$ contain
\begin{widetext}
\begin{align}
 2 \Bs_2 (\w_{2i_1}\w_{2i_2}\cdots) &=0 ,\ \ \ \ p_n\se{2} \w_{2n}^2, \ \ \ \
X_n = \Bs_2 \w_{2k} \text{ if } n=2k+1, \ \ \ \ 
X_n^2 = p_k \text{ if } n=2k,
\\
 \Bs_2  w(I) \Bs_2  w(J) &=
\sum_{k\in I} \Bs_2  \w_{2k}\ \Bs_2  \w[(I-\{k\})\cup J-(I-\{k\})\cap J]\ p[(I-\{k\})\cap J] ,
\nonumber 
\end{align}
\end{widetext}
where $I=\{i_1,i_2,\cdots\}$, $\w(I)=\w_{2i_1}\w_{2i_2}\cdots$.  and
$p(I)=p_{i_1}p_{i_2}\cdots$.  Many last kind of the equivalence relations are
trivial identities.  The first non-trivial equivalence relations appears at
dimension 14:
\begin{align}
&
 \Bs_2 (\w_4\w_2) \Bs_2  \w_6 = 
 \Bs_2  \w_4 \Bs_2  (\w_6 \w_2)
+\Bs_2  \w_2 \Bs_2  (\w_6 \w_4)
,
\nonumber\\
&
 \Bs_2 (\w_4\w_2) \Bs_2  (\w_4\w_2) = 
 p_1\Bs_2  \w_4 \Bs_2  \w_4 
+p_2\Bs_2  \w_2 \Bs_2  \w_2
.
\end{align}
We see that there are no effective equivalence relations of the last kind for
dimensions less than 14. 
So for low dimensions, 
\begin{align}
\label{HBSOZ}
 H^0(\cB SO_\infty;\Z) &=\Z,
\nonumber \\
 H^1(\cB SO_\infty;\Z) &=0,
\nonumber \\
 H^2(\cB SO_\infty;\Z) &=0,
\nonumber\\
 H^3(\cB SO_\infty;\Z) &=\Z_2=\{\Bs_2  \w_2\},
\nonumber\\
 H^4(\cB SO_\infty;\Z) &=\Z=\{p_1\},
\nonumber\\
 H^5(\cB SO_\infty;\Z) &=\Z_2=\{\Bs_2  \w_4\},
\\
 H^6(\cB SO_\infty;\Z) &=\Z_2=\{\Bs_2  \w_2\Bs_2  \w_2\},
\nonumber\\
 H^7(\cB SO_\infty;\Z) &=\Z_2^{\oplus 3}=\{ \Bs_2  \w_6 ,\Bs_2  (\w_2\w_4),  p_1\Bs_2  \w_2  \},
\nonumber\\
 H^8(\cB SO_\infty;\Z) &=\Z^{\oplus 2}\oplus \Z_2=\{p_1^2, p_2, \Bs_2  \w_2 \Bs_2  \w_4\}.
\nonumber 
\end{align}
In the above, we also list the basis (or generators) of cohomology classes.
Using \eqn{HalmH}, 
the above allows us to obtain
\begin{align}
\label{HBSOU1}
 H_\text{a}^0(\cB SO_\infty;\RZ) &=0,
\nonumber \\
 H_\text{a}^1(\cB SO_\infty;\RZ) &=0,
\nonumber\\
 H_\text{a}^2(\cB SO_\infty;\RZ) &=\Z_2= \{ \frac12 \w_2 \},
\nonumber\\
 H_\text{a}^3(\cB SO_\infty;\RZ) &=\Z= \{ \om_3 \},
\nonumber\\
 H_\text{a}^4(\cB SO_\infty;\RZ) &=\Z_2= \{ \frac12 \w_4 \},
\\
 H_\text{a}^5(\cB SO_\infty;\RZ) &=\Z_2= \{ \frac12 \w_2\w_3) \},
\nonumber\\
 H_\text{a}^6(\cB SO_\infty;\RZ) &=\Z_2^{\oplus 3}= \{ \frac12 \w_6 , \frac12 \w_2\w_4, \frac12 p_1 \w_2  \},
\nonumber\\
 H_\text{a}^7(\cB SO_\infty;\RZ) &=\Z^{\oplus 2}\oplus \Z_2= \{ \om_3 p_1, \om_7 ,  
\frac12 \w_3 \w_4 \}.
\nonumber 
\end{align}
where $\om_{4n-1}$ is a $\R/\Z$-valued almost-cocycle  on $\cB SO_\infty$ (the gravitational Chern-Simons term)
\begin{align}
 \dd \om_{4n-1} = p_n.
\end{align}
The above basis give rise to the basis in \eqn{HBSOZ} through the natural map
$\Bs$: $H^d(\cB G,\RZ)\to H^{d+1}(\cB G,\Z)$.  

% http://mathoverflow.net/questions/185990

\section{The ring of $H^*(\cB O_\infty;\Z)$}
\label{HBO}

The ring $H^{*}(\cB O_n;\Z_2)$ is given by
\begin{align}
 H^{*}(\cB O_n;\Z_2)=\Z_2[\w_1,\w_2,\cdots,\w_n].
\end{align}
and the ring $H^{*}(\cB O_\infty;\Z)$ is given by\cite{B8283}
\begin{align}
 H^{*}(\cB O_\infty;\Z)=\Z[\{p_i\}, \{\Bs_2 (\w_1^\eps\w_{2i_1}\w_{2i_2}\cdots)\}]/ \sim ,
\end{align}
where $\Z[\{p_i\}, \{\Bs_2 (\w_1^\eps\w_{2i_1}\w_{2i_2}\cdots)\}]$ is a
polynomial ring with integer coefficients, generated by $p_i$ and
$\Bs_2(\w_1^\eps\w_{2i_1}\w_{2i_2}\cdots)$, $\eps=0,1$, 
$0<i\leq \toZ{\frac n2}$,
and $0<i_1<i_2<\cdots \leq \toZ{\frac n2}$.
Here $p_i\in H^{4i}(\cB O_\infty;\Z)$ is the
Pontryagin class of dimension $4i$ and $\w_i\in H^{i}(\cB O_\infty;\Z_2)$ is the
Stiefel-Whitney class of dimension $i$.  

To obtain the ring $H^{*}(\cB O_\infty;\Z)$ from a polynomial ring generated by
$p_i$ and $\Bs_2(\w_1^\eps\w_{2i_1}\w_{2i_2}\cdots)$, we need to quotient out
certain equivalence relations $\sim$.  The  equivalence relations $\sim$
contain
\begin{widetext}
\begin{align}
 2 \Bs_2 (\w_{2i_1}\w_{2i_2}\cdots) &=0 , \ \ \ 
 \Bs_2 (\w_1\w_n) =0 ,\ \ \ p_n\se{2} \w_{2n}^2, \ \ \
 (\Bs_2 \w_{n})^2\big|_{n=\text{even}} = p_{n/2}\Bs_2 \w_1 ,
\\
 \Bs_2  \w(I) \Bs_2  \w(J) &=
\sum_{k\in I} \Bs_2  \w_{2k}\ \Bs_2  \w[(I-\{k\})\cup J-(I-\{k\})\cap J]\ p[(I-\{k\})\cap J] ,
\nonumber 
\end{align}
\end{widetext}
where $I=\{\frac{\eps}{2},i_1,i_2,\cdots\}$,
$\w(I)=\w_1^\eps\w_{2i_1}\w_{2i_2}\cdots$,  and $p(I)=\w_1^\eps
p_{i_1}p_{i_2}\cdots$.  Many last kind of the equivalence relations are trivial
identities.  The non-trivial equivalence relations for dimension 9 and less are
given by:
\begin{align}
 [\Bs_2 (\w_1\w_2) ]^2 &=  (\Bs_2 \w_2)^2 \Bs_2 \w_1  + (\Bs_2 \w_1)^2 p_1 ,
\nonumber\\
 \Bs_2 (\w_1\w_2) \Bs_2 \w_4 &=  \Bs_2 \w_2 \Bs_2 (\w_1 \w_4) +\Bs_2 \w_1 \Bs_2 (\w_2 \w_4)
\end{align}
So for low dimensions, 
\begin{align}
\label{HBOZ}
 H^0(\cB O_\infty;\Z) &=\Z,
\nonumber \\
 H^1(\cB O_\infty;\Z) &=0, 
\nonumber \\
 H^2(\cB O_\infty;\Z) &=\Z_2=\{\Bs_2 \w_1\},
\nonumber\\
 H^3(\cB O_\infty;\Z) &=\Z_2=\{\Bs_2  \w_2\},
\nonumber\\
 H^4(\cB O_\infty;\Z) &=\Z_2\oplus \Z=\{(\Bs_2 \w_1)^2,p_1\},
\nonumber\\
 H^5(\cB O_\infty;\Z) &=\Z_2^{\oplus 2}=\{\Bs_2  \w_4,\Bs_2 \w_1\Bs_2  \w_2\},
\\
 H^6(\cB O_\infty;\Z) &=\Z_2^{\oplus 3}=\{(\Bs_2  \w_2)^2,(\Bs_2 \w_1)^3, p_1\Bs_2  \w_1\},
\nonumber\\
 H^7(\cB O_\infty;\Z) &=\Z_2^{\oplus 5}
=\{ \Bs_2  \w_6 ,  \Bs_2  (\w_2\w_4), 
\nonumber\\
&  p_1\Bs_2  \w_2 ,(\Bs_2 \w_1)^2\Bs_2  \w_2 ,\Bs_2 \w_1\Bs_2  \w_4 \},
\nonumber\\
 H^8(\cB O_\infty;\Z) &= \Z_2^{\oplus 5} \oplus \Z^{\oplus 2}
=\{  \Bs_2   (\w_1\w_2\w_4),(\Bs_2 \w_1)^4 ,
\nonumber\\
& \Bs_2  \w_2 \Bs_2  \w_4,\Bs_2 \w_1 (\Bs_2  \w_2)^2, p_1(\Bs_2  \w_1)^2, p_1^2, p_2\}.
\nonumber 
\end{align}
In the above, we also list the basis (or generators) of cohomology classes.
%Using the universal coefficient theorem
%\begin{align}
%\label{ucf}
% H^n(X,\M) &\simeq 
%H^n(X;\Z) \otimes_{\Z} \M \oplus \Tor(H^{n+1}(X;\Z),\M)  ,
%\end{align}
Due to the relation 
\eqn{HalmH},
the above allows us to obtain 
\begin{align}
\label{HBOU1}
 H_\text{a}^0(\cB O_\infty;\RZ) &=\R/\Z, 
\nonumber \\
 H_\text{a}^1(\cB O_\infty;\RZ) &=\Z_2=\{\frac12 \w_1\},
\nonumber\\
 H_\text{a}^2(\cB O_\infty;\RZ) &=\Z_2=\{\frac12  \w_2\},
\nonumber\\
 H_\text{a}^3(\cB O_\infty;\RZ) &=\Z_2\oplus \Z=\{\frac12 \w_1\Bs_2 \w_1,\om_3\},
\nonumber\\
 H_\text{a}^4(\cB O_\infty;\RZ) &=\Z_2^{\oplus 2}=\{\frac12  \w_4,\frac12 \w_1\Bs_2  \w_2\},
\\
 H_\text{a}^5(\cB O_\infty;\RZ) &=\Z_2^{\oplus 3}=\{\frac12 \w_2\Bs_2  \w_2,\frac12 \w_1(\Bs_2 \w_1)^2, \frac12 \w_1 p_1\},
\nonumber\\
 H_\text{a}^6(\cB O_\infty;\RZ) &=\Z_2^{\oplus 5}
=\{ \frac12  \w_6 ,  \frac12 \w_2\w_4, \frac12 \w_2 p_1,
\nonumber\\
&
\frac12 \w_2 (\Bs_2 \w_1)^2 , \frac12 \w_1\Bs_2  \w_4 \},
\nonumber\\
 H_\text{a}^7(\cB O_\infty;\RZ) &= \Z_2^{\oplus 5} \oplus \Z^{\oplus 2}
=\{  \frac12 \w_1\w_2\w_4,\frac12 \w_1 (\Bs_2 \w_1)^3 ,
\nonumber\\
 \frac12 \w_2 \Bs_2  \w_4, &\frac12 \w_1 (\Bs_2  \w_2)^2, \frac12 \w_1 p_1 \Bs_2  \w_1, \om_3 p_1, \om_7\}.
\nonumber 
\end{align}

\section{Relations between cocycles and Stiefel-Whitney classes
on a closed manifold}
\label{Rswc}

The cocycles and the Stiefel-Whitney classes on a closed manifold satisfy
many relations.  In this section, we will show how to generate those relations.

\subsection{Introduction to Stiefel-Whitney classes}

The Stiefel-Whitney classes $\w_i \in H^i(M^D;\Z_2)$ is defined for an $O_n$
vector bundle on a $d$-dimensional space with $n\to \infty$.  If the
$O_\infty$ vector bundle on $d$-dimensional space, $M^D$, happen to be the
tangent bundle of $M^D$ direct summed with a trivial $\infty$-dimensional
vector bundle, then the corresponding Stiefel-Whitney classes are referred as
the Stiefel-Whitney classes of the manifold $M^D$.

The Stiefel-Whitney classes of manifold behave well under the connected sum of
manifolds.  Let 
\begin{align}
\w(M) = 1+\w_1(M)+\w_2(M)+\cdots
\end{align}
be the total Stiefel-Whitney class of a manifold $M$.
If we know $\w(M)$ and $\w(N)$,  then we can obtain $\w(M\#N)$:
\begin{align}
\label{SWsum}
\w(M\#N)\se{2,\dd}\w(M)+\w(N) -1.
\end{align}
Under the product of manifolds, we have
\begin{align}
\w(M\times N)\se{2,\dd}\w(M)\w(N) .
\end{align}

The Stiefel-Whitney numbers are non-oriented cobordism invariant.  All the
Stiefel-Whitney numbers of a smooth compact manifold vanish iff the manifold is
the boundary of some smooth compact manifold.  Here the manifold can be
non-orientable.

The Stiefel-Whitney numbers and Pontryagin numbers are oriented cobordism
invariant.  All the Stiefel-Whitney numbers and Pontryagin numbers  of a smooth
compact orientable manifold vanish iff the manifold is the boundary of some
smooth compact orientable manifold.

\subsection{Relations between Stiefel-Whitney classes of the tangent bundle}

For generic $O_\infty$ vector bundle, the Stiefel-Whitney classes are all
independent. However, the Stiefel-Whitney classes for a manifold (\ie for the
tangent bundle) are not independent and satisfy many relations.

To obtain those relations, we note that,  for any $O_\infty$ vector bundle,
the total Stiefel-Whitney class $\w=1+\w_1+\w_2+\cdots$ is related to the
total Wu class $u=1+u_1+u_2+\cdots$ through the total Steenrod square
\cite{W5008}:
\begin{align}
 \w\se{2,\dd}Sq(u),\ \ \ Sq=1+\Sq^1+\Sq^2+ \cdots .
\end{align}
Therefore, 
$\w_n\se{2,\dd}\sum_{i=0}^n \Sq^i (u_{n-i})$.
The Steenrod squares have the following properties:
\begin{align}
\Sq^i(x_j) &\se{2,\dd}0, \  i>j, \ \ 
\Sq^j(x_j) \se{2,\dd}x_jx_j,  \ \  \Sq^0=1,
\end{align}
for any $x_j\in H^j(M^D;\Z_2)$.
Thus
\begin{align}
u_n\se{2,\dd}\w_n+\sum_{i=1, 2i\leq n} \Sq^i (u_{n-i}).
\end{align}
This allows us to compute $u_n$ iteratively, using Wu formula
\begin{align}
\label{WuF}
\Sq^i(\w_j) &\se{2,\dd}0, \ \ i>j, \ \ \ \ \
\Sq^i(\w_i) \se{2,\dd}\w_i\w_i, 
\\
 \Sq^i(\w_j) &\se{2,\dd} \w_i\w_j+\sum_{k=1}^i 
\frac{(j-i-1+k)!}{(j-i-1)!k!}
\w_{i-k} \w_{j+k},\ \ i<j ,
\nonumber\\
\Sq^1(\w_j) &\se{2,\dd} \w_1\w_j + (j-1) \w_{j+1},
\nonumber 
\end{align}
and the Steenrod relation 
\begin{align}
\label{Sqrel}
	\Sq^n(xy)\se{2,\dd}\sum_{i=0}^n \Sq^i(x)\Sq^{n-i}(y).
\end{align}
We find
\begin{align}
u_0&\se{2,\dd}1, 
\ \ \ \ \
u_1\se{2,\dd}\w_1, 
\ \ \	 \ \
u_2\se{2,\dd}\w_1^2+\w_2, 
	\nonumber\\
u_3&\se{2,\dd}\w_1\w_2, 
\ \ \ \ \
u_4\se{2,\dd}\w_1^4+\w_2^2+\w_1\w_3+\w_4, 
	\\
u_5&\se{2,\dd}\w_1^3\w_2+\w_1\w_2^2+\w_1^2\w_3+\w_1\w_4, 
	\nonumber\\
u_6&\se{2,\dd}\w_1^2\w_2^2+\w_1^3\w_3+\w_1\w_2\w_3+\w_3^2+\w_1^2\w_4+\w_2\w_4, 
	\nonumber\\
u_7&\se{2,\dd}\w_1^2\w_2\w_3+\w_1\w_3^2+\w_1\w_2\w_4, 
	\nonumber\\
u_8&\se{2,\dd}\w_1^8+\w_2^4+\w_1^2\w_3^2+\w_1^2\w_2\w_4+\w_1\w_3\w_4+\w_4^2
	\nonumber\\
	&\ \ \ \ 
	+\w_1^3\w_5 +\w_3\w_5+\w_1^2\w_6+\w_2\w_6+\w_1\w_7+\w_8. 
	\nonumber
\end{align}
We note that the Steenrod squares form an algebra:
\begin{align}
 \Sq^a\Sq^b
&=\sum_{j=0}^{[a/2]} \bpm b-j-1 \\ a-2j \\ \epm  \Sq^{a+b-j} \Sq^j, 
\nonumber\\
&=\sum_{j=0}^{[a/2]} \frac{(b-j-1)!}{(a-2j)!(b-a+j-1)!} \Sq^{a+b-j} \Sq^j, 
\nonumber\\
& 0<a<2b.
\end{align}
which leads to the relation $\Sq^1\Sq^1=0$.

If the $O_\infty$ vector bundle on $d$-dimensional space, $M^D$, happen to be
the tangent bundle of $M^D$, then the corresponding Wu class and the
Steenrod square satisfy 
\begin{align}
\label{SqWu}
\Sq^{D-j}(x_j)\se{2,\dd}u_{D-j} x_j,  \text{ for any } x_j \in H^j(M^D;\Z_2) .
\end{align}
We can generate many relations for cocycles and Stiefel-Whitney classes on a
manifold using the above result:
\begin{enumerate}
\item If we choose $x_j$ to be a combination of Stiefel-Whitney classes, plus
the Sq operations them, the
above will generate many relations between Stiefel-Whitney classes.
\item If we choose $x_j$ to be a combination of Stiefel-Whitney classes and cocycles, plus
the Sq operations them, the
above will generate many relations between Stiefel-Whitney classes and cocycles.
\end{enumerate}
As an application, we note that
$\Sq^i(x_j)\se{2,\dd}0$ if $i>j$. Therefore $u_ix_{D-i}\se{2,\dd}0$ for any $x_{D-i}
\in H^{D-i}(M^D;\Z_2)$ if $i>D-i$.  Since $\Z_2$ is a field and according to the Poincar\'e duality, this implies that
$u_i\se{2,\dd}0$ for  $2i>D$.  
Also $\Sq^n\cdots \Sq^m(u_i)\se{2,\dd}0$ if $2i>D$.  This also gives
us relations among  Stiefel-Whitney classes.  

%http://mathoverflow.net/questions/40539)

\subsection{Relations between Stiefel-Whitney classes and a $\Z_2$-valued
1-cocycle in 3-dimensions}
\label{Rswc3D}

On a 3-dimensional manifold, we can find
many relations between Stiefel-Whitney classes:\\
(1) $u_2\se{2,\dd}\w_1^2+\w_2\se{2,\dd}0$.\\
(2) $u_3\se{2,\dd}\w_1\w_2\se{2,\dd}0$.\\
(3) $\Sq^1(u_2)\se{2,\dd}0$. Using $\Sq^1(\w_i)\se{2,\dd}\w_1\w_i+(i+1)\w_{i+1}$, we find that
$\Sq^1(\w_1^2+\w_2)\se{2,\dd}\Sq^1(\w_1)\w_1+\w_1\Sq^1(\w_1)+\Sq^1(\w_2)\se{2,\dd}\w_1\w_2+\w_3\se{2,\dd}0$.\\
This gives us three relations
\begin{align}
\label{wrel3}
 \w_1^2\se{2,\dd}\w_2,\ \ \w_1\w_2\se{2,\dd}\w_3\se{2,\dd}0.
\end{align}
Let $a^{\Z_2}$ be a $\Z_2$-valued 1-cocycle. 
We can also find a relation between the
Stiefel-Whitney classes and $a^{\Z_2}$: 
\begin{align}
\label{warel3}
\w_1(a^{\Z_2})^2\se{2,\dd}\Sq^1((a^{\Z_2})^2)\se{2,\dd}2(a^{\Z_2})^3\se{2,\dd}0.
\end{align}

There are six possible 3-cocycles
that can be constructed from the Stiefel-Whitney classes and the 1-cocycle
$a^{\Z_2}$:
\begin{align}
& (\w_1)^3, && \w_1 \w_2, && \w_3,
\nonumber\\
& (a^{\Z_2})^3, && \w_1 (a^{\Z_2})^2, && \w_1^2 a^{\Z_2}.
\end{align}
From the above relations, we see that only two of them are non-zero:
\begin{align}
& (a^{\Z_2})^3, && \w_1^2 a^{\Z_2}.
\end{align}

\subsection{Relations between Stiefel-Whitney classes and a $\Z_2$-valued
1-cocycle in 4-dimensions}
\label{Rswc4D}

The relations between the  Stiefel-Whitney classes for
4-dimensional manifold can be listed:\\
(1) $u_3\se{2,\dd}\w_1\w_2\se{2,\dd}0$.\\
(2) $u_4\se{2,\dd}\w_1^4+\w_2^2+\w_1\w_3+\w_4\se{2,\dd}0$.\\
(3) $\Sq^1(u_3)\se{2,\dd}0$, which implies
$\Sq^1(\w_1\w_2)\se{2,\dd}\Sq^1(\w_1)\w_2+\w_1\Sq^1(\w_2)\se{2,\dd}\w_1^2\w_2
+\w_1^2\w_2+\w_1\w_3\se{2,\dd}\w_1\w_3\se{2,\dd}0$,\\
which can be summarized as
\begin{align}
\label{wrel4}
 \w_1\w_2\se{2,\dd}0,\ \ \w_1\w_3\se{2,\dd}0,\ \ \w_1^4+\w_2^2+\w_4\se{2,\dd}0.
\end{align}
We also have many relations between
the  Stiefel-Whitney classes and $a^{\Z_2}$:\\
(1) $\Sq^1((a^{\Z_2})^3)\se{2,\dd}(a^{\Z_2})^4\se{2,\dd}\w_1(a^{\Z_2})^3$.\\
(2) $\Sq^1(\w_1(a^{\Z_2})^2)\se{2,\dd}\w_1^2(a^{\Z_2})^2
\se{2,\dd}\w_1[\w_1(a^{\Z_2})^2]$.\\
(3) $\Sq^1(\w_1^2a^{\Z_2})\se{2,\dd}\w_1^2(a^{\Z_2})^2\se{2,\dd}\w_1^3a^{\Z_2}$.\\
(4) $\Sq^1(\w_2a^{\Z_2})\se{2,\dd}(\w_1\w_2+\w_3)a^{\Z_2}+\w_2(a^{\Z_2})^2
\se{2,\dd}\w_1\w_2a^{\Z_2}$, which implies that
$\w_3a^{\Z_2}\se{2,\dd}\w_2(a^{\Z_2})^2$.\\
(5) $\Sq^2((a^{\Z_2})^2)\se{2,\dd}(a^{\Z_2})^4\se{2,\dd}(\w_1^2+\w_2)(a^{\Z_2})^2$.\\
(6) $\Sq^2(\w_1a^{\Z_2})\se{2,\dd}\w_1^2(a^{\Z_2})^2\se{2,\dd}(\w_1^2+\w_2)\w_1a^{\Z_2}\se{2,\dd}\w_1^3a^{\Z_2}$, which is the same as (2).\\
To summarize
\begin{align}
\label{warel4}
& \w_1^2(a^{\Z_2})^2\se{2,\dd}\w_1^3a^{\Z_2},
&& (a^{\Z_2})^4 \se{2,\dd}\w_1 (a^{\Z_2})^3, 
\\
&  \w_2 (a^{\Z_2})^2\se{2,\dd} \w_3 a^{\Z_2}, &&
(a^{\Z_2})^4 + \w_1^2(a^{\Z_2})^2  +\w_2(a^{\Z_2})^2\se{2,\dd}0
.
\nonumber 
\end{align}

There are nine 4-cocycles that can be constructed from Stiefel-Whitney classes
and a 1-cocycle $a^{\Z_2}$:
\begin{align}
& (a^{\Z_2})^4, &&
 \w_1 (a^{\Z_2})^3, &&
 \w_1^2 (a^{\Z_2})^2,
\nonumber\\
& \w_2 (a^{\Z_2})^2, &&
 \w_1^3 a^{\Z_2}, &&
 \w_3 a^{\Z_2},
\nonumber\\
& \w_1^4, &&
  \w_2^2, &&
  \w_4 .
\end{align}
Only four of them are independent
\begin{align}
 \w_1^4,\ \  \w_2^2,\ \ \w_3 a^{\Z_2},\ \ \w_1^3 a^{\Z_2}.
\end{align}

%\subsection{Relations between Stiefel-Whitney classes and a $\Z_2$-valued
%2-cocycle in 4-dimensions}
%
%\label{bSWrel4}
%
%There are two relations between
%the  Stiefel-Whitney classes and a $\Z_2$-valued
%2-cocycle $b^{\Z_2}$:\\
%(1) $\Sq^1(\w_1b^{\Z_2})\se{2,\dd} \w_1^2 b^{\Z_2} + \w_1 \Bs_2b^{\Z_2} \se{2,\dd}\w_1^2b^{\Z_2}$, which implies $\w_1 \Bs_2 b^{\Z_2}\se{2,\dd}0$.\\
%(2) $\Sq^2(b^{\Z_2})\se{2,\dd} (b^{\Z_2})^2\se{2,\dd}(\w_1^2+\w_2)b^{\Z_2}$.\\
%There are seven 4-cocycles that can be constructed from Stiefel-Whitney classes
%and a $\Z_2$-valued 2-cocycle $b^{\Z_2}$:
%\begin{align}
%& (b^{\Z_2})^2, &&
% \w_1 \Bs_2 b^{\Z_2}, &&
% \w_1^2 b^{\Z_2}, &&
% \w_2 b^{\Z_2},
%\nonumber\\
%& \w_1^4, &&
%  \w_2^2, &&
%  \w_4 .
%\end{align}
%So the following four  4-cocycles are independent
%\begin{align}
% \w_1^4,\ \  \w_2^2,\ \ \w_2 b^{\Z_2},\ \ \w_1^2 b^{\Z_2}.
%\end{align}
%

\section{Relation between Pontryagin classes and Stiefel-Whitney classes}
\label{PandSW}

%Ok, a more general statement holds. It is apparently due to Wu, but I found it
%in [this paper by Emery Thomas][1], Theorem C.  
%[1]: http://www.jstor.org/stable/1993484

There is result due to Wu that relate  Pontryagin classes and Stiefel-Whitney classes (see \Ref{T6067} Theorem C):\\
Let $B$ be a vector bundle over a manifold $X$, $\w_i$ be its Stiefel-Whitney
classes and $p_i$ its Pontryagin classes. Let $\rho_4$ be the reduction modulo
4 and $\theta_2$ be the embedding of $\mathbb{Z}_2$ into $\mathbb{Z}_4$ (as
well as their induced actions on cohomology groups). 
Then
\begin{align}
\cP_2(\w_{2i}) \se{4} p_i + 2 \Big( \w_1 Sq^{2i-1} \w_{2i} + \sum_{j = 0}^{i-1} \w_{2j} \w_{4i-2j} \Big),
\end{align}
% http://www.encyclopediaofmath.org/index.php/Pontryagin_square
where $\cP_2$ is the Pontryagin square,
which maps $x\in H^{2n}(X,\Z_2)$ to $\cP_2(x) \in H^{4n}(X,\Z_4)$.  The
Pontryagin square has a property that $ \cP_2(x)\se{2} x^2$.
%http://www.encyclopediaofmath.org/index.php/Pontryagin_square
Therefore
\begin{align}
\label{piw2i}
\cP_2(\w_{2i})\se{2} \w_{2i}^2 \se{2} p_i.
\end{align}

\section{Spin and Pin structures}
\label{spinstructure}

Stiefel-Whitney classes can determine when a manifold can have a spin
structure.  The  spin structure is defined only for orientable manifolds.  The
tangent bundle for an orientable manifold $M^d$ is a $SO_d$ bundle.  The group
$SO_d$ has a central extension to the group ${Spin}(d)$.  Note that
$\pi_1(SO_d)=\Z_2$. The group ${Spin}(d)$ is the double covering of the
group  $SO_d$.  A spin structure on $M^D$ is a ${Spin}(d)$ bundle, such
that under the group reduction ${Spin}(d) \to SO_d$, the ${Spin}(d)$
bundle reduces to the $SO_d$ bundle.  Some manifolds cannot have such a
lifting from $SO_d$ tangent bundle to the ${Spin}(d)$ spinor bundle.  The
manifolds that have such a lifting is called spin manifold.  A manifold is a
spin manifold iff its first and second Stiefel-Whitney class vanishes
$\w_1=\w_2=0$.  

For a non-orientable manifold $N^d$, the tangent bundle is a $O_d$ bundle.
The non-connected group $O_d$ has two nontrivial central extensions (double
covers) by $Z_2$ with different group structures, denoted by ${Pin}^+(d)$
and  ${Pin}^-(d)$.  So the $O_d$ tangent bundle has two types of lifting
to a ${Pin}^+$ bundle and a ${Pin}^-$ bundle, which are called
${Pin}^+$ structure and ${Pin}^-$ structure respectively.  The
manifolds with such liftings are called ${Pin}^+$ manifolds or
${Pin}^-$ manifolds.  We see that the concept of ${Pin}^\pm$
structure applies to both orientable and non-orientable manifolds.  A manifold
is a  ${Pin}^+$ manifold iff $\w_2=0$.  A manifold is a  ${Pin}^-$
manifold iff $\w_2+\w_1^2=0$.  If a manifold $N^d$ does admit ${Pin}^+$
or ${Pin}^-$ structures, then the set of isomorphism classes of
${Pin}^+$-structures (or ${Pin}^-$-structures) can be labled by
elements in $H^1(N^d;\Z_2)$. For example $\R P^4$ admits two
${Pin}^+$-structures and no ${Pin}^-$-structures since $\w_2(\R
P^4)=0$ and $\w_2(\R P^4)+\w_1^2(\R P^4)\neq 0$.

From \eqn{SWsum}, we see that $M\# N$ is Pin$^+$ iff both $M$ and $N$ are
Pin$^+$. Similarly, $M\# N$ is Pin$^-$ iff both $M$ and $N$ are Pin$^-$.

\section{Higher group as simplicial complex}
\label{hgroup}

\subsection{Higher group and its classifying space}

Given a \emph{topological space} $K$, we can triangulate it and use the
resulting \emph{complex} $\cK$ to model it.  If $K$ is connected, we can choose
the complex $\cK$ to have only one vertex.  We can even choose the  one-vertex
complex to be a \emph{simplicial set}.  Such a simplicial set is called a
\emph{higher group} if various Kan conditions are satified and the
corresponding space $K$ is called the \emph{classifying space} of the higher
group. More precisely, $\cK$ is an $n$-group ($n\in  \{1, 2, \dots\} \sqcup
\{\infty\}$), if $\cK$ satisfies Kan conditions  $\Kan(m, j)$, i.e. the
natural horn projection $\cK_m\xrightarrow{p^m_j}
\Lambda^m_j(\cK)$ is surjective, for all $0\le j \le m$; and strict Kan
conditions $\Kan(m, j)!$, i.e. $\cK_m\xrightarrow{p^m_j} \Lambda^m_j(\cK)$ is isomorphic, for
all $0\le j \le m$ and $m\ge n+1$.  Here, $\Lambda^m_j(\cK)$ denotes the set of
$(m, j)$-horns in $\cK$.
%, **a picture**.   
We will use $\cG$ to
denote a higher group (\ie a simplicial set), and use $B\cG \equiv K$ to denote
the classifying space (\ie the topological space modeled by the  simplicial
set).

Let us describe an explicit construction for such a higher group $\cG$. As
a simplicial set, $\cG$ is described by a set of vertices
$[\cG]_0$, a set of links $[\cG]_1$, a set of triangles  $[\cG]_2$, etc.  The
complex $\cG$ is formally described by
\begin{equation}\label{eq:nerveM}
\xymatrix{ 
[\cG]_0 & 
[\cG]_1 \ar@<-1ex>[l]_{d_0, d_1}\ar[l] & 
[\cG]_2 \ar@<-1ex>_{d_0, d_1 , d_2}[l] \ar@<1ex>[l] \ar[l] & 
[\cG]_3 \ar@<-1ex>[l]_{d_0, ..., d_3} \ar@<1ex>[l]_{\cdot} & 
[\cG]_4 \ar@<-1ex>[l]_{d_0, ..., d_4} \ar@<1ex>[l]_{\cdot}  ,
}
\end{equation}
where $d_i$ are the face maps, describing how the $n$-simplices are attached to
a $(n-1)$-simplex. 

As the set of vertices, $[\cG]_0 = \{pt\}$, \ie there is only one vertex. An
link in $[\cG]_1$ is labeled simply by a label $a_{01}$ whose end points are
both this point $pt$.  Such labels from a group $G$.  An triangle in $[\cG]_2$
is labeled by its three links $a_{01},a_{12},a_{02}$, and possibly an
additional label $b_{012}$.  Such additional labels form an Abelian group
$\Pi_2$.  Thus an triangle is labeled by $(a_{01},a_{12},a_{02}; b_{012})$.  We
introduce a compact notation
\begin{align}
 s[012] = (a_{01},a_{12},a_{02}; b_{012})
\end{align}
to denote such a triangle.  Similarly, an generic $d$-simplex is labeled by
a label-set $s[0\cdots d]=(a_{ij};b_{ijk};\cdots)$.

We see that a higher group $\cG$ in this model consists the data of a
collection of groups $G,\Pi_2,\Pi_3,\cdots$, where $G$ can be non-Abelian and
$\Pi_i$'s are Abelian.  Both $G$ and $\Pi_i$ can be discrete or continuous.  We
denote such a higher group as $\cG(G,1;\Pi_2,2;\Pi_3,3;\cdots)$.  With such a
labeling of the simplices, such as $s[012]=(a_{01}, a_{12}, a_{02}; b_{012})$,
the face map $d_i$ can be expressed simply
\begin{align}
 d_0 (a_{01},a_{12},a_{02};b_{012}) &= a_{12},
\nonumber\\
 d_1 (a_{01},a_{12},a_{02};b_{012}) &= a_{02},
\nonumber\\
 d_2 (a_{01},a_{12},a_{02};b_{012}) &= a_{01},
\end{align}
or
\begin{align}
\label{dms}
 d_m s[0\cdots d] = s[0\cdots \hat m \cdots d]
\end{align}
where $\hat m$ means that the $m$ index is removed.

However, in order for the label-set $s[0\cdots d]$ to label a $d$-simplex in
complex $\cG(G,1,\Pi_2,2;\Pi_3,3;\cdots)$, the labels $a_{ij}$, etc., 
in the set $s[0\cdots d]$ must satisfy certain conditions.  Those conditions
determine the structure of a higher group
$\cG(G,1;\Pi_2,2;\Pi_3,3;\cdots)$.  Such constructed  higher group is a
triangulation of a topological space $K$. Different higher groups give us a
classification of homotopy types of topological spaces.

From our labeling of the simplices $s[0\cdots d]=(a_{ij};b_{ijk};\cdots)$, we
can also introduce the canonical cochains on the higher group
$\cG(G,1,\Pi_2,2;\Pi_3,3;\cdots)$.  The canonical $G$-valued 1-cochain $a$ is
given by its evaluation on 1-simplices $s[01]=a_{01}$:
\begin{align}
 \<a, a_{01}\> = a_{01},\ \ \ \ a_{01}\in G.
\end{align}
The canonical $\Pi_2$-valued 2-cochain $b$ is given by its evaluation on
2-simplices $s[012]=(a_{01},a_{12},a_{02};b_{012})$:
\begin{align}
\<b,(a_{01},a_{12},a_{02};b_{012})\> = b_{012},\ \ \ \ 
b_{012}\in \Pi_2.
\end{align}
The canonical $\Pi_n$-valued $n$-cochain $x^n$ can be defined in a similar
fashion.

The conditions satisfied by the labels $a_{ij}$, $b_{ijk}$, etc.,  in the set
$s[0\cdots d]$ can be expressed as the conditions on those canonical cochains.
In other words, we start with a chain complex of groups \[ G \xleftarrow{q_2}
\Pi_2 \xleftarrow{q_3} \Pi_3 \xleftarrow{q_4} \dots \xleftarrow{q_k} \Pi_k, \]
and group actions $G \xrightarrow{\alpha_j} Aut(\Pi_j)$, where $q_i$ are
$G$-equivariant with $G$ acting on $G$ trivially, and
\begin{equation}\label{eq:alpha}
\alpha_j(a_{pq} a_{qr} a_{pr}^{-1})=id.
\end{equation}

Then the structure and the definition of a higher group $\cG_{n_3, \dots,
n_{k+1}}(G, 1; \Pi_2, 2; \dots; \Pi_k,k)$ can be formulated via the conditions on the
canonical cocycles inductively: given $k-1$ cocycles 
\begin{align}
n_3 &\in Z^3(G, (\Pi_2^0)^{\alpha_2}), 
\nonumber\\
n_4 &\in Z^4(\cG_{n_3}(G, 1; \Pi_2, 2), (\Pi_3^0)^{\alpha_3}), 
\nonumber\\
\dots
\\
n_{k+1} & \in Z^{k+1}(\cG_{n_3, \dots, n_k}(G, 1; \Pi_2, 2;
\dots; \Pi_{k-1}, k-1), (\Pi_{k}^0)^{\alpha_k})
\nonumber 
\end{align}
where $\Pi_j^0:=\ker q_j \subset  \Pi_j$ for $j=2, \dots, k$, 
\begin{align}
\label{Xd}
\begin{split}
X_d & :=\{ s[0\dots  d]=(x^1_{01}, x^1_{02}, \dots, x^1_{d-1d} ; \\
& x^2_{012}, \dots, x^2_{(d-2)(d-1)d}; \dots; x^d_{0\dots d} ) | \\
 &x^1_{..}\in G,  x^j_{..} \in \Pi_j, \\
& \dd_{\alpha_{j}} x^j = q_{j+1} (x^{j+1}) + n_{j+1}(x^1; x^2; \dots;
  x^{j-1}), \\ 
&
\forall j=2, 3, \dots, d, \text{ and } \;
\dd x^1=q_2(x^2).  \}
\end{split}
\end{align}
is a $k$-group. Here we take all $\Pi_{\ge k+1}=0$ (thus $q_{\ge
  k+1}=0$) and all $n_{\ge 
  k+2}=0$ in the general definition of $X_d$. Here 
\begin{align}
    \dd_{\alpha_j} x^j (s[0\dots j+1] :&= \alpha_j(a_{01})\cdot x^j(s[1\dots j+1])
    \nonumber \\
    &- x^j(s[02\dots j+1]) + \dots
\end{align}
Equation \eqref{eq:alpha} guarantees that
$d_{\alpha_{j}} \circ d_{\alpha_j}=0$. 

We notice that $\cG_{n_3}(G, 1; \Pi_2, 2)$ is the 2-group constructed via
cocycles $n_3$ with equations $dx^1 = q_2(x^2)$ and $\dd_{\alpha_2} x^2 =
n_3(x^1)$. But it is not in contradiction with the equation $\dd_{\alpha_2} x^2
= q_3(x^3)+ n_3(x^1)$ in the definition of $X_d$. First of all, $n_3$ is a
cocycle, therefore it is possible that $\dd_{\alpha_2} x^2 = n_3(x^1)$ has
solutions. Secondly, $\dd_{\alpha_2} x^2 = q_3(x^3)+ n_3(x^1)$ is describing
another set of solutions in $X_d$, which is also possible to be there: why?  if
we apply $\dd_{\alpha_2}$ to it, we have $\dd_{\alpha_2} q_3 (x^3)=0$, but this
holds naturally, since $\dd_{\alpha_2} q_3 (x^3) = q_3 (\dd_{\alpha_3} x^3) =
q_3(n_4(x^1, x^2))=0$, no matter we have $\dd_{\alpha_3} x^3 = q_4 (x^4) +
n_4(x^1, x^2)$ or $\dd_{\alpha_3} x^3 =n_4(x^1, x^2)$. Thirdly, from both
equations, we have $q_2(n_3(x^1))=q_2(\dd_{\alpha_2} x^2)=\dd q_2(x^2)=\dd \dd
x^1 = 0$, which is also fine since $n_3$ takes value in $\Pi_2^0 =\ker q_2$.
We thus can further understand these equations, which are not contradict to
each other, inductively for higher $k$'s.

Now let us prove that what we construct satisfies desired Kan
conditions, therefore is a higher group. Notice that the horn space $\Lambda_{j}^m(X)$ has the
same $(m-2)$-skeleton as $X_m$, thus to verify the Kan condition
$\Kan(m, j)$, we only need to take care of $(m-1)$-faces. Since there
is no non-trivial $\ge k+1$ faces, it is clear that $\Kan(\ge k+2,
j)!$ are satisfied.  Then $\Kan(k+1, j)!$ are satisfied for $0\le j \le k+1$ because the
following equation,
\begin{equation}\label{eq:last-one}
\dd_{\alpha_k} x^k = n_{k+1} (x^1; x^2; \dots; x^{k-1}),
\end{equation}
implies that as long as we know any $k+1$ out of $k+2$ $k$-faces in
the $(k+1)$-simplex $s[01\dots k+1]$, then the other one is determined
uniquely. Similarly, $\Kan(m+1, j)$ are satisfied for $0\le j \le m <
k$ because the following equation, 
\[
\dd_{\alpha_m} x^m =q_{m+1}(x^{m+1})+  n_{m+1} (x^1; x^2; \dots; x^{m-1}),
\]
implies that any $m+1$ out the $m+2$ $m$-faces in the $(m+1)$-simplex
$s[01\dots m+1]$ determines the other one up to a choice of
$q_{m+1}(x^{m+1})$. Thus we can always fill the $(m+1, j)$-horn and we
have unique filling if and only if $q_{m+1}=0$. 

Moreover, if two sets of canonical two chains $n^X_j$'s and $n^Y_j$'s
differ by coboundaries valued in $\ker q_j$, then they define weak
equivalent $k$-groups. More precisely, this is an inductive process: if $n_3^Y-n_3^X=\dd n_2'$, and
$n_2' \in \ker q_2$, then we let $f_2: x^2 \mapsto x^2+ n_2(x_1)$ and
$f_1=f_0=id$; further using this truncated simplicial homomorphism,  if $f^*n_4^Y-n_4^X=\dd n_3'$, and $n_3' \in \ker q_3$,
then we let $f_3: x^3\mapsto x^3 + n'_3(x_1, x_2)$; further using $f_0,
\dots, f_3$, if $f^*n_5^Y-n_5^X=\dd n_4'$, and $n_4'\in \ker q_4$, then
we let $f_5: x^5 \mapsto x^5 + n'_4(x_1, x_2, x_3)$; ... in the end,
we obtain a simplicial homomorphism $f$ made up by automorphisms $f_j$ of $ G^{j}\times \Pi_2^{(^j_2)}
\times \Pi_3^{(^j_3)} \times \dots \times \Pi_j $. This simplicial
morphism is a weak equivalence $f: X\to Y$ of higher groups. Here
$X$ is the higher group defined by canonical chains $n^X_j$ and $Y$
the one defined by $n^Y_j$. This is because $f$ introduces isomorphisms
on homotopy groups of $X$ and $Y$. Notice that $X$ and $Y$ both have
the same homotopy groups: $\pi_0=0$, $\pi_1=G/\Im q_2$,
$\pi_2=\Pi_2/\Im q_3$, $\dots$. The construction above makes
sure that $f$ is a simplicial homomorphism, and it gives rise to
isomorphisms when passing to homotopy groups. 

Therefore, if we fix other canonical cocycles, up to weak equivalence, we may take $n_{k+1}\in H^{k+1}(\cG_{n_3, \dots, n_{k}} (G, 1; \Pi_2, 2; \dots; \Pi_{k-1}, k-1),
(\Pi_k^0)^{\alpha_{k}})$, and if $q_k=0$, we may further assume that  $n_{k+1}\in H^{k+1}(\cG_{n_3, \dots, n_{k}} (G, 1; \Pi_2, 2; \dots; \Pi_{k-1}, k-1),
{\Pi_k}^{\alpha_{k}})$.

\subsection{3-group}

In the following, we discuss a 3-group $\cG(G,1,\Pi_2,2;\Pi_3,3)$ in more
details.  The missing labels 
$\Pi_n,n|_{n>3}$ mean that 
$\Pi_n=|_{n>3} 0$. 
In order for
$(a_{01},a_{12},a_{02};b_{012})$ to label a triangle in the complex,
$a_{01},a_{12},a_{02}$ must satisfy
\begin{align}
\label{aaa}
(\del a)_{012}\equiv a_{01}a_{12} a_{02}^{-1} =q_2(b_{012}) 
.
\end{align}
In terms of canonical cochains, the above condition can be rewritten as
\begin{align}
 \del a = q_2(b).
\end{align}
Here $q_2$ is a group homomorphism $q_2: \Pi_2 \to G$.  So only $a_{01}$ and
$a_{12}$, $b_{012}$ are independent.  The triangles in the complex are
described by independent labels
\begin{align}
 s[012]=[a_{01},a_{12};b_{012}].
\end{align}
Therefore the set of triangles is given by $G^{\times 2} \times \Pi_2$.  If
$a_{01},a_{12},a_{02}$ do not satisfy the above condition, then the three links
$a_{01},a_{12},a_{02}$ simply do not bound a triangle (\ie there is a hole).

Similarly, for a tetrahedron $s[0123]$,
the labels $a_{ij}$ in the label-set $s[0123]$
all satisfy \eqn{aaa} if we replace 012 by $i<j<k$.
There an additional condition
\begin{align}
\label{bn3}
(\dd_{\al_2} b)_{0123} 
&\equiv
\al_2(a_{01})\cdot b_{123}- b_{023}+ b_{013}- b_{012} 
\nonumber\\
&=q_3(c_{0123})+n_3(a_{01},a_{12},a_{23}),
\end{align}
where $q_3$ is a group homomorphism $q_3: \Pi_3 \to \Pi_2$,  $\al_2$ is a group
homomorphism $\al_2: G \to \text{Aut}(\Pi_2)$, and $n_3 \in
Z^3(\cG(G,1),(\Pi_2^0)^{\al_2})$.  In terms of canonical cochains, the above can be
rewritten as
\begin{align}
 \dd_{\al_2} b = q_3(c)+n_3(a)
\end{align}
We see that a tetrahedron are described by
independent labels
\begin{align}
 s[0123]=[a_{01},a_{12},a_{23};b_{012},b_{013},b_{023};c_{0123}].
\end{align}
Therefore the set of tetrahedrons is given by $G^{\times 3} \times
\Pi_2^{\times 3}\times \Pi_3$.

For a 4-simplex $s[01234]$, the labels $a_{ij}$ and $b_{ijkl}$ in the label-set
$s[01234]$ all satisfy \eqn{aaa} and  \eqn{bn3}.  There an additional
condition
\begin{align}
\label{cn4}
&\ \ \ \
(\dd_{\al_3} c)_{01234} 
\nonumber\\
&\equiv
\al_3(a_{01})\cdot c_{1234}- c_{0234}+ c_{0134}- c_{0124}+  c_{0123}
\nonumber\\
&=n_4(a_{01},a_{12},a_{23},a_{34}, b_{012}, b_{013}, b_{023})
\end{align}
where $\al_3$ is a group
homomorphism $\al_3: G \to \text{Aut}(\Pi_3)$ and $n_4 \in
Z^4(\cG(G,1;\Pi_2,2),(\Pi_3^0)^{\al_3})$ is a closed cochain.  
In terms of canonical cochains, the above can be
rewritten as
\begin{align}
 \dd_{\al_3} c = n_4(a,b) .
\end{align}

In general, the 3-group $\cG(G,1;\Pi_2,2;\Pi_3,3)$ has the following sets of
simplices:
\begin{widetext}
\begin{equation}\label{eq:nerve2}
\xymatrix{ pt & G \ar@<-1ex>[l]_{d_0, d_1}\ar[l] & G^{\times 2} \times
  \Pi_2 \ar@<-1ex>_{d_0, ..., d_2}[l] \ar@<1ex>[l] \ar[l] & G^{\times
    3} \times \Pi_2^{\times 3} \ar@<-1ex>[l]_{d_0, ..., d_3}
  \ar@<1ex>[l]_{\cdot} & G^{\times 4} \times \Pi_2^{\times 6} \times \Pi_3 \ar@<-1ex>[l]_{d_0, ..., d_4}
  \ar@<1ex>[l]_{\cdot}& G^{\times 5} \times \Pi_2^{\times 10} \times
  \Pi_3^{\times 10} \dots \ar@<-1ex>[l]_{d_0, ..., d_5} \ar@<1ex>[l]_{\cdot} }
\end{equation}
\end{widetext}
The $d$-simplices form a set $G^{\times d} \times \Pi_2^{\times
(^d_2)} \times \Pi_3^{\times (^d_3)}$. 
The $d$-simplices in $G^{\times d}\times \Pi_2^{(^d_2)}\times \Pi_3^{\times
(^d_3)}$ are labeled by $\{a_{ij}, b_{ijk},c_{ijkl}\}$, $i,j,k,l=0,1,\cdots,d$,
that satisfy the conditions \eqn{aaa} (after replacing $012$ by $i<j<k$),
\eqn{bn3} (after replacing $0123$ by $i<j<k<l$) and \eqn{cn4} (after
replacing $01234$ by $i<j<k<l<m$).

We find that
3-groups $\cG(G;\Pi_2;\Pi_3)$ are classified by the following data
\begin{align}
  G;\ \Pi_2, q_2, \al_2, n_3;\ \Pi_3, q_3, \al_3, n_4
\end{align}
where $G$ is a group, $\Pi_2,\Pi_3$ are Abelian groups,
$\al_2,\al_3$ are group actions $\al_2: G\to \text{Aut}(\Pi_2)$ and $\al_3:
G\to \text{Aut}(\Pi_3)$, $n_3 \in Z^3(\cG(G,1),
(\Pi_2^0)^{\al_2})$, and $n_4 \in Z^4(\cG(G;\Pi_2);
(\Pi_3^0)^{\al_3})$. When $n_3, n_4$ differ by a coboundary valued in
$\ker q_2$, $\ker q_3$ respectively, the 3-groups are weak equivalent.

\subsection{3-group cocycle}

In the following, we give an explicit description of 3-group cocycles, which
are the cocycles on the complex $\cG(G,1;\Pi_2,2;\Pi_3,3)$.  First, a $d$-dimensional
3-group cochain $\nu_d$ with value $\M$ is a function $\om_d: G^{\times
d}\times \Pi_2^{(^d_2)}\times \Pi_3^{(^d_3)} \to \M$.  Then, using the face map
\eqn{dms}, we can define the differential operator $\dd$ acting on the
3-group cochains as the following:
\begin{align}
\label{cccond3}
(\dd \om_d)( s[0\cdots d+1] )
=\sum_{m=0}^{d+1} (-)^m \om_d(s[1\cdots \hat m \cdots d+1]),
\end{align}
With the above definition of $\dd$ operator, we can define the 3-group cocycles
as the 3-group cochains that satisfy $\dd\om_d =0$.  Two different 3-group
cocycles $\dd\om_d$ and $\dd\om_d'$ are equivalent if they are different by a
3-group coboundary $\dd \nu_{d-1}$.  The set of equivalent classes of
$d$-dimensional 3-group cocycles is denoted by $H^{d}(\cG(G,1;\Pi_2,2;\Pi_3,3),
\M)$, which in fact forms an Abelian group.

In the above, we gave a quite general definition of $k$-group.
In more standard definition, $q_i$ is chosen to be $q_i=0$.
Such $q_i=0$ $k$-group will be denoted by
$\cB(G,1;\Pi_2,2;\Pi_3,3;\cdots)$.
Its homotopy group are given by 
\begin{align}
 \pi_n\big(\cB(G,1;\Pi_2,2;\Pi_3,3;\cdots)\big)=\Pi_n,\ \ \
\Pi_0=G.
\end{align}

Usually, we can use the canonical cochains $a,b,c$ \etc to construct the
cocycles on $\cG(G,1;\Pi_2,2;\Pi_3,3)$.  For example, on a 3-group
$\cB_0(G,1;\Pi_2,2;\Pi_3,3)$ defined via its canonical cochains: $ \del a =1,\
\dd b =0, \  \dd c=0$, a 3-group cocycle $\om_d$ will be a cocycle on
$\cB_0(G,1;\Pi_2,2;\Pi_3,3)$. The expressions $b, b^2, \Sq^3 c$, \etc are also
cocycles on $\cB_0(G.1;\Pi_2,2;\Pi_3,3)$.

\subsection{Continuous group}

%In the above discussion, we have assumed that $\Pi_1=G,\Pi_2,\Pi_3,\cdots$ are
%all finite groups.  When $\Pi_1=G$ is a continuous group, the definition of
%$\cG(G,1;\Pi_2,2;\cdots)$ need to be modified.
%The set of $d$-simplex $X_d$ in $\cG(G,1;\Pi2,2;\cdots)$ now is given by
%\begin{align}
%\label{XdC}
%\begin{split}
%X_d & :=\{ s[0\dots  d]=(x^1_{01}, x^1_{02}, \dots, x^1_{d-1d} ; \\
%& x^2_{012}, \dots, x^2_{(d-2)(d-1)d}; \dots; x^d_{0\dots d} ) | \\
% &x^1_{..}\in G,  x^j_{..} \in \Pi_j, \\
%& \dd_{\alpha_{j}} x^j = q_{j+1} (x^{j+1}) + n_{j+1}(x^1; x^2; \dots;
%  x^{j-1}), \\ 
%&
%\forall j=2, 3, \dots, d, \text{ and } \;
%\pi_\text{dis}(\dd x^1)=q_2(x^2)  \} .
%\end{split}
%\end{align}
%Compare to \eqn{Xd}, we have modified the condition
%$\dd x^1=q_2(x^2)$ to $\pi_\text{dis}(\dd x^1)=q_2(x^2)$, where
%$\pi_\text{dis}$ is the group homomorphism
%\begin{align}
% G \xrightarrow{\pi_\text{dis}} G/G_0
%\end{align}
%and $G_0$ is the connected component of $G$ that contain identity.

The above discussions apply to both discrete and continuous groups.  However,
in order to construct principle bundle or higher  principle bundle on space-time $M$, it is not enough to consider only strict simplicial homomorphisms
from space-time complex $\cM$ to $\mathcal{G} (G, 1; \Pi_2, 2; \dots)$ when $G$
is a continuous group. The reason is that, for example, in the case of
$\mathcal{G}(SU_2, 1) =\mathcal{B}SU_2$, strict simplicial homomorphisms
$\phi_\text{strict}: \cM \to \mathcal{B}SU_2$ can only produce trivial principal
$SU_2$-bundles on $M=|\cM|$, which is the geometric realization of $\cM$. We thus need to allow generalised morphisms
$\phi_\text{gen}$ from $\cM$ to $\mathcal{G} (G, 1; \Pi_2, 2; \dots)$, so that
their pullback can produce non-trivial higher principal bundles on $\cM$. 

Let us explain this via an example: a generalized morphism $\cM
\xrightarrow{\phi} \mathcal{B} SU_2$ consists of a zig-zag, $\cM
\xleftarrow{\chi} \tilde{\cM} \xrightarrow{\tilde{\phi}} \mathcal{B} SU_2$,
where both $\chi$ and $\tilde{\phi}$ are strict simplicial homomorphisms and
$\chi$ is a weak equivalence. Here, we define $X\to Y$ between simplicial
topological spaces being a weak equivalence if and only if their geometric
realization $|X|$ and $|Y|$ are weakly homotopy equivalent (namely all homotopy
groups are the same). Homotopy equivalence clearly implies weak homotopy
equivalence. This coincides with usual weak equivalence between simplicial sets
when both $X$ and $Y$ are simplicial sets (taking discrete topology).   Then to
present an $SU_2$-principal bundle $P$ on $\cM$, we take a good cover $\{U_i
\}$ (that is, all sorts of finite intersections $\cap U_i$ are contractible),
where $P$ is trivial on each $U_i$. Then we take $\tilde{\cM}$ to be the Cech
groupoid $\sqcup_i U_i \Leftarrow \sqcup_{ij} U_i \cap U_j \dots$,  and
$\tilde{\phi}$ is determinate by $\tilde{\phi}_1$ with $\tilde{\phi}_1(x)=
a_{ij}(x)$, where $x$ is a link in $U_{ij}$ and $a_{ij} \in G=SU_2$ labels the
links (or transition functions to glue $P$) in $\mathcal{B} SU_2$.  We take the
so called abstract nerve $N(\tilde{\cM})$ of the covering simplicial space
$\tilde{\cM}$, which is constructed as following: $N_0:=N(\tilde{\cM})_0$ is
the index set $I$ of the cover $\{U_i\}$. We denote a vertex by $v_i$ with
$i\in I$. A $d$-simplex $s[0, \dots, d]$ is a set $\{ v_{i_0}, \dots,
v_{i_d}\}$ with $d\ge 0$ and $i_0, \dots, i_d \in I$, such that $U_{i_0} \cap
\dots \cap U_{i_d} \neq \emptyset$. It is clear that there is a map $\tilde{M}
\xrightarrow{\chi'} N(\tilde{M})$ by mapping all points in $U_{i_0} \cap \dots
\cap U_{i_d} $ exactly back to the simplex $s[0, \dots, d]$.   As long as $M$
is paracompact and $\{U_i\}$ is a good cover, as we assumed, Borsuk Nerve
Theorem ensures that $M$ and $|N(\tilde{\cM})|$ are homotopy equivalent.
Segal\cite{S6805} proved that in $|\tilde{\cM}|$ and $M$ are homotopy
equivalent. Thus $|N(\tilde{\cM})|$ and $\tilde{\cM}$ are homotopy equivalent.
Thus $\chi'$ is a weak equivalence. At the same time, as both $\cM$ and $
N(\tilde{\cM})$ are simplicial sets, the homotopy equivalence between their
geometric realization implies the homotopy equivalence between themselves,
which in turn can be expressed as a zig-zag of weak equivalences. Thus we have
a chain of weak equivalences $\mathcal{M} \xleftarrow{\cong} L
\xrightarrow{\cong} N(\tilde{\cM}) \xleftarrow{\cong} \tilde{\cM}$. Then
pulling back $L$ along $\chi'$, we arrive at $\hat \cM$, which may be thought
as a refinement of $\tilde \cM$.  Then  we have weak equivalences $\cM
\xleftarrow{\cong} \hat \cM \xrightarrow{\cong} \tilde \cM$. Thus using $\hat
\cM$ we can then realise $P$ as a generalised morphism, $\cM \xleftarrow{\cong}
\hat \cM \rightarrow \mathcal{B} SU_2$. 

Thus in our article, when we talk about homomorphisms between simplicial
objects, we understand them as this correct version of morphisms, namely
generalised morphisms when $G$ is continuous or strict simplicial homomorphisms
when $G$ is discrete.

\section{Calculate $a^3 {\smile\atop 1} a^3$} \label{Sq2a3}

Let $a$ be a 1-cochain.  We have
\begin{align}
&\ \ \ \
 \<a^3\hcup{1} a^n, (012\cdots n+2)\> \se{2}\sum \<a^3, A_1\cup A_3\> \<x_n,  A_2\>
\nonumber\\
&\se{2} 
 (a^3)_{0,n,n+1,n+2} (a^n)_{01\cdots n}
+(a^3)_{0,1,n+1,n+2} (a^n)_{12\cdots n+1}
\nonumber\\
&\ \ \ \ \ \ \
+(a^3)_{0,1,2,n+2} (a^n)_{23\cdots n+2}
\nonumber\\
&\se{2}
a_{01}a_{12}\cdots a_{n+1,n+2} (a_{0n} +a_{1,n+1}+a_{2,n+2}) .
\end{align} 
Thus
\begin{align}
a^3\hcup{1} a^n
& \se{2}
(a\hcup{1} a^n) a^2+ a(a\hcup{1} a^n) a+ a^2(a\hcup{1} a^n)
\end{align}

When $a$ is a $\Z_2$-valued 1-cocycle, the above allows us to obtain
\begin{align}
 a^3\hcup{1} a^3 
\se{2}
(a\hcup{1} a^3) a^2+ a(a\hcup{1} a^3) a+ a^2(a\hcup{1} a^3) .
\end{align}
Using \eqn{cupkrel}, we find
$a\hcup{1} a^3 \se{2} a^3 \hcup{1} a $, and
\begin{align}
&
a^3 \hcup{1} a \se{2} 
(a\hcup{1}a)a^2 +a(a\hcup{1}a)a +a^2 (a\hcup{1}a)
\se{2} a^3
\end{align}
This allows us to show
\begin{align}
\label{a3a5}
  a^3\hcup{1} a^3 =\Sq^2 a^3 \se{2} a^5 .
\end{align}
More generally, we can show that for $\Z_2$-valued 1-cocycles $a_1$ and $a_2$,
\begin{align}
\label{Sq2xy}
 \Sq^2 (a_1a_2^2) \se{2} a_1 a_2^4.
\end{align}

\section{Group extension and trivialization}

\label{cenext}

Consider an extension of a group $H$
\begin{align}
 A \to G \to H
\end{align}
where $A$ is an Abelian group with group multiplication given by
$x+y \in A$ for $x,y\in A$. 
Such a group extension is denoted by $G=A\gext H$.  
It is
convenient to label the elements in $G$ as $(h,x)$, where $h\in H$ and $x
\in A$.  The group multiplication of $G$ is given by
\begin{align}
 (h_1,x_1)(h_2,x_2)=(h_1h_2,x_1+\al(h_1)\circ x_2+e_2(h_1,h_2)) .
\end{align}
where $e_2$ is a function 
\begin{align}
 e_2: H\times H \to A,
\end{align}
and $\al$ is a function
\begin{align}
 \al: H \to \text{Aut}(A).
\end{align}
We see that group extension is defined via $e_2$ and $\al$.
The associativity 
\begin{align}
&
[(h_1,x_1)(h_2,x_2)](h_3,x_3)
=(h_1,x_1)[(h_2,x_2)](h_3,x_3)]
\end{align}
requires that
\begin{align}
&
x_1+\al(h_1)\circ x_2+e_2(h_1,h_2)+\al(h_1h_2)\circ x_3+e_2(h_1h_2,h_3)
\nonumber\\ 
&=\al(h_1)\circ x_2+\al(h_1)\al(h_2)\circ x_3+\al(h_1)\circ e_2(h_2,h_3)]
\nonumber\\ & \ \ \ \
+x_1+e_2(h_1,h_2h_3)
\end{align}
or
\begin{align}
 \al(h_1)\al(h_2) = \al(h_1h_2)
\end{align}
and
\begin{align}
& e_2(h_1,h_2)-e_2(h_1,h_2h_3)+e_2(h_1h_2,h_3)
\nonumber\\ & 
-\al(h_1)\circ e_2(h_2,h_3)=0.
\end{align}
Such a $e_2$ is a group 2-cocycle $e_2\in H^2(\cB H;A_\al) $, where $H$ has a
non-trivial action on the coefficient $A$ as described by $\al$.  Also, $\al$
is a group homomorphism $\al: H \to \text{Aut}(A)$.  We see that the $A$
extension from $H$ to $G$ is described by a group 2-cocycle $e_2$ and a
homomorphism $\al$.  Thus we can more precisely denote the group extension by
$G=A\gext_{e_2,\al} H$.

Note that the homomorphism $\alpha: H \to \text{Aut}(A)$ is in fact the
action by conjugation in $G$,
\begin{align}
  (h,0)(\one, x)=(h,\alpha(h)\circ x)&=(\one,\alpha(h)\circ x)(h,0),\nonumber\\
  \Rightarrow (h,0)(\one,x)(h,0)^{-1}&=(\one,\alpha(h)\circ x).
\end{align}
Thus, $\alpha$ is trivial if and only if $A$ lies in the center of $G$. This
case is called a central extension, where the action $\alpha$ will be omitted.

Our way to label group elements in $G$:
\begin{align}
 g=(h,x) \in G
\end{align}
defines two projections of $G$:
\begin{align}
 \pi: G\to H, \ \ \ \pi(g)=h,
\nonumber\\
 \si: G\to A, \ \ \ \si(g)=x.
\end{align}
$\pi$ is a group homomorphism while $\si$ is a generic function.
Using the two projections, $g_1g_2=g_3$ can be written as
\begin{align}
&\ \ \ \
 (\pi(g_1),\si(g_1)) (\pi(g_2),\si(g_2))
\nonumber \\
&=[\pi(g_1)\pi(g_2),\si(g_1)+\al(\pi(g_1))\circ \si(g_2)+e_2(\pi(g_1),\pi(g_2)) ]
\nonumber \\
&= (\pi(g_3),\si(g_3))= (\pi(g_1g_2),\si(g_1g_2))
\end{align}
We see that the group cocycle $e_2(h_1,h_2)$ in $H^2(\cB H;A) $
can be pullback to give a group cocycle $e_2(\pi(g_1),\pi(g_2))$ in $H^2(\cB G;A) $, and such a pullback is a coboundary
\begin{align}
\label{e2sig}
e_2(\pi(g_1),\pi(g_2)) = -\si(g_1) +\si(g_1g_2) -\al(\pi(g_1))\circ \si(g_2),
\end{align}
\ie an element in $B^2(\cB G; A_\al)$,
where $G$ has a non-trivial action on the coefficient $A$ as described by $\al$. 

The above result can be put in another form. Consider the homomorphism
\begin{align}
 \vphi: \cB G \to \cB H
\end{align}
where $G=A\gext_{e_2}H$, and $e_2$ is a $A$-valued 2-cocycle on
$\cB H$.  The homomorphism $\vphi$ sends an link of  $\cB G$ labeled by
$a^{G}_{ij}\in G$ to an link of  $\cB H$ labeled by
$a^{H}_{ij}=\pi(a^{G}_{ij})\in H$.  The pullback of $e_2$ by $\vphi$, $\vphi^*
e_2$, is always a coboundary on $\cB G$

The above discussion also works for continuous group, if we only consider a
neighborhood near the group identity $\one$.  In this case, $e_2(h_1,h_2)$ and
$\al(h)$ are continuous functions on such a neighborhood.  But globally,
$e_2(h_1,h_2)$ and $\al(h)$ may not be continuous functions.

\bibliography{../../bib/wencross,../../bib/all,../../bib/publst,./local} 
\end{document}